\newcommand*\xbar[1]{%
  \hbox{%
    \vbox{%
      \hrule height 0.5pt % The actual bar
      \kern0.5ex%         % Distance between bar and symbol
      \hbox{%
        \kern-0.1em%      % Shortening on the left side
        \ensuremath{#1}%
        \kern-0.1em%      % Shortening on the right side
      }%
    }%
  }%
}

\documentclass[12pt]{report}

\usepackage{times}
\usepackage[T1]{fontenc}

\usepackage[letterpaper]{geometry}
 \usepackage[doublespacing]{setspace}
 \usepackage{tocloft}
 \usepackage[rm, tiny, center, compact]{titlesec}
 \usepackage{indentfirst}
 \usepackage{etoolbox}
\usepackage{pict2e}% to improve picture environment
\usepackage{tocvsec2}
 \usepackage[titletoc]{appendix}
 \usepackage{appendix}
 \usepackage{tamuconfig}
\usepackage{color}
\usepackage{amsmath,amscd}%gives simple commutative diagrams but no diagonal arrows

\usepackage{amsmath, amsthm, amssymb} %math and symbols
\usepackage{amsfonts} %this is for things such as \mathbb{}
\usepackage{latexsym}
\usepackage{amsmath,amscd}%gives simple commutative diagrams but no diagonal arrows
\usepackage{graphicx}

\usepackage{makeidx}%to make index
\usepackage[english]{babel}%babel is to change languagues. to load more than one language, use as e.g.
\usepackage{changepage}  %to change margins
\usepackage{color}
\usepackage{pict2e}% to improve picture environment
\usepackage{hyperref}

\usepackage{rotating}

\usepackage{amsmath, amsthm}
\usepackage{amsfonts}

\usepackage{graphicx}

\DeclareGraphicsExtensions{.png}

\graphicspath{ {./graphic/} }

\usepackage{afterpage}

\usepackage{footnote}

\usepackage{longtable}

\usepackage{setspace}

\DeclareMathOperator{\real}{Re}
\DeclareMathOperator{\Hom}{Hom}

\DeclareMathOperator{\coker}{coker}

\DeclareMathOperator{\image}{Im}

\theoremstyle{remark}

\theoremstyle{definition}

\begin{document}

%The title of your document goes here.
%Spacing may need to be adjusted if your title is long
%and pushes the copyright off the page.
\renewcommand{\tamumanuscripttitle}{Topological defects in string theory orbifolds with target spaces $\mathbb{C}/\mathbb{Z}_n$ and $S^1/\mathbb{Z}_2$ }

%Type only Thesis, Dissertation, or Record of Study.
\renewcommand{\tamupapertype}{ Dissertation}

%Your full name goes here, as it is in university records. Check your student record on Howdy if there is any mismatch.
\renewcommand{\tamufullname}{Yaniel Cabrera}

%The degree title goes here. See the OGAPS site for more info.
\renewcommand{\tamudegree}{Doctor of Philosophy}
\renewcommand{\tamuchairone}{Melanie Becker}

% Uncomment out the next line if you have co-chairs.  You will also need to edit the titlepage.tex file.
%\newcommand{\tamuchairtwo}{Additional Chair Name}
\renewcommand{\tamumemberone}{Christopher Pope}
\newcommand{\tamumembertwo}{Teruki Kamon}
\newcommand{\tamumemberthree}{Stephen Fulling}
\renewcommand{\tamudepthead}{Peter McIntyre}

%Type only May, August, or December.
\renewcommand{\tamugradmonth}{August}
\renewcommand{\tamugradyear}{2017}
%Your department name goes here.
\renewcommand{\tamudepartment}{Physics}

%%%%%%%%%%%%%%%%%%%%%%%%%%%%%%%%%%%%%%%%%%%%%%%%%%%
%
%  New template code for TAMU Theses and Dissertations starting Fall 2016.  
%
%
%  Author: Sean Zachary Roberson
%	 Version 3.17.01
%  Last updated 9/12/2016
%
%%%%%%%%%%%%%%%%%%%%%%%%%%%%%%%%%%%%%%%%%%%%%%%%%%%

%%%%%%%%%%%%%%%%%%%%%%%%%%%%%% 
%% TITLE PAGE
%% The values get updated automatically.  Please do not make changes to this file other than adding/deleting committee members where necessary.
%%%%%%%%%%%%%%%%%%%%%%%%%%%%%%

\providecommand{\tabularnewline}{\\}

\begin{titlepage}
\begin{center}
\MakeUppercase{\tamumanuscripttitle}
\vspace{4em}

A \tamupapertype

by

\MakeUppercase{\tamufullname}

\vspace{4em}

\begin{singlespace}

Submitted to the Office of Graduate and Professional Studies of \\
Texas A\&M University \\

in partial fulfillment of the requirements for the degree of \\
\end{singlespace}

\MakeUppercase{\tamudegree}
\par\end{center}
\vspace{2em}
\begin{singlespace}
\begin{tabular}{ll}
 & \tabularnewline
& \cr
% If you have Co-Chairs comment out the 'Chair of Committee' line below and uncomment the 'Co-Chairs of Committee' line.
Chair of Committee, & \tamuchairone\tabularnewline
%Co-Chairs of Committee, & \tamuchairone\tabularnewline & \tamuchairtwo\tabularnewline
Committee Members, & \tamumemberone\tabularnewline
 & \tamumembertwo\tabularnewline
 & \tamumemberthree\tabularnewline
Head of Department, & \tamudepthead\tabularnewline

\end{tabular}
\end{singlespace}
\vspace{3em}

\begin{center}
\tamugradmonth \hspace{2pt} \tamugradyear

\vspace{3em}

Major Subject: \tamudepartment \par
\vspace{3em}
Copyright \tamugradyear \hspace{.5em}\tamufullname 
\par\end{center}
\end{titlepage}
\pagebreak{}

 % This is simply a file that formats and adds your titlepage, please do not edit this unless you have a specific need. .
%%%%%%%%%%%%%%%%%%%%%%%%%%%%%%%%%%%%%%%%%%%%%%%%%%%
%
%  New template code for TAMU Theses and Dissertations starting Fall 2016.  
%
%  Author: Sean Zachary Roberson
%	 Version 3.17.01
%  Last updated 1/10/2017
%
%%%%%%%%%%%%%%%%%%%%%%%%%%%%%%%%%%%%%%%%%%%%%%%%%%%
%%%%%%%%%%%%%%%%%%%%%%%%%%%%%%%%%%%%%%%%%%%%%%%%%%%%%%%%%%%%%%%%%%%%%
%%                           ABSTRACT 
%%%%%%%%%%%%%%%%%%%%%%%%%%%%%%%%%%%%%%%%%%%%%%%%%%%%%%%%%%%%%%%%%%%%%

\chapter*{ABSTRACT}
\addcontentsline{toc}{chapter}{ABSTRACT} % Needs to be set to part, so the TOC doesnt add 'CHAPTER ' prefix in the TOC.

\pagestyle{plain} % No headers, just page numbers
\pagenumbering{roman} % Roman numerals
\setcounter{page}{2}

We study conformal defects in two important examples of string theory orbifolds. First, we show that topological defects in the language of Landau-Ginzburg models carry information about the RG flow between the non-compact orbifolds $\mathbb{C}/\mathbb{Z}_d$. Such defects are shown to correctly implement the bulk-induced RG flow on the boundary. Secondly, we study what the possible conformal defects are between the $c=1$ bosonic 2D conformal field theories with target space $S^1/\mathbb{Z}_2$. The defects cataloged here are obtained from boundary states corresponding to D-branes in the $c=2$ free theory with target space $S^1/\mathbb{Z}_2 \times S^1/\mathbb{Z}_2$. Via the unfolding procedure, such boundary states are later mapped to defects between the circle orbifolds. Furthermore, we compute the algebra of the topological class of defects at different radii.

\pagebreak{}

%%%%%%%%%%%%%%%%%%%%%%%%%%%%%%%%%%%%%%%%%%%%%%%%%%%
%
%  New template code for TAMU Theses and Dissertations starting Fall 2016.  
%
%  Author: Sean Zachary Roberson
%	 Version 3.17.01
%  Last updated 1/10/2017
%
%%%%%%%%%%%%%%%%%%%%%%%%%%%%%%%%%%%%%%%%%%%%%%%%%%%

%%%%%%%%%%%%%%%%%%%%%%%%%%%%%%%%%%%%%%%%%%%%%%%%%%%%%%%%%%%%%%%%%%%%%%
%%                           DEDICATION
%%%%%%%%%%%%%%%%%%%%%%%%%%%%%%%%%%%%%%%%%%%%%%%%%%%%%%%%%%%%%%%%%%%%%
\chapter*{DEDICATION}
\addcontentsline{toc}{chapter}{DEDICATION}  % Needs to be set to part, so the TOC doesnt add 'CHAPTER ' prefix in the TOC.

\begin{center}
\vspace*{\fill}
To my family.
\vspace*{\fill}
\end{center}

\pagebreak{}

%%%%%%%%%%%%%%%%%%%%%%%%%%%%%%%%%%%%%%%%%%%%%%%%%%%
%
%  New template code for TAMU Theses and Dissertations starting Fall 2016.
%
%  Author: Sean Zachary Roberson
%	 Version 3.17.01
%  Last updated 1/10/2017
%
%%%%%%%%%%%%%%%%%%%%%%%%%%%%%%%%%%%%%%%%%%%%%%%%%%%

%%%%%%%%%%%%%%%%%%%%%%%%%%%%%%%%%%%%%%%%%%%%%%%%%%%%%%%%%%%%%%%%%%%%%%
%%                           ACKNOWLEDGMENTS
%%%%%%%%%%%%%%%%%%%%%%%%%%%%%%%%%%%%%%%%%%%%%%%%%%%%%%%%%%%%%%%%%%%%%
\chapter*{ACKNOWLEDGMENTS}
\addcontentsline{toc}{chapter}{ACKNOWLEDGMENTS}  % Needs to be set to part, so the TOC doesnt add 'CHAPTER ' prefix in the TOC.

\indent 

I would like to give thanks to my Ph.D. adviser Dr. Melanie Becker for being a great mentor during my time at Texas A\&M University, and all the invaluable advice she has given me. I am also very grateful to Dr. Daniel Robbins for being an integral collaborator in the projects composing this dissertation. 
I would also like to give thanks to my committee members Dr. Christopher Pope,  Dr. Teruki Kamon, and Dr. Stephen A. Fulling for participating in my preliminary and final defense exams. Also a special thanks to Dr. Katrin Becker for agreeing to be part of my defense exam. I am grateful to Ilka Brunner for her helpful feedback on the Landau-Ginzburg work.

I extend my thanks to Ning Su with whom I shared many insightful discussions and previous work. I would like to take this moment to thank Sebastian Guttenberg, Jakob Palmkvist, Andy Royston, and William D. Linch III, Yaodong Zhu, Sunny Guha, and Zhao Wang  who have been great colleagues. Special thanks to Ilarion V. Melnikov for  introducing me to the beautiful theory of defects.

\pagebreak{}
%\include{data/contributors}
%\include{data/nomenclature}

%%%%%%%%%%%%%%%%%%%%%%%%%%%%%%%%%%%%%%%%%%%%%%%%%%%
%
%  New template code for TAMU Theses and Dissertations starting Fall 2016.
%
%  Author: Sean Zachary Roberson 
%	 Version 3.17.01
%  Last updated 1/10/2017
%
%%%%%%%%%%%%%%%%%%%%%%%%%%%%%%%%%%%%%%%%%%%%%%%%%%%
%%%%%%%%%%%%%%%%%%%%%%%%%%%%%%%%%%%%%%%%%%%%%%%%%%%%%%%%%%%%%%%%%%%%%%
%%       TABLE OF CONTENTS
%%%%%%%%%%%%%%%%%%%%%%%%%%%%%%%%%%%%%%%%%%%%%%%%%%%%%%%%%%%%%%%%%%%%%
% single-space sections in Table of Contents  - commented in version 1.7
%\renewcommand{\cftsecafterpnum}{\vskip0.5\baselineskip}
%\renewcommand{\cftsubsecafterpnum}{\vskip0.5\baselineskip}
%\renewcommand{\cftsubsubsecafterpnum}{\vskip0.5\baselineskip}
%%%%%%%%%%%%%%%%%%%%%%%%%%%%%%%%%%%%%%%%%%%%%%%%%%%

\phantomsection
\addcontentsline{toc}{chapter}{TABLE OF CONTENTS}  

\begin{singlespace}
\renewcommand\contentsname{\normalfont} {\centerline{TABLE OF CONTENTS}}

\setcounter{tocdepth}{4} % This puts \subsubsection[]{×} in your List of Tables.  The default is 3.

%%%%%%%%%%%%%  Adds Page above the page number in TOC
\setlength{\cftaftertoctitleskip}{1em}
\renewcommand{\cftaftertoctitle}{%
\hfill{\normalfont {Page}\par}}

\tableofcontents

%\addtocontents{toc}{\protect\afterpage{~\hfill\normalfont{Page}\par\medskip}}
\end{singlespace}

\pagebreak{}

%%%%%%%%%%%%%%%%%%%%%%%%%%%%%%%%%%%%%%%%%%%%%%%%%%%%%%%%%%%%%%%%%%%%%%
%%                           LIST OF FIGURES
%%%%%%%%%%%%%%%%%%%%%%%%%%%%%%%%%%%%%%%%%%%%%%%%%%%%%%%%%%%%%%%%%%%%%

\phantomsection
\addcontentsline{toc}{chapter}{LIST OF FIGURES}  

\renewcommand{\cftloftitlefont}{\center\normalfont\MakeUppercase}

\setlength{\cftbeforeloftitleskip}{-12pt} %% Positions the LOF title vertically to match the chapter titles
\renewcommand{\cftafterloftitleskip}{12pt}

\renewcommand{\cftafterloftitle}{%
\\[4em]\mbox{}\hspace{2pt}FIGURE\hfill{\normalfont Page}\vskip\baselineskip}

\begingroup

\begin{center}
\begin{singlespace}
%% These values make the lof table entries appear double spaced between.
\setlength{\cftbeforechapskip}{0.4cm}
\setlength{\cftbeforesecskip}{0.30cm}
\setlength{\cftbeforesubsecskip}{0.30cm}
\setlength{\cftbeforefigskip}{0.4cm}
\setlength{\cftbeforetabskip}{0.4cm} 

\listoffigures

\end{singlespace}
\end{center}

%\pagebreak{}

%%%%%%%%%%%%%%%%%%%%%%%%%%%%%%%%%%%%%%%%%%%%%%%%%%%%%%%%%%%%%%%%%%%%%%
%%                           lIST OF TABLES
%%%%%%%%%%%%%%%%%%%%%%%%%%%%%%%%%%%%%%%%%%%%%%%%%%%%%%%%%%%%%%%%%%%%%%
%
%\phantomsection
%\addcontentsline{toc}{chapter}{LIST OF TABLES}  

\renewcommand{\cftlottitlefont}{\center\normalfont\MakeUppercase}

\setlength{\cftbeforelottitleskip}{-12pt} %% Positions the LOT title vertically to match the chapter titles

%Note that the similar parameter in the LOF is 12pt; this
%is intentional to make the spacing between the headers
%and the first entry look consistent.
\renewcommand{\cftafterlottitleskip}{1pt}

\renewcommand{\cftafterlottitle}{%
\\[4em]\mbox{}\hspace{2pt}TABLE\hfill{\normalfont Page}\vskip\baselineskip}

\begin{center}
\begin{singlespace}

%% These values make the lot table entries appear double spaced between.
\setlength{\cftbeforechapskip}{0.4cm}
\setlength{\cftbeforesecskip}{0.30cm}
\setlength{\cftbeforesubsecskip}{0.30cm}
\setlength{\cftbeforefigskip}{0.4cm}
\setlength{\cftbeforetabskip}{0.4cm}

%\listoftables 

\end{singlespace}
\end{center}
\endgroup
\pagebreak{}  % Need this for the pagenumbering to be correct.   % This is simply a file that formats and adds your toc, lof, and lot, please do not edit this unless you have a specific need.

%%%%%%%%%%%%%%%%%%%%%%%%%%%%%%%%%%%%%%%%%%%%%%%%%%%
%
%  New template code for TAMU Theses and Dissertations starting Fall 2016.
%
%  Author: Sean Zachary Roberson 
%	 Version 3.17.01
%  Last updated 1/10/2017
%
%%%%%%%%%%%%%%%%%%%%%%%%%%%%%%%%%%%%%%%%%%%%%%%%%%%

%%%%%%%%%%%%%%%%%%%%%%%%%%%%%%%%%%%%%%%%%%%%%%%%%%%%%%%%%%%%%%%%%%%%%%
%%                           SECTION I
%%%%%%%%%%%%%%%%%%%%%%%%%%%%%%%%%%%%%%%%%%%%%%%%%%%%%%%%%%%%%%%%%%%%%

\pagestyle{plain} % No headers, just page numbers
\pagenumbering{arabic} % Arabic numerals
\setcounter{page}{1}

\chapter{\uppercase {Introduction}}\label{intro}

Conformal field theories in two dimensions (2D) initially encompassed the study of  theories which are invariant under mappings which transform the metric by an overall scale factor. In local coordinates, this transformation is given by
\begin{equation}
g'_{\rho \sigma}(x') \frac{\partial x'^\rho}{\partial x^\mu} \frac{\partial x'^\sigma}{\partial x^\nu} = \Omega(x)g_{\mu\nu}(x),
\end{equation}
where $x'=f(x)$, with $f:(U,x)\rightarrow (V,x')$ and $U, V\subset \mathbb{R}^2$. Differently from higher dimensions, in 2D the set of all such local conformal transformations form an infinite algebra: the Virasoro algebra. The aesthetics and power of 2D CFTs are greatly derived from the fact that these conformal mappings are holomorphic (and antiholomorphic) functions on the complex plane. Works by Belavin, Polyakov, and Zamolodchikov \cite{belavin} developed much of the general formalism for 2D CFTs. 

Later on, consideration was given to 2D CFTs on surfaces with boundaries such as the upper-half plane and the infinite strip. Most of the seminal work on systems with boundaries was developed by John Cardy \cite{cardy}. The study of boundary CFTs (BCFTs) has been an important field of research in 2D theories with wide applications to D-branes in string theory \cite{Recknagel:1998ih, Brunner:1999, Gaberdiel:2001zq, Gaberdiel:2008rk}. The introduction of boundaries to the worldsheet has the consequence of reducing the amount of conformal symmetries allowed; and more interestingly, it mixes the holomorphic and antiholomorphic degrees of freedom. The reduction of conformal symmetries follows from the need to consider only those transformations which preserve the boundary. A deeper consequence of boundaries is the appearance of new elements in the Hilbert space called boundary states which are non-existent for a CFT purely living in the bulk. Such boundary states form the boundary Hilbert space of the CFT. Corresponding fields which reside on the boundary have their own OPE algebra. OPEs can also be taken between bulk fields and the fields living on the boundary. That is, boundaries give new physics.

After boundaries, the next step was taken by Affleck \cite{affleck} by studying defects. Differently from boundaries, the defects considered by Affleck were curves with field content to either side. One can think of a boundary as a defect where one of the theories is the trivial (empty) theory. Again, new types of fields and corresponding states appear with the introduction of defects. Very importantly, defects can be mapped to boundaries where the bulk theory is the tensor product of the two theories flanking the original defect. One way to think of these defects is as boundary conditions consistent with respect to both theories.

\begin{figure}[h]
\centering
\includegraphics[width=100mm,scale=0.5]{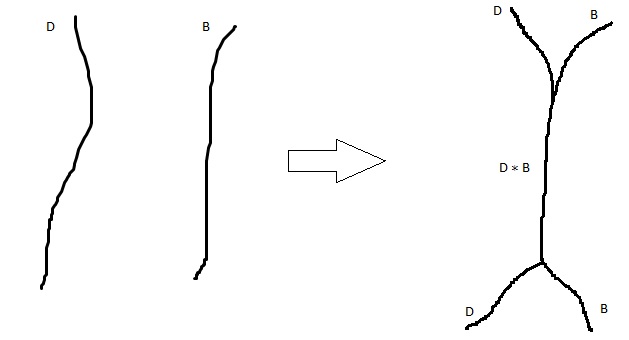}
\caption{Fusion of defects}
\label{fig:tamu-fig1}
\end{figure}

\begin{figure}[h]
\centering
\includegraphics[width=100mm,scale=0.5]{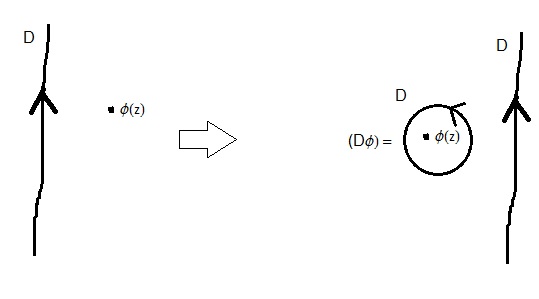}
\caption{Defect action on field insertions}
\label{fig:tamu-fig2}
\end{figure}

More recently, 2D theories have been shown to contain a rich class of new defects \cite{bachas02,bachas07,fuchs07,gaiotto12,Quella:2002CT, brunner03, brunner07, Konechny:2015qla, Graham:2003nc, Fuchs:2015ska}. In this sense, a \emph{defect} is a one-dimensional object in 2D theories, and more generally a submanifold of positive co-dimension in higher dimensional spaces. These objects are also defects in the sense of those considered by Affleck since they are one-dimensional curves separating two theories. But that is where the similarities end. More than simply domain walls, or consistent boundary conditions, this new type of defects have the following properties:
\begin{itemize}
\item Defects have a binary operation called fusion where two defects are brought together to form a third defect as shown in Figure \ref{fig:tamu-fig1}. For two defects $D$ and $B$, we denote this operation by $(D,B)=D*B$.
\item Via the fusion operation defects form representations of the symmetries present in the theory.
\end{itemize}
\begin{itemize}
\item  Defects encode information about dualities and mappings between the theories to either side of the defect.
\item A defect $D$ gives rise to a linear map $\widehat D: \mathcal{H}_1 \rightarrow \mathcal{H}_2$ between the Hilbert spaces of two theories separated by $D$ \cite{frohlich09}. By first choosing an orientation for the defect,  we can move $D$ across a field insertion and pinch the defect to wrap it around the insertion point, as shown in Figure \ref{fig:tamu-fig2}.
\end{itemize}
 
An important class of defects are those called \emph{topological defects} which commute with the field insertion of the energy-momentum tensor. That is, a defect $D$ is topological if 
\begin{equation}
\widehat D T_1(z) = T_2(z)\widehat D,
\end{equation}
where $\widehat D: \mathcal{H}_1 \rightarrow \mathcal{H}_2$ is the representation of the defect $D$ as an operator intertwining the Hilbert spaces of the adjacent CFTs. In this case, the defect can be deformed through the worldsheet without affecting the values of the correlation functions, as long as it does not cross an operator insertion point. Hence the name ``topological''. To see that this is the case, we note that commuting with the energy-momentum tensor $T(z)$ means that the defect commutes with the Virasoro generators (which are the elements of the infinite conformal algebra in 2D). Topological defects form a subset of a larger class of interest called \emph{conformal defects} whose elements commute with the difference of the holomorphic and antiholomorphic components of the energy-momentum tensor \cite{fuchs07},
\begin{equation}
\widehat D (T_1(z)- \xbar T_1(z)) = (T_2(z)- \xbar T_2(z)) \widehat D.
\end{equation}

The behavior of boundary degrees of freedom under the renormalization group (RG) flow represents a problem in both string theory and condensed matter physics which is not fully understood (see \cite{dorey00}, \cite{keller07}, \cite{hori04}, \cite{fredenhagen06} and references therein). A new approach consists of utilizing defects to bring the RG flow from the bulk to the boundary. This technique was exploited in \cite{brunner07a} within the framework of Landau-Ginzburg (LG) models to study the boundary RG flow between the two-dimensional orbifolds $\mathcal{M}_{d-2}/\mathbb{Z}_d$, where $\mathcal{M}_{d-2}$ are the supersymmetric minimal models.  RG flow defects were also constructed in \cite{gaiotto12} between consecutive Virasoro minimal models in the bulk.

Defects in LG models have a general description in terms of matrix factorizations which allows us to construct examples of boundaries and defects. Also, the language of matrix factorization contains a general operation called the tensor product of matrix factorizations which gives a recipe to compute the fusion of any two LG defects \cite{brunner07,brunner07a, enger05}. The theory of defects in LG models is versatile because it provides direct information on other theories which are not necessarily LG models. This fact follows because LG models can be mapped to other interesting theories via different RG flows or mirror symmetry \cite{hori00, vafa01}. In this article we are particularly interested in the non-compact orbifold $\mathbb{C}/\mathbb{Z}_d$. This orbifold is not target-space supersymmetric, but it exhibits $\mathcal{N}=2$ worldsheet supersymmetry. 

In LG models, defects are topological provided that a topological twist has been performed \cite{brunner07}. There are two types of twists that render $N=2$ theories topological \cite{hori00a} and they are called A-twist and B-twist. In the presence of boundaries or defects, only half of the total $(2,2)$ supersymmetry is preserved, and similar to the topological twist there are two ways to break half of the supersymmetry. The remaining symmetry is called A-type or B-type depending on which supersymmetric charges are kept. The generators for each respective supersymmetry are
\begin{equation}\label{abgenerators}
\begin{split}
&(A) \ \ \ \ \xbar Q_A:=\xbar Q_+ + \operatorname{e}^{i\alpha}Q_-, \ \ \ \ Q_A:=Q_+ +\operatorname{e}^{-i\alpha}\xbar Q_-,\\
&(B) \ \ \ \ \xbar Q_B:=\xbar Q_+ +\operatorname{e}^{i\beta}\xbar Q_-, \ \ \ \ Q_B:= Q_+ +\operatorname{e}^{-i\beta} Q_-,
\end{split}
\end{equation}
where $\alpha$ and $\beta$ are real numbers, and $Q_\pm$, $\xbar Q_\pm$ are the generators of the full $\mathcal{N}=(2,2)$ supersymmetry.

A topological A(B)-twist must be done alongside A(B)-type supersymmetry if the boundaries and defects are to be supersymmetric and topological. In this dissertation, it is assumed throughout that the LG models are already topological. In each case, there is a BRST-operator $Q_A$ or $Q_B$ which characterizes the physical degrees of freedom at the boundary.

In this dissertation, the machinery of matrix factorizations for defects is applied to the non-compact orbifold $\mathbb{C}/\mathbb{Z}_d$ which is the archetype for string theory on
\begin{equation}\label{arch}
\mathbb{R}^{d-1,1}\times\mathbb{R}^{10-d}/G ,
\end{equation}
where $G$ is some discrete $SO(d-10)$ subgroup \cite{harvey01}. This important model is linked to the LG language in two ways that are exploited in this note. First, by introducing superspace variables the fermionic string theory on $\mathbb{C}/\mathbb{Z}_d$ can be viewed as the orbifold of a LG model with zero superpotential. And second, we also go from the $\mathbb{C}/\mathbb{Z}_d$ theory to a twisted LG model description using mirror symmetry as discussed in \cite{vafa01}.

In this dissertation we extend the work of \cite{brunner07a} which describes the boundary RG flow in LG models and supersymmetric minimal models in terms of topological defects. Our work generalizes the results of \cite{brunner07a} to the non-supersymmetric case of the non-compact $\mathbb{C}/\mathbb{Z}_n$ theories. The orbifold $\mathbb{C}/\mathbb{Z}_d$ is physically relevant because it is the simplest model to study  tachyon condensation \cite{adams01}; in (\ref{arch}), the tachyons are closed strings localized at the fixed points of the orbifold group action. Techniques to study the RG flow in these models have been considered in \cite{vafa01, harvey01}.

To study the problem at hand, we consider the $\mathbb{C}/\mathbb{Z}_d$ orbifold theory on the upper-half plane $\Sigma = \left\{(x^0,x^1)\in \mathbb{R}^2 \ | \ x^0 \geq 0\right\}$ with B-type supersymmetry. Inserting the identity defect at $x^0=y>0$, we can perturb the theory over $x\geq y$. Letting the perturbation drive the theory to the IR we obtain a setup describing the IR theory in the bulk while near the boundary we still have the UV theory, with a defect $D$ sitting at the interface $x^0=y$. The next step is to take the RG flow to the boundary via the limit $y\rightarrow 0$. In terms of defects, this limit gives the fusion of the boundary $B$ and the defect $D$.

Aside from the approach described above of utilizing matrix factorizations to represent defects, we also study defects as linear operators between the Hilbert spaces of the given theories. In order to talk about defects in the operator representation, we must first deal with boundary states in product theories. The theories of conformal boundaries and that of defects are intertwined. Not only are boundaries a class of trivial defects, but via the folding trick defects are mapped to boundaries \cite{affleck}. The folding trick is a powerful tool in QFT where one considers two theories, say $\text{CFT}_1$ and $\text{CFT}_2$, separated by a defect or domain wall. Then the theory $\text{CFT}_1\oplus\text{CFT}_2$ in the presence of the defect is equivalent to the folded theory $\text{CFT}_1\otimes\xbar{\text{CFT}}_2$, where the bar means interchanging the left and right movers \cite{affleck, bachas07}. The defect is mapped to a boundary in the $\text{CFT}_1\otimes\xbar{\text{CFT}}_2$ theory. The folding trick also applies to non-CFT theories such as general LG models \cite{brunner07}.

In this dissertation we study the boundary states of the $c=2$ bosonic theory taking value in $(\mathbb{S}^1/\mathbb{Z}_2)^2$ and apply the unfolding procedure in order to obtain defects between the orbifolds $S^1_R/\mathbb{Z}_2$, where $R$ is the radius of the circle. Unfolding is the inverse of the folding trick. In this procedure elements of the boundary theory $\mathcal{H}^{1\otimes 2}_\partial$ of some bulk theory $\mathcal{H}_1 \otimes \mathcal{H}_2$ are mapped to defects between the theories with Hilbert spaces $\mathcal{H}_1$ and $\mathcal{H}_2$. This method was applied in \cite{bachas02, bachas07} in order to study the defects between bosonic $S^1$-valued theories. Defects in the context of the compact boson were also considered in \cite{fuchs07} but without the unfolding map. Here we rely on the unfolding procedure due to consistency. In general it is difficult to have a criteria describing a consistent set of defects. By construction, the criteria met in this dissertation is that the defects map to consistent boundary states in the product theory.

Our work on topological defects for the
superstring on $\mathbb{C}/\mathbb{Z}_d$ provides a novel approach to the question of tachyon condensation. One of the main reasons supersymmetry enters string theory is to attain stable, tachyon-free spacetimes. As previously mentioned, the model $\mathbb{C}/\mathbb{Z}_d$ is non-supersymmetric and contains closed string tachyons (the $\mathcal{N}=(2,2)$ supersymmetry present is on the worldsheet only). The question of tachyon condensation is a difficult one which has been studied in a few cases \cite{vafa01, harvey01, adams01}. At the worldsheet level, the process of tachyon condensation is due to RG flows generated by perturbations of the starting point CFT \cite{adams01}. So far not much is known about these bulk RG flows and the present methods to describe them are laborious and complicated. The new defects presented in this dissertation provide a new perspective on the bulk and boundary RG flows for the  $\mathbb{C}/\mathbb{Z}_d$ models as well as a simpler way to do computations without the need for regularization schemes.

More technically, our results for $\mathbb{C}/\mathbb{Z}_d$ show that matrix factorizations of $W(X)=0$ are well defined objects which can successfully describe defects for the non-compact orbifold in question. Furthermore, it is shown that the RG flow between the $\mathbb{C}/\mathbb{Z}_d$ models can be described in terms of these defects. An important application of our work, and one we show here, is the description of the boundary RG flow of these theories in terms of the fusion product of boundaries and RG defects.  

The significance of our work on the compact orbifold $S^1/\mathbb{Z}_2$ comes from the present dearth of non-trivial theories whose spectrum of consistent defects is known: until this work, the compact boson was the only theory whose spectrum of defects was written down. Taken together, our work on the $S^1/\mathbb{Z}_2$ theory and those of \cite{bachas07, fuchs07} on the compact boson will provide a very complete picture of the possible defects between elements of the same branches of $c=1$, 2D CFT phase space \cite{Ginsparg:1988ui}. Very importantly, since we have covered the part of the spectrum which contains the twisted degrees of freedom, using our results it should be doable to build defects between the different branches of the $c=1$ phase space. That is, defects between $S^1/\mathbb{Z}_2$ and $S^1$. It is important to emphasize that the D-branes studied in this work carry their own weight aside from their utility to derive defects. D-branes in compact orbifolds have not been well studied in the literature \cite{Brunner:1999} so our work provides more examples in the boundary CFT formalism.

The dissertation is structured as follows. In chapter 2, defects and boundary conditions are developed for the Type I superstring on a worldsheet with boundaries and taking values in the cone $\mathbb{C}/\mathbb{Z}_d$. We start by reviewing $\mathcal{N}=(2,2)$ theories in the presence of boundaries. The introduction of a boundary reduces the supersymmetry and we are left with either A-type or B-type supersymmetry which are subsets of the full $\mathcal{N}=(2,2)$ symmetry. We review the algebraic language of matrix factorizations suitable for B-type boundaries and defects. The geometrical description of wave-front trajectories for A-type D-branes (A-branes) for supersymmetric sigma models is also reviewed with an emphasis on LG models. Both descriptions are related by mirror symmetry mappings.

Proceeding the reviews, a superspace description of $\mathbb{C}/\mathbb{Z}_d$ as a LG model with zero superpotential is developed to obtain a description of boundary conditions and defects in terms of matrix factorizations of $W(X)=0$. We show that suitable defects exist such that they divide the UV and IR theories. In the case of $\mathbb{C}/\mathbb{Z}_d$ we can keep track of both RG endpoints by means of the chiral ring. The addition of chiral terms to the Lagrangian to induce an RG flow also produces deformations to the chiral ring of the theory. The resulting chiral ring at each endpoint of the flow characterizes the theory in the UV or IR. Lastly, we show that RG defects can be used to work out the boundary RG flows of these theories. We work with the mirror theories of the non-compact orbifolds which are orbifolded LG theories with non-zero superpotentials. The B-type boundary conditions have a dual description in terms of A-branes. The action of the RG B-type defects on the B-type boundaries is compared with the RG flow as described by the dual A-type branes. This comparison demonstrates that indeed the posited RG defects enforce the RG flow on the boundary without a need for regularization techniques.

In chapter 3 we move away from matrix factorizations as descriptions of D-branes and instead the BCFT formalism is used. We start with a review of Affleck and Oshikawa's boundary theory construction for the single boson taking values in $S^1/\mathbb{Z}_2$ \cite{affleck}. Following this method, we work out possible boundary states for the free bosonic theory on $S^1/\mathbb{Z}_2 \times  S^1/\mathbb{Z}_2$. In the BCFT formalism, D-branes are represented as coherent states which solve conformal boundary operator equations. In the free theory, this problem can be reduced to searching for elements of the Hilbert space which are consistent with boundary conditions of the bulk fields.

Lastly, in chapter 4 the D-branes in the product theory are mapped to conformal defects between the $S^1/\mathbb{Z}_2$ bosonic theories. In this chapter we review the unfolding map used by \cite{bachas07} which gives a correspondence between D-branes (i.e., boundary states) and defects. The unfolding map is employed here as a direct way to obtain the possible spectrum of classes of defects. From the spectrum of defects, those defects which are transmissive or totally reflective are identified. We finish this chapter by computing the fusions of those transmissive defects which are topological. These products show that the topological defects form a closed algebra. We conclude in chapter 5 with a summary of the work presented here and some remarks on possible future directions.

%%%%%%%%%%%%%%%%%%%%%%%%%%%%%%%%%%%%%%%%%%%%%%%%%%%%%%%

%%%%%%%%%%%%%%%%%%%%%%%%%%%%%%%%%%%%%%%%%%%%%%%%%%%
%
%  New template code for TAMU Theses and Dissertations starting Fall 2016.
%
%  Author: Sean Zachary Roberson
%	 Version 3.17.01
%  Last updated 1/10/2017
%
%%%%%%%%%%%%%%%%%%%%%%%%%%%%%%%%%%%%%%%%%%%%%%%%%%%

%%%%%%%%%%%%%%%%%%%%%%%%%%%%%%%%%%%%%%%%%%%%%%%%%%%%%%%%%%%%%%%%%%%%%%%
%%%                           SECTION II
%%%%%%%%%%%%%%%%%%%%%%%%%%%%%%%%%%%%%%%%%%%%%%%%%%%%%%%%%%%%%%%%%%%%%%

%\chapter{\uppercase{Defects between $\mathbb{C}/\mathbb{Z}_n$ orbifolds}}

%\section[hello]{hello\footnote{another author}}

\chapter[DEFECTS BETWEEN $\mathbb{C}/\mathbb{Z}_n$ ORBIFOLDS]{DEFECTS BETWEEN $\mathbb{C}/\mathbb{Z}_n$ ORBIFOLDS \footnote{Reprinted with permission from ``Defects and boundary RG flows in $\mathbb{C}/\mathbb{Z}_d$'' by M. Becker, Y. Cabrera, and D. Robbins, 2017, \emph{JHEP}  2017 : 7, Copyright [2017] by the authors.}}

This chapter develops a description of topological defects for the supersymmetric string with target space $\mathbb{C}/\mathbb{Z}_n$. This language is used to find defects which successfully encode the RG flow between these theories.  

\section{$\mathcal{N}=(2,2)$ supersymmetry on $\mathbb{R}^2$}

In this section we review the $\mathcal{N}=(2,2)$ supersymmetric string in 1+1 dimensions with complex-valued free fields, as well as general aspects of the $(2,2)$ algebra. We follow the conventions of \cite{hori02} but we will restrict to definitions and concepts which are necessary for this work.

The action for the RNS supersymmetric model is 
\begin{equation}\label{rns}
S= \displaystyle \int_{\mathbb{R}^2} d^2x \left(|\partial_0 \phi|^2 - |\partial_1 \phi|^2 + i\bar \psi_-(\partial_0 +\partial_1)\psi_- i\bar \psi_+(\partial_0 -\partial_1)\psi_+\right),
\end{equation}
where $\phi$ is a scalar and the fields $\psi_\pm$ form a Dirac fermion. The bar symbol means complex conjugation in this context. The action is left invariant under the following supersymmetry transformations
\begin{equation}
\delta \phi = \epsilon_+\psi_--\epsilon_-\psi_+ \ \ , \ \ \delta\psi_\pm= \pm i \bar \epsilon _\mp (\partial_0\pm\partial_1)\phi,
\end{equation}
\begin{equation}
\delta \bar\phi = -\bar\epsilon_+\bar \psi_-+ \bar\epsilon_-\bar \psi_+ \ \ , \ \ \delta\bar \psi_\pm= \mp i  \epsilon _\mp (\partial_0\pm\partial_1)\bar\phi,
\end{equation}
where the $\epsilon$ parameters are fermionic. The above symmetries give rise to two left and two right conserved supercharges, $Q_\pm$ and $\xbar Q_\pm$ respectively. In terms of the bulk fields, these charges are given by
\begin{equation}
Q_\pm =  \displaystyle \int dx^1 (\partial_0 \pm \partial_1)\bar\phi \psi_\pm,
\end{equation}
\begin{equation}
\xbar Q_\pm=  \displaystyle\int dx^1\bar \psi_\pm (\partial_0 \pm \partial_1)\phi .
\end{equation}
The RNS model in Eq. (\ref{rns}) also exhibits two additional $U(1)$ symmetries given by the following transformations
\begin{equation}\label{u1vComponent}
e^{i\alpha F_V}:(\phi,\psi_\pm,\bar\psi_\pm) \rightarrow (\phi,e^{-i\alpha}\psi_\pm,e^{i\alpha}\bar\psi_\pm) ,
\end{equation}
\begin{equation}\label{u1aComponent}
e^{i\alpha F_A}:(\phi,\psi_\pm,\bar\psi_\pm) \rightarrow (\phi,e^{\pm i\alpha}\psi_\pm,e^{\mp i\alpha}\bar\psi_\pm).
\end{equation}
The first is called \emph{vector R-rotation} and the second is called \emph{axial R-rotation}; the respective conserved supercharges are 
\begin{equation}
F_V= \displaystyle \int dx^1(\bar \psi_-\psi_-+\bar \psi_+ \psi_+) \ \ , \ \ F_A= \displaystyle \int dx^1(-\bar \psi_-\psi_-+\bar \psi_+ \psi_+).
\end{equation}

Under the usual canonical quantization relations
\begin{equation}
[\phi(x^1),\partial_0 \bar \phi(y^1)]=\delta(x^1-y^1) \ \ \ , \ \ \ \left\{ \psi_\pm(x^1),\bar\psi_\pm(y^1) \right\}=\delta(x^1-y^1),
\end{equation}
the supersymmetry charges together with the Hamiltonian $H$, the momentum generator $P$, and the angular momentum generator $M$ satisfy the following algebra
\begin{equation}
\left\{ Q_\pm, \xbar Q_\pm\right\} = H\pm P, \ \ \ [iM, Q_\pm] =\mp Q_\pm, \ \ \ [iM, \xbar Q_\pm] =\mp \xbar Q_\pm,
\end{equation}
\begin{equation}
\left\{ Q_\pm,  Q_\pm\right\}=\left\{ \xbar Q_\pm,  \xbar Q_\pm\right\}=0,
\end{equation}
\begin{equation}
\left\{ Q_+,  Q_-\right\}=\left\{ \xbar Q_+,  \xbar Q_-\right\}=0,
\end{equation}
\begin{equation}
\left\{ \xbar Q_+,  Q_-\right\}=\left\{  Q_+,  \xbar Q_-\right\}=0.
\end{equation}
Including the $U(1)_V$ and $U(1)_A$ R-generators the algebra extends by means of the relations,
\begin{equation}
[iF_V,Q_\pm]=-iQ_\pm, \ \ \ [iF_V,Q_\pm]=iQ_\pm,
\end{equation}
\begin{equation}
[iF_V,Q_\pm]=\mp iQ_\pm, \ \ \ [iF_V,Q_\pm]=\pm iQ_\pm.
\end{equation}

\section{Superspace formalism and the Landau-Ginzburg model}

This section constitutes a review of the superspace approach to supersymmetric theories including the one of Eq. (\ref{rns}) for the RNS string. Superspace formalism is obtained via the introduction of four new fermionic variables aside of the bosonic spacetime coordinates. The new worldsheet manifold, which is called ``superspace'', gives rise to functionals which are manifestly invariant under the $\mathcal{N}=(2,2)$ supersymmery.

The local chart of superspace is given by the coordinates $(x^0,x^1,\theta^\pm,\bar \theta^\pm)$ where the $\theta^\pm$ and $\bar \theta^\pm$ are fermionic coordinates, that is,
\begin{equation}
\theta^a \theta^b = -\theta^b\theta^a, \ \ \ \bar\theta^a \bar \theta^b = -\bar \theta^b\bar \theta^a, \ \ \ \theta^a \bar \theta^b = -\bar \theta^b\theta^a.
\end{equation}
These fermionic coordinates are complex, with the bar and non-bar pairs being complex conjugate to each other.

In superspace, the supercharges have the following representation as differential operators,
\begin{equation}
Q_\pm = \frac{\partial}{\partial \theta^\pm}+ i\bar \theta^\pm \partial_\pm,
\end{equation}
\begin{equation}
\xbar Q_\pm = -\frac{\partial}{\partial\bar \theta^\pm}- i \theta^\pm \partial_\pm.
\end{equation}
In the above expressions, $\partial_\pm$ are spacetime derivatives with respect to the lightcone coordinates $x^\pm:=x^0\pm x^1$,
\begin{equation}
\partial_\pm =\frac{1}{2}(\partial_0\pm \partial_1).
\end{equation}
Since the pure fermionic derivatives $\partial/\partial \theta^\pm$ and $\partial/\partial \bar \theta^\pm$ do not anticommute with the supersymmetry operators, in superspace one uses the following covariant superderivatives,
\begin{equation}
D_\pm:=\frac{\partial}{\partial \theta^\pm}- i\bar \theta^\pm \partial_\pm,
\end{equation}
\begin{equation}
D_\pm:=-\frac{\partial}{\partial \theta^\pm}+i\bar \theta^\pm \partial_\pm.
\end{equation}
The superderivatives and supersymmetry generators obey the anticommutation rules,
\begin{equation}
\left\{ D_\pm,\xbar D_\pm\right\}=2i\partial_\pm  , \ \ \ \left\{ Q_\pm,\xbar Q_\pm\right\}=-2i\partial_\pm,
\end{equation}
\begin{equation}
\left\{ Q_\pm, Q_\pm\right\}=\left\{\xbar Q_\pm, \xbar Q_\pm\right\}=0,
\end{equation}
\begin{equation}
\left\{ D_\pm, Q_\pm\right\}=\left\{ \xbar D_\pm, \xbar Q_\pm\right\}=\left\{  D_\pm, \xbar Q_\pm\right\}=0.
\end{equation}
The general $\mathcal{N}=(2,2)$ variation is then given by
\begin{equation}\label{fullvar}
\delta_{\epsilon,\bar \epsilon}=\epsilon_+Q_ - -\epsilon_-Q_+ -\bar \epsilon_+\xbar Q_-+\bar\epsilon_-\xbar Q_+,
\end{equation}
where the infinitesimal parameters $\epsilon_\pm$ and $\bar \epsilon_\pm$ are fermionic. The most general elements in the representation of the supersymmetry algebra are called \emph{superfields}. That is, a field $\mathcal{F}$ on superspace is called a {superfield} if it transforms as $\mathcal{F}+\delta_{\epsilon,\bar\epsilon}\mathcal{F}$ under the action of the supersymmetry algebra. A superfield $\mathcal{F}$ is fermionic if $\left\{\mathcal{F},\theta^a\right\}=0$ and bosonic $[\mathcal{F},\theta^a]=0$. In our work, all bulk superfields will be bosonic while those restricted to the boundary will be of either type. Instead of generic superfields, we will restrict to those which are \emph{chiral}, \emph{antichiral}, or \emph{twisted chiral superfields}. We call a superfield $X$ {chiral} if 
\begin{equation}\label{chiraldef}
\xbar D_\pm X=0,
\end{equation}
and a superfield $\widetilde X$ twisted chiral if
\begin{equation}\label{twisteddef}
\xbar D_+\widetilde X=D_-\widetilde X=0.
\end{equation}
The solutions to Eq. (\ref{chiraldef}) and Eq. (\ref{twisteddef}) are 
\begin{equation}\label{chiralexpansion}
 X=\phi(y)+\theta^+\psi_+( y)+\theta^-\psi_-( y)+\theta^+ \theta^-F( y),
\end{equation}
\begin{equation}\label{twistexpansion}
\widetilde X=\tilde\phi(\tilde y)+\theta^+\bar\chi_+(\tilde y)+\bar\theta^-\chi_-(\tilde y)+\theta^+\bar \theta^-E(\tilde y),
\end{equation}
respectively. In these expansions we used the usual notation where $\tilde y^\pm=x^\pm \mp i \theta^\pm \bar \theta^\pm$ and $ y^\pm=x^\pm - i \theta^\pm \bar \theta^\pm$.

The two $U(1)$ symmetries whose action on the component fields is given in Eq. (\ref{u1vComponent}) and Eq. (\ref{u1aComponent}) is carried over to superspace via the following actions on superfields
\begin{equation}\label{u1v}
e^{i\alpha F_V}:\mathcal{F}(x^\mu,\theta^\pm,\bar\theta^\pm) \rightarrow e^{i\alpha q_V}\mathcal{F} (x^\mu,e^{-i\alpha}\theta^\pm,e^{i\alpha}\bar\theta^\pm) ,
\end{equation}
\begin{equation}\label{u1a}
e^{i\beta F_A}: \mathcal{F}(x^\mu,\theta^\pm,\bar\theta^\pm) \rightarrow e^{i\beta q_A} \mathcal{F} (x^\mu,e^{\mp i\beta}\theta^\pm,e^{\pm i\beta}\bar\theta^\pm).
\end{equation}

Since will be working directly we supersymmetric actions, here we list all possible supersymmetry invariant functionals which can be constructed in superspace. For the superfields $\mathcal{F}_i$, chiral superfields $X_i$, and twisted chiral superfields $\widetilde X_i$ as above, one can construct the following $\mathcal{N}=(2,2)$-invariant functionals 
\begin{equation}
S_D[\mathcal{F}_i]=\displaystyle \int d^2x d^4\theta \ K(\mathcal{F}_i)= \displaystyle \int d^2x d\theta^+ d\theta^- d\bar\theta^- d\bar\theta ^+K(\mathcal{F}_i)\ ,
\end{equation}
\begin{equation}\label{Fcomplex}
S_F[X_i]=\displaystyle \int d^2x d^2\theta \ W(X_i)= \displaystyle \int d^2x d\theta^+ d\theta^- W(X_i)\big |_{\bar \theta ^\pm=0}\ ,
\end{equation}
\begin{equation}
\widetilde S_F[\widetilde X_i]=\displaystyle \int d^2x d^2\tilde\theta \ \widetilde W(\widetilde X_i)= \displaystyle \int d^2x d\bar \theta^- d\theta^+ \widetilde W(\widetilde X_i)\big |_{\bar \theta ^+=\theta^-=0} \ ,
\end{equation} %EXPLAIN MEASURE
where $K=K(\mathcal{F}_i)$ is a smooth real-valued function, and  $W=W(X_i) $ and $\widetilde W = \widetilde W(\widetilde X_i)$ are holomorphic functions. In the above functionals, the measures are $d^4\theta :=d\theta^+d\theta^-d\bar\theta^-d\bar \theta^+$ and $d^2\theta:=d\theta^-d\theta^+$. The function $K$ is usually called the K\"ahler potential, and $W$ the superpotential. In this work we are mainly concerned with with LG models which are defined for chiral fields $X_i$ by the action
\begin{equation}\label{LGdef}
S[X_i,\xbar X_{\bar i}] := S_D[X_i,\xbar X_{\bar i}]+\real  S_F[X_i].
\end{equation}
We restrict to a LG model with a single chiral field $X$ and the K\"ahler potential
\begin{equation}\label{kahler}
K(X,\xbar X) = \xbar X X.
\end{equation}

A basic result that we will exploit later in this work is that the RNS string as given in Eq. (\ref{rns}) corresponds to a LG model with vanishing superpotential. That is, in the superspace formalism the RNS action is given by
\begin{equation}\label{LGzero}
S_D=\displaystyle \int d^2xd^4\theta\ \xbar X X,
\end{equation}
where $X$ is a chiral superfield. To obtain this result one uses a Taylor expansion of the chiral field over the $\theta$ variable,
\begin{equation}
\begin{split}
X&=\phi(y^\pm)+\theta^a\psi_a(y^\pm)+\theta^+\theta^- F(y^\pm)\\
&=\phi(x^\pm) - i\theta^+ \bar \theta^+\partial_+ \phi(x^\pm) - i\theta^- \bar \theta^-\partial_- \phi(x^\pm) - \theta^+\theta^- \bar \theta^- \bar \theta^+ \partial_+\partial_- \phi(x^\pm)\\
&\ \ +\theta^+\psi_+ (x^\pm) - i\theta^+\theta^-\bar \theta ^-\partial_- \psi_+ (x^\pm) - i\theta^-\theta^+\bar \theta^+\partial_+\psi_- (x^\pm) +\theta^+\theta^- F(x^\pm).
\end{split}
\end{equation}
Inserting the above series expansion into the LG model in Eq. (\ref{LGzero}) and the integrating out the fermionic variables, one obtains
\begin{equation}\label{}
S= \displaystyle \int_{\mathbb{R}^2} d^2x \left(|\partial_0 \phi|^2 - |\partial_1 \phi|^2 + i\bar \psi_-(\partial_0 +\partial_1)\psi_- i\bar \psi_+(\partial_0 -\partial_1)\psi_+ +|F|^2\right),
\end{equation}
which it is the action of the RNS string up to the last term $|F|^2$. Noting that $F=0$ is the equation of motion of the field $F$, we see that indeed we have recovered the supersymmetric RNS action of Eq.(\ref{rns}).  We will utilize this result to work with the supersymmetric $\mathbb{C}/\mathbb{Z}_n$ models in terms of the LG formalism.

Under the action of the axial $R$-symmetry defined in Eq. (\ref{u1a}), the LG functional in Eq. (\ref{LGzero}) is invariant if the chiral field has zero $U(1)_A$ weight. This invariance holds when the superpotential is a monomial, that is,
\begin{equation}
S_F = \displaystyle\int d^2x  d^2\theta \  X^k.
\end{equation}
Furthermore, the LG action is invariant under vector $R$-symmetry if the $X$ is assigned $U(1)_V$ weight $2/k$ in order for the $F$-term to be invariant. 

The  action given in Eq. (\ref{LGzero}) can be generalized to admit a set of chiral superfields $X_i$, $i=1,\dots, n$, taking values on a manifold $M$. The K\"ahler potential can be generalized to any differentiable real-valued function and we assume that
\begin{equation}
g_{i\bar j}:=\partial_i \partial_{\bar j}K(X_i,\xbar X_{\bar i}),
\end{equation}
is positive-definite which defines a K\"ahler metric on $M$. In this case the Lagrangian density is given by \cite{hori00}
\begin{equation}
\begin{split}
\mathcal{L}= &-g_{i\bar j} \partial_\mu \phi^i \partial^\mu \bar \phi^{\bar j} +i g_{i\bar j}\bar \psi_-^{\bar j}(D_0+D_1)\psi_-^i +i g_{i\bar j}\bar \psi_+^{\bar j}(D_0-D_1)\psi_+^i\\
&+R_{i\bar j k\bar l} \psi_+^i \psi_-^k \bar \psi_- ^{\bar j} \psi_+ ^{\bar l} + g_{i\bar j} (F^i - \Gamma^i _{jk}\psi^j_+ \psi^k_-) (\bar F^{\bar j} -\bar{\Gamma}^{\bar j} _{\bar k \bar l}\bar\psi^{\bar k}_- \bar \psi^{\bar l}_+),
\end{split}
\end{equation}
where $R_{i\bar j k\bar l}$ is the Riemann curvature of the K\"ahler metric and $D_\mu\psi^i_\pm:= \partial _\mu\psi^i +\partial_\mu\phi^j \Gamma^i_{jk}\psi^k_\pm$.  Under the inclusion of the real-valued F-term $\real S_F[X_i]$, where $S_F$ is given in Eq. (\ref{Fcomplex}), the equations of motion of spin-2 fields $F^i$ and  $\xbar F^{\bar i}$ are given by
\begin{equation}
F^i = \Gamma^i _{jk}\psi^j_+\psi^k _- - g^{i\bar l} \partial _{\bar l}\xbar W,
\end{equation}
\begin{equation}
\xbar F^{\bar i} = \bar \Gamma^{\bar i} _{\bar j\bar k}\bar \psi^{\bar j}_+\bar \psi^{\bar k} _- - g^{\bar i l} \partial _{ l} W.
\end{equation}
Using the above two equations, the action in components for the LG model on the K\"ahler manifold $M$ has the form
\begin{align}
S=&\displaystyle \int_{\mathbb{R}^2} d^2x ( -g_{i\bar j} \partial_\mu \phi^i \partial^\mu \bar \phi^{\bar j} +i g_{i\bar j}\bar \psi_-^{\bar j}(D_0+D_1)\psi_-^i +i g_{i\bar j}\bar \psi_+^{\bar j}(D_0-D_1)\psi_+^i\nonumber\\
&+R_{i\bar j k\bar l} \psi_+^i \psi_-^k \bar \psi_- ^{\bar j} \psi_+ ^{\bar l} -\frac{1}{4}g^{\bar i j}\partial_{\bar i} W\partial _j W - \frac{1}{2}D_i\partial_j W\psi_+^i\psi_-^j - \frac{1}{2}D_{\bar i} \partial_{\bar j} \xbar W\bar\psi_+^{\bar i}\bar \psi_-^{\bar j} ).
\end{align}

%%%%%%%%%%%%%%%%%%%%%%%%%%%%
\section{Supersymmetry preserving boundaries}

The previous section contains a treatment of sigma models on worldsheets without boundaries. In this section we review the same theory but in the presence of a nontrivial boundary which gives rise to D-branes. D-branes are boundary conditions for open strings or equivalently sources for emission and absorption of closed strings. These D-branes break the $\mathcal{N}=(2,2)$ supersymmetry to a $(1,1)$ supersymmetry. The remaining supersymmetry is called \emph{A-type} or \emph{B-type} depending on which supercharges are preserved by the D-brane. Our work on defects focuses on the B-type but we will need to extensively use the A-type to make comparisons so this section includes a review of both cases.

We start with the $(2,2)$ supersymmetry on the upper half-plane $\Sigma= \mathbb{R}\times [0,\infty)$ as our model theory.  At the boundary the fields satisfy boundary conditions which usually relates the left- and right-moving modes. These conditions relate the left and right fermionic variables $\theta^\pm$ and $\bar \theta^\pm$ in one of the two ways \cite{hori00}
\begin{equation}\label{boundarytype}
\begin{split}
&(A) \ \ \ \ \theta^+ +\operatorname{e}^{i\alpha}\bar\theta^-=0, \ \ \bar\theta^+ +\operatorname{e}^{-i\alpha}\theta^-=0,\\
&(B) \ \ \ \ \theta^+ -\operatorname{e}^{i\beta}\theta^-=0, \ \ \bar \theta^+ -\operatorname{e}^{-i\beta}\bar\theta^-=0.\\
\end{split}
\end{equation}
In theories with the \emph{A-boundary} or \emph{B-boundary} the following supercharges are conserved, respectively,
\begin{equation}\label{abgenerators}
\begin{split}
&(A) \ \ \ \ \xbar Q_A:=\xbar Q_+ + \operatorname{e}^{i\alpha}Q_-, \ \ Q_A:=Q_+ +\operatorname{e}^{-i\alpha}\xbar Q_-,\\
&(B) \ \ \ \ \xbar Q_B:=\xbar Q_+ +\operatorname{e}^{i\beta}\xbar Q_-, \ \ Q_B:= Q_+ +\operatorname{e}^{-i\beta} Q_-.
\end{split}
\end{equation}
The theory with A-boundary and conserved charges $(Q_A,\xbar Q_A)$ is called \emph{A-type supersymmetric};  when considering $(Q_B,\xbar Q_B)$ and B-boundary, the theory is called \emph{B-type supersymmetric}. For simplicity we take $\alpha$ and $\beta$ to be zero in the above equations. The results can always be generalized to the non-zero case via the appropriate $U(1)$ rotation.

%%%%%%%%%%%%%%%%%%%%%%%%%%%%
\subsection{A-type supersymmetry}

Specializing to A-boundary means that the boundary of our theory is preserved by the following operators
\begin{align}
\xbar D_A&:= \xbar D_++D_-=\frac{\partial}{\partial\bar \theta}+i\theta\partial_0 ,\\
 D_A&:=  D_++\xbar D_-=\frac{\partial}{\partial \theta}-i\bar\theta\partial_0 ,\\
\xbar Q_A&:= \xbar Q_++Q_-=-\frac{\partial}{\partial\bar \theta}-i\theta\partial_0, \\
 Q_A&:=  Q_++\xbar Q_-=\frac{\partial}{\partial \theta}+i\bar\theta\partial_0. 
\end{align}
So the general A-type variation \index{A-type supersymmetry} is given by
\begin{equation}\label{avariation}
\delta_A=\epsilon \bar Q_A -\bar \epsilon Q_A.
\end{equation}
Observe that the full $(2,2)$ variation preserves A-type boundary conditions if $\epsilon_+=\bar \epsilon_-=:\epsilon$ and $\bar \epsilon_+= \epsilon_-=:\bar \epsilon$. The A-type variation of the twisted F-term  $S_{\widetilde F}$ is given by
\begin{equation}\label{varytwistedf}
\delta_AS_{\widetilde F}=2i\epsilon\displaystyle\int_{\partial \Sigma}dtd\theta\ \widetilde W(\widetilde X)\big |_{\bar\theta=0}.
\end{equation}
Similar expression for the antiholomorphic part, just as above but everything conjugated. To obtain this result we start with the twisted F-term
\begin{equation}
S_{\widetilde F}=\displaystyle\int_\Sigma d^2xd\bar\theta^-d\theta^+\ \widetilde W(\widetilde X)\big |_{\bar\theta^+=\theta^-=0},
\end{equation}
whose argument is a twisted chiral field $\widetilde X$:  $\xbar D_+\widetilde X= D_- \widetilde X=0$ \cite{hori00}. We have the variation,
\begin{equation}\label{avariationsteps}
\begin{split}
\delta_A S_{\widetilde F}&=\displaystyle\int_\Sigma d^2xd\bar\theta^-d\theta^+\ \delta_A \widetilde W(\widetilde X)\big |_{\bar\theta^+=\theta^-=0}\\
&=\displaystyle\int_\Sigma d^2xd\bar\theta^-d\theta^+\ \left( [\epsilon(\xbar Q_++Q_-)-\bar \epsilon (Q_++\xbar Q_- )] \widetilde W(\widetilde X)\right)\big |_{\bar\theta^+=\theta^-=0}\\
&=\displaystyle\int_\Sigma d^2xd\bar\theta^-d\theta^+\  2i \epsilon\left(  \bar\theta^-\partial _- \widetilde W -\theta^+\partial _+\widetilde W\right)\big |_{\bar\theta^+=\theta^-=0},
\end{split}
\end{equation}
where we used the following results
\begin{eqnarray}
&&Q_+\widetilde W\big |_{\bar\theta^+=\theta^-=0} =\frac{\partial}{\partial \theta^+}\widetilde W \big |_{\bar\theta^+=\theta^-=0}\\
&&\xbar Q_-\widetilde W\big |_{\bar\theta^+=\theta^-=0}=-\frac{\partial}{\partial \bar\theta^-}\widetilde W \big |_{\bar\theta^+=\theta^-=0}\\
&& \xbar Q_+\widetilde W\big |_{\bar\theta^+=\theta^-=0}=-2i\theta^+\partial_+\widetilde W \big |_{\bar\theta^+=\theta^-=0}\\
&&  Q_-\widetilde W\big |_{\bar\theta^+=\theta^-=0}=2i\bar\theta^-\partial_-\widetilde W \big |_{\bar\theta^+=\theta^-=0}.
\end{eqnarray}
To obtain the last two equations we used 
\begin{align}
\xbar Q_+=\xbar D_+-2i\theta^+\partial_+\ , \ \ \  Q_+= D_-+2i\bar\theta^-\partial_-,
\end{align}
and the fact that $\widetilde W(\widetilde X)$ itself is twisted chiral since holomorphic functions of $\widetilde X$ are also twisted chiral. Now we use the expansion of a twisted chiral field,
\begin{equation}
\widetilde \Phi=\tilde\phi(\tilde y)+\theta^+\bar\chi_+(\tilde y)+\bar\theta_-\chi_-(\tilde y)+\theta^+\bar \theta^-E(\tilde y),
\end{equation}
where $\tilde y^\pm=x^\pm \mp i \theta^\pm \bar \theta^\pm$. Inserting this expansion into Eq. (\ref{avariationsteps}), we get

\begin{equation}
\delta_A S_{\widetilde F}=2i\epsilon\displaystyle \int_{\partial \Sigma}dt \ (\bar \chi_+(t)-\chi_-(t))= 2i\epsilon\displaystyle \int_{\partial \Sigma}dtd\theta \ \widetilde W\big |_{\bar \theta=0}.
\end{equation}
The last equality follows by using the same chiral expansion for the twisted superpotential, and restricting to the A-boundary.

Note that the A-type variation of the F-term is not as nice:
\begin{equation}
\delta_A S_F=-2i\displaystyle \int_{\partial \Sigma} dt (\bar \epsilon \psi_+(t)+ \epsilon \psi_-(t)).
\end{equation}
Thus, we use work with B-type supersymmetry when dealing with LG models and A-type for twisted LG functionals.

With the above observation in mind, we find the sufficient and necessary requirements for A-supersymmetry. The characterization will be a geometrical description of D-branes which in this case are called A-branes. We obtain the A-type classification following the work of \cite{Hori0005}. \index{A-branes}
We consider the supersymmetric sigma model with superpotential $W$ on $\Sigma =\mathbb{R}\times[0,\infty)$ with variables $(x^0,x^1)$ and with an $n$-dimensional target space $M$ which we assume to be a K\"ahler manifold. The action in components is 
\begin{equation}\label{sigma}
\begin{split}
S=&\displaystyle \int_\Sigma d^2x \left\{ -g_{i\bar j}\partial^\mu\phi^i\partial_\mu \xbar \phi ^{\bar j}+\frac{i}{2}g_{i\bar j} \xbar \psi_-^{\bar j}(\overleftrightarrow D_0+\overleftrightarrow D_1) \psi_-^{i} + \frac{i}{2}g_{i\bar j} \xbar \psi_+^{\bar j}(\overleftrightarrow D_0-\overleftrightarrow D_1) \psi_+^{i} \right. \\
& \left. -\frac{1}{4} g^{\bar j i}\partial_{\bar j}\xbar W \partial_i W -\frac{1}{2}(D_i\partial_j W)\psi_+^i\psi_-^j-\frac{1}{2}(D_{\bar i}\partial_{\bar j}\xbar W)\xbar \psi_-^{\bar i}\xbar \psi_+^{\bar j}+R_{i\bar k j \bar l}\psi_+^{ i}\psi_-^{ j}\xbar \psi_-^{\bar k}\xbar \psi_+^{\bar l}\right\},
\end{split}
\end{equation}
where $\xbar \psi^{\bar j}\overleftrightarrow D_\mu \psi^i :=\xbar \psi^{\bar j} D_\mu \psi^i - ( D_\mu \xbar \psi^{\bar j}) \psi^i$ and $D_\mu \psi^i :=\partial_\mu \psi^i+\partial_\mu\phi^j\Gamma^i_{jk}\psi^k$.

Under a general variation one obtains the Euler-Lagrange equations for the fields $X=(\phi^I,\psi_\pm ^I)$ plus  boundary conditions needed for the vanishing of the boundary integral. These constraints on the boundary $\partial \Sigma$ are
\begin{equation}\label{bdyconstraint1}
g_{IJ}\delta\phi^I\partial_1\phi^J=0
\end{equation}
\begin{equation}\label{bdyconstraint2}
g_{IJ}(\psi_-^I\delta\psi_-^J-\psi_+^I\delta\psi_+^J)=0
\end{equation}
Since $\phi:\partial\Sigma \hookrightarrow\gamma$, the vector $\delta\phi^I$ is tangent to $\gamma$. Hence, by the constraint (\ref{bdyconstraint1}) $\partial\phi^I$ is normal to $\gamma$. Under the general supersymmetry action, the action varies as
\begin{equation}\label{sigmavar}
\begin{split}
\delta &S = \frac{1}{2}\displaystyle \int_{\partial\Sigma} dx^0  \left\{\epsilon_+(-g_{i\bar j} (\partial_0+\partial_1)\xbar\phi^{\bar j}\psi_-^i+\frac{i}{2}\xbar\psi_+^{\bar i}\partial_{\bar i}\xbar W) +\epsilon_-(-g_{i\bar j} (\partial_0-\partial_1)\xbar\phi^{\bar j}\psi_+^i \right. \\
&\left. -\frac{i}{2}\xbar\psi_-^{\bar i}\partial_{\bar i}\xbar W) + \xbar \epsilon_+ (g_{i\bar j} \xbar \psi_-^{\bar j}(\partial_0+\partial_1)\phi^{ i}+\frac{i}{2}\psi_+^{ i}\partial_{ i} W)  + \xbar \epsilon_- (g_{i\bar j} \xbar \psi_+^{\bar j}(\partial_0-\partial_1)\phi^{ i}-\frac{i}{2}\psi_-^{ i}\partial_{ i} W) \right\}.
\end{split}
\end{equation}

Before stating the requirements for a D-brane to be an A-brane we restrict to the conditions for invariance of the theory under the $\mathcal{N}=1$ subalgebra. The $\mathcal{N}=1$ subalgebra is obtained by taking $\epsilon_\pm=\pm i\epsilon$ with $\epsilon\in\mathbb{R}$. One can see that under this restriction one has both A- and B-supersymmetry. To see this algebra, one takes the general A-type generator $\delta_A=\epsilon_+(Q_-+\xbar Q_+)-\bar \epsilon _+(\xbar Q_-+Q	_+)$
and sets $\epsilon_+=i\epsilon$, $\bar \epsilon_+=\epsilon_+^*$. In this case, we have $\delta^*=i\epsilon(Q+\xbar Q)$. The anticommutator is then
\begin{equation}
\frac{1}{(i\epsilon)^2}\left\{\delta^*,\delta^*\right\}=4H.
\end{equation}

Let $\gamma\subset M$ be the submanifold containing the image of $\partial \Sigma$. The supersymmetric sigma model (\ref{sigma}) is invariant under the $\mathcal{N}=1$ subalgebra of A-supersymmetry if  $\psi^J_{-}+\psi^J_{+}$ and $\psi^J_{-}-\psi^J_{+}$ are tangent and normal to $\gamma$ respectively; and $\image W$ is constant along $\gamma$. To obtain this result, one first restricts $\bar\epsilon_\pm$ and $\epsilon_\pm$ to this case in the boundary contribution (\ref{sigmavar}). The one can rewrite the integral using the two results below:
\begin{equation}
\begin{split}
&\xbar\psi_+^{\bar i}\partial_{\bar i}\xbar W - \psi_-^{ i}\partial_{ i} W+\xbar\psi_-^{\bar i}\partial_{\bar i}\xbar W-\psi_+^{ i}\partial_{ i}\xbar W\\
&=(\xbar \psi _+^{\bar i}+\xbar \psi _-^{\bar i})\partial_{\bar i}\xbar W- ( \psi _+^{ i}+ \psi _-^{ i})\partial_{ i} W\\
&=-( \psi _+^{ I}+ \psi _-^{ I})\partial_{ I}(W-\xbar W),
\end{split}
\end{equation}
and
\begin{equation}
\begin{split}
&-g_{i\bar j} (\partial_0+\partial_1)\xbar\phi^{\bar j}\psi_-^i +  g_{i\bar j}\xbar\psi_+ ^{\bar j} (\partial_0-\partial_1)\phi^{ i} +g_{i\bar j} (\partial_0-\partial_1)\xbar\phi^{\bar j}\psi_+^i - g_{i\bar j} \xbar\psi_-^{\bar j} (\partial_0+\partial_1)\phi^{ i}\\
&= -g_{IJ}\partial_0\phi^I(\psi_-^J-\psi_+^J)-g_{IJ}\partial_1\phi^I(\psi^J_-+\psi_+^J).
\end{split}
\end{equation}
Denoting $\delta^*S$ the variation of the action under the $\mathcal{N}=1$ generators, and using the steps above, we have
\begin{align}\label{n1var}
\delta^*S&=\frac{i\epsilon}{2}\displaystyle \int_{\partial\Sigma}dx^0 \ \left\{ -g_{IJ}\partial_0\phi^I(\psi_-^J-\psi_+^J)-g_{IJ}\partial_1\phi^I(\psi^J_-+\psi_+^J)\right. \nonumber\\
&\left. -\frac{i}{2}( \psi _+^{ I}+ \psi _-^{ I})\partial_{ I}(W-\xbar W)\right\}.
\end{align}
We see that $\psi^J_-+\psi_+^J$ is the $\mathcal{N}=1$ variation of $\phi^I$ and hence tangent to $\gamma$. Therefore the second term vanishes because $\partial_1\phi^I$ is normal. Then the rest of the integral vanishes if $\psi^J_--\psi_+^J$ is normal to $\gamma$ and $\image W$ is constant along $\gamma$. This is because the vector $\partial_0\phi^I$ is tangent to $\gamma$ and  each term vanishes independently since $\partial_0\phi^I$ and $\psi^J_-+\psi_+^J$ are generally independent.

Note that not only is the supersymmetric sigma model action invariant under $\mathcal{N}=1$ provided the above boundary conditions, but the boundary conditions themselves are also invariant, that is,
\begin{equation}
\delta^* (\psi^J_-+\psi_+^J) =-2\epsilon \partial_0\phi^J,
\end{equation}
\begin{equation}
\delta^* (\psi^J_--\psi_+^J) =-2\epsilon \partial_1\phi^J+2\epsilon g^{JI}\partial_I\image W,
\end{equation}
in the background $\psi^I_\pm=0$. Thus the boundary conditions are invariant when $\image W$ is locally constant on $\gamma$. 

Now we proceed to general case when A-supersymmetry is preserved to geometrically describe A-branes. We first state the result for the case $e^{i\alpha}=1$ which is the trivial phase in the A-type generators in Eq. (\ref{abgenerators}). A D-brane wrapped on $\gamma$ preserves A-supersymmetry iff $\gamma$ is Lagrangian submanifold of $M$ with respect to the K\"ahler form, and $W(\gamma)\subset \mathbb{C}$ is a straight line parallel to the real axis, and invariant under the gradient flow of $\real W$.

The general supersymmetry action on the bosons is $\delta \phi^i=\epsilon_+\psi_-^i-\epsilon_-\psi^i_+$ so the A-type action is
\begin{equation}
\delta \phi^i=\epsilon_+\psi_-^i-\bar\epsilon_+\psi^i_+.
\end{equation}
We decompose $\epsilon_+$ into its real and imaginary part $\epsilon_+=\epsilon_1+i\epsilon_2$. The equation above is then
\begin{equation}\label{reime}
\delta \phi^i=\epsilon_1(\psi_-^i-\psi^i_+)+i\epsilon_2 (\psi^i_-+\psi_+^i).
\end{equation}
Observing that $\phi$ parametrizes $\gamma$ and that the vector $\delta\phi^i$ is tangent to $\gamma$ we see that $\epsilon(\psi_-^i-\psi^i_+)$ and $i\epsilon(\psi_-^i+\psi^i_+)$ are the holomorphic components of vectors tangent to $\gamma$, where $\epsilon\in\mathbb{R}$. Yet in from Eq. (\ref{n1var}) in the $\mathcal{N}=1$ case we see that $i\epsilon(\psi_-^i-\psi^i_+)$ and $i\epsilon(\psi_-^i+\psi^i_+)$ are normal and tangent to $\gamma$ respectively when $\epsilon\in\mathbb{R}$. Therefore the $\mathcal{N}=2$ supersymmetry requires that map $f: T_X M \rightarrow T_X M$ which multiplies the holomorphic component of vectors by $i=\sqrt{1}$ interchanges tangent and normal vectors to $\gamma\subset M$. This means that $T_X\gamma \cong (T_X\gamma)^\perp$ in $T_X M$ which implies $ \dim\gamma=1/2\dim M$. 

The target manifold $M$ is a sympletic manifold $(M,\omega)$ where $\omega$ is the K\"ahler form 
\begin{equation}
\omega(A,B):=g(JA,B),
\end{equation}
 for $A, B \in T_X M$, where $J:=i\operatorname{diag}(1_{n},-1_n)$ is the complex structure \index{complex structure} on $M$ compatible with $g$, e.g. $g(JA,JB)=g(A,B)$. To see that 
$\omega$ is a sympletic 2-form:
\begin{equation}
\begin{split}
\omega(A,B)&=g(JA,B)=g_{m\bar n}(iA^mB^{\bar n}-iA^{\bar n}B^m)\\
&=-g_{m\bar n}(iB^mA^{\bar n} - iB^{\bar n}A^m)\\
&=-\omega(B,A).
\end{split}
\end{equation}
Thus we can write $\omega=ig_{m\bar n}dz^m\wedge dz^{\bar n}$ which means $dw=0$ because
\begin{equation}
\begin{split}
dw&=\partial_lg_{m\bar n}dz^m\wedge dz^l\wedge dz^{\bar n}+\partial_{\bar o}g_{m\bar n}dz^m\wedge dz^{\bar n}\wedge dz^{\bar o}\\
&=\partial_l \partial_m\partial_{\bar n} K(X,\xbar X) dz^m\wedge dz^l\wedge dz^{\bar n} + \partial_{\bar o} \partial_m\partial_{\bar n} K(X,\xbar X) dz^m\wedge dz^{\bar n}\wedge dz^{\bar o}\\
&=0.
\end{split}
\end{equation}
The K\"ahler form is non-degenerate by definition.  Hence indeed sympletic. Now it is easy to show that $\omega$ vanishes on $\gamma$. Writing $A=(A^n,A^{\bar n})$ for $A\in T_X M$, $JA =(iA^n, -iA^{\bar n})$ but this is the same map $f$ which multiplies the holomorphic component by $i$ since $V^{\bar n}=V^{n*}$ for all $V \in T_X M$ so $f(A)=\tilde A =(\tilde A^n, \tilde A^{\bar n})=(iA^n, -iA^{\bar n})$. Let $A,B \in T_X \gamma$, then $\omega(A,B)=g(f(A),B)=0$ since $f:T_X\gamma\rightarrow (T_X\gamma)^\perp$. Thus $\gamma$ is a Lagrangian submanifold of $M$.

Now we proceed to show that $\image W$ is constant on $\gamma$. As noted above  the vector $\epsilon(\psi^i_--\psi^i_+)$ is tangent to $\gamma$ and since we require A-type boundary invariance, we have $\delta_A[\epsilon(\psi^i_--\psi^i_+)]$ is tangent to $\gamma$ as well. In particular for $\psi^i_\pm =0$, we have
\begin{equation}\label{parallelvec}
\delta_A[\epsilon(\psi^i_--\psi^i_+)]=2i\epsilon\epsilon_1 (\partial_0\phi^i)+ 2i\epsilon\epsilon_2(i\partial_1\phi^i)+ 2i\epsilon\epsilon_2F^i.
\end{equation}
The coefficients $i\epsilon\epsilon_i$ are real: $(i\epsilon\epsilon_i)^\dag=-i\epsilon^\dag\epsilon_i^\dag=-i\epsilon\epsilon_i=i\epsilon_i\epsilon$ where we used $\epsilon, \epsilon_i \in \mathbb{R}$. As stated, $\partial_0\phi^i$ and $\partial_1\phi^i$ are tangent and normal to $\gamma$ respectively. Multiplying by $i$ the holomorphic component $\partial_1\phi^i$ makes this vector, that is $(\partial_1\phi^i, \partial\bar\phi^{\bar i})$, an element of tangent space as well. Then by Eq. (\ref{parallelvec}) $F^i$ is tangent to $\gamma$ as well. In the subspace defined by $\psi^i_\pm=0$ we have \begin{equation}
F^i=-\frac{1}{2}g^{i\bar j}\partial_{\bar j}\xbar W.
\end{equation}
This means that the gradient of $\real W$
\begin{equation}
\operatorname{grad}(\real W)=g^{i\bar j}\partial_{\bar j}(W+\xbar W)\partial_i+g^{\bar j i}\partial_{i}(W+\xbar W)\partial_{\bar j},
\end{equation}
 is tangent to $\gamma$. 

Note that $\operatorname{grad}(\real W)=-iJ\operatorname{grad}(\image W)$  so $\operatorname(\image W)$ is normal to $\gamma$ as it was also required by the $\mathcal{N}=1$ case. To see this result one writes explictly,
\begin{equation}
J\operatorname{grad}(\image W)=i\begin{pmatrix} 1_n & 0_n \\ 0_n & -1_n \end{pmatrix} \begin{pmatrix} -g^{i\bar j}\partial_{\bar j} \xbar W \\ g^{\bar j i}\partial_{i}  W \end{pmatrix} =-i\begin{pmatrix} g^{i\bar j}\partial_{\bar j} \xbar W \\ g^{\bar j i}\partial_{i}  W \end{pmatrix} = -i \operatorname{grad}(\real W).
\end{equation}
Thus the flow of $\operatorname{grad}\real W$ is along $\gamma$ and $\image W$ is constant along the  flow (one checks that $v^I\partial_I \image W =0$ where $v^i =g^{i\bar j}\partial_{\bar j}$), hence $W(\gamma)$ is invariant under the flow of $\operatorname{grad}\real W$.

The above generalizes to the case when we take $e^{i\alpha}\neq 1$. 

\subsubsection{Wave-front trajectories}

For our purposes, the most relevant example of D-branes preserving A-supersymmetry are those wrapped on the submanifold defined by the action of the gradient of $\real W$ on a non-degenerate critical point of the superpotential $W$. Since $\operatorname{grad}[\real W] (\image W)=0$ every point on this manifold has the same value $\image W$ as the critical point.

 For definiteness, let $X_*$ be a critical point of $W$ of order $n=1$, and let $f_X(t)=f(t,X)$ be the global flow (also called the local one-parameter group action) \index{global flow} generated by $\operatorname{grad}[\real W]$. In general a global flow is a continuous map $f:[0,1)\times M \rightarrow M$ which satisfies $f(0,X)=X$, $f(t,f(s,X))=f(t+s,X)$. Here we are interested on such a flow that satisfies
\begin{equation}\label{flowdef}
f'_X(t)=\operatorname{grad}[\real W]_{f_X(t)}.
\end{equation}
And define 
\begin{equation}\label{gammadef}
\gamma_{X_*} :=\left\{ X\in M \big | \lim_{t\rightarrow -\infty} f_X(t)=X_*\right\},
\end{equation}
then the claim is D-branes wrapped on $\gamma_{X_*}$ are a A-branes. So we have to check that this submanifold is Lagrangian whose image in the $W$-plane is parallel to the real axis.
\begin{equation}
\begin{split}
\operatorname{grad}[\real W](\image W) &=g^{IJ}\partial_J(W+\xbar W)\partial_I (W-\xbar W)\\
&=g^{i\bar j}\partial_{\bar j}\xbar W\partial_i W - g^{\bar j i}\partial_iW \partial_{\bar j}\xbar W\\
&=|\partial W|^2-|\partial W|^2\\
&=0.
\end{split}
\end{equation}
Therefore $\image W$ is constant along $\operatorname{grad}[\real W]$ and thus $W(\gamma_{X_*})$ is a ray starting at the critical value $w_*:= W(X_*)$. 

Now we need to show that $\gamma_{X_*}$ is middle dimensional. Recall that if $z_0$ is a critical point of $f:\mathbb{C}\rightarrow\mathbb{C}$ of order $m-1$, then there exists a change of coordinates near $z_0$ and $f(z_0)$ such that $f$ has the form $f(\xi)=\xi^m+f(z_0)$. Thus near $X_*$ we write
\begin{equation}\label{wlocal}
W=W(X_*)+\sum_{i=1}^n z_i^2+ o(z_i^3).
\end{equation}

If the change of variables brings the metric into the standard form $ds^2=\delta_{a\bar b}dz^a\otimes d\bar z^{\bar b}$, then in the local coordinates near $0\in\mathbb{C}$ the flow equation (\ref{flowdef}) becomes
\begin{equation}
{z}_a'(t) =\delta^{a\bar b} \partial_{\bar b} \sum_{k=1}^{n-1}\bar z_k^2+\cdots,
\end{equation}
after inserting Eq. (\ref{wlocal}) into the flow equation (\ref{flowdef}). The dots are higher order terms which which we can make arbitrarily small by considering smaller neighborhoods near $0\in\mathbb{C}$ which is equivalent to $t\rightarrow -\infty$. The solution is $z_a(t)= r_a e^t+\cdots$ with $r_a\in\mathbb{R}$ which is required to solve $z'(t)\sim\bar z(t)$. Thus, near 0 the (or equivalently in small neighborhood of $X_*$) the submanifold $\gamma_{X_*}$ is an $n$-dimensional real manifold. Since the flow $f(t,X)$ defines $\gamma_{X_*}$ we see that $\dim\gamma_{X_*}=n=1/2\dim M$. If the metric was not in the standard form, we would only obtain a different submanifold but also $n$-dimensional. 

Now we are left to show that the induced symplectic form vanishes on $\gamma_{X_*}$. The first step to show that $\omega$ is invariant along the gradient of $\real W$. This holds if $\mathcal{L}_v \omega=0$, $v:= \operatorname{grad}[\real W]$. By Cartan's formula,  the right-hand side is $i_vd\omega+d i_v\omega$ where $i$ is the interior product. The K\"ahler form $\omega$ is closed so the first term does not contribute. The second term is zero by showing that $i_v\omega$ is exact.
\begin{equation}
\begin{split}
i_v\omega&=i_vig_{i\bar j}dz^i\wedge d\bar z^{\bar j}=ig_{i\bar j}(v^i \bar z^{\bar j}-\bar v^{\bar j}dz^i)=i(\delta^{\bar k}_{\bar j} \partial_{\bar k}\xbar W d\bar z^{\bar j}-\delta^{ k}_{ j} \partial_{ k} W d z^{ j})=id(\xbar W -W).
\end{split}
\end{equation}

Now let $X\in\gamma_{X_*}$ and $v_1, v_2 \in  T_X\gamma_{X_*}$. Considering $\omega 
_{f_X(t)}\in T_{f_X(t)}^*\gamma_{X_*}$, we write $\omega 
_X\in T_{X}^*\gamma_{X_*}$ as the pullback $\omega_X =(f^*_t \omega)_X:=f^*_t( \omega_{f_X(t)})$ since $\omega$ is invariant along the flow $f_t$ generated by the vector field $v$. Therefore, $\omega_X (v_1,v_2)= f^*_t( \omega_{f_X(t)})(v_1,v_2)= \omega_{f_X(t)}(f_{t*} v_1,f_{t*}v_2)$. In the limit $t\rightarrow -\infty$ the right-hand side is zero since the vectors $f_{t*}v_i\rightarrow 0$. To see this one takes the limit of $f_{t*}v_i(g)= v_i (g\circ f_t)$ where $g$ is any function on $M$. The function $g\circ f_t$ is a constant function in the limit. Thus $\omega_X (v_1,v_2)=0 $ since it is independent of the parameter $t$. Thus $\gamma_{X_*}$ is a Lagrangian submanifold of $(M,\omega)$.

To summarize, the we have shown that D-branes wrapped on $\gamma_{X_*}$ as defined in Eq. (\ref{gammadef}) are A-branes which are mapped to $W(X_*)+\mathbb{R}^{\geq 0}$, where $X_*$ is a critical point of $W$.

\subsection{A-branes in Landau-Ginzburg models}\label{abraneslg}

 Below we provide the wave-front trajectory description for LG models with polynomial superpotentials. This application of the geometrical description of A-branes will be used later when following the action of the RG flow on boundary degrees of freedom. We first consider the case $W=X^{k+2}$ with $k$ a non-negative integer.  Then $W$ has only one critical point $X_*=0$. As noted above we know that $\gamma_{0}$ (defined in (\ref{gammadef})) is the pre-image of the set $[0,\infty)\subset\mathbb{C}$. Explicitly, A-branes wrap the submanifold
\begin{equation}
\gamma_{0}=\left\{ r \exp\left(  \frac{2\pi ni}{k+2}\right) : r\in[0,\infty) \ , \ n \in \left\{ 0,\dots,k+1\right\} \right\}\subset\mathbb{C}.
\end{equation}

Using submanifolds of $\mathbb{C}$ which asymptote to $\gamma_0$, we can also describe the A-branes of LG theories with more general superpotentials of the type
\begin{equation}\label{deformedlg}
W_{ \lambda}(X)=X^{k+2}+\sum_{j=0}^{k-1} \lambda_j X^{j+2}.
\end{equation}
We have observed that a constant term does not contribute to the fermionic integral of the Lagrangian so it can be shifted away. A linear term does not introduce any new branch points. So we have the freedom to gauge it away and thus always translating one of the critical points to the origin.

In the most general case, $\lambda_j\neq0$ for all $j$, and $W_{ \lambda}$ has $k+1$ non-degenerate critical points which are isolated. In this case we have $k+1$ possible Lagrangian submanifolds to wrap the A-branes, corresponding to each of the critical points. We assume that $\image w_i\neq \image w_j$ for $i\neq j$, where $w_j:=W_{ \lambda}(X_{*j})$ are the critical values. This assumption eliminates the possibility of having overlapping images in the $W$-plane of the submanifolds $\gamma_{i}$ corresponding to the $X_{*j}$ critical points.

The A-branes of the deformed theory are curves asymptoting to $\mathcal{L}_{n_1}\cup \mathcal{L}_{n_2}$, $n_1\neq n_2$, where $\mathcal{L}_{n_j} \subset \gamma_{0}$ are slices corresponding to each value of $n_i \in\left\{ 0,\dots,k+1\right\}$. This claim follows by noting that for large $X$, $W_\lambda$ approaches the non-deformed $W$ since the leading term $X^{k+2}$ in $W_\lambda$ dominates. So $W^{-1}_\lambda$ is close to $W^{-1}$ in this regime. Now, let $X_{*j}$ be one of the critical points of the deformed potential. By assumption it is of order one so locally near $X_{*j}$ and its image, $W_{\lambda}$ is biholomorphically equivalent to a quadratic map. Thus the preimage of $w_j+\mathbb{R}^{\geq 0}$ near $w_j$ forms two wave-front trajectories starting at $X_{*j}$. As noted, these curves approach some $\mathcal{L}_{n_1}$ and $\mathcal{L}_{n_2}$. The curves intersect at the branch points only (consider $W_\lambda$ as a branched cover) which means $n_1\neq n_2$. For non-generic values of the $\lambda_j$, the branch points can be degenerate. Then the A-brane associated with one of these points, say $X_*$, will asymptote  $\mathcal{L}_{n_1}\cup\cdots \cup \mathcal{L}_{n_{o(X_*)+1}}$, where $o(X_*)$ is the order the critical point $X_*$.

Following the work of \cite{brunner07a} we can depict the A-brane description above for the Landau-Ginzburg models by compactifying the $X$-plane to the disk $D$. The resulting graph contains the critical points $X_{*i}$ in the interior of the disk; cyclically ordered preimages 
$B^1,$ $\dots,$ $B^{k+2}$  of $\infty\in W$-plane on the boundary of the disk $\partial D$; and $({o(X_{*i})+1})$-many segments $\gamma_i^a$ connecting the point $X_{*i}$ to that many of the $B^a$. We define $\Gamma_i := \cup_a  \gamma_i^a$ and $\Gamma := \cup_i \Gamma_i$. We call the graph formed by $\Gamma$ and the boundary $\partial D$ the \emph{schematic representation} of the superpotential. The two graphs below are examples of schematic representations for A-branes in LG models with superpotentials $W=X^4$ and $W=X^4+\lambda X^3$.

\setlength{\unitlength}{7mm}
\begin{picture}(10,10)(-6,-6)

\put(0,0){\circle{6}}
\put(0,0){\line(0,5){3}}
\put(0,0){\line(0,-5){3}}
\put(0,0){\line(5,0){3}}
\put(0,0){\line(-5,0){3}}
\put(0,0){\circle*{.2}}
\put(3,0){\circle*{.2}}
\put(-3,0){\circle*{.2}}
\put(0,3){\circle*{.2}}
\put(0,-3){\circle*{.2}}
\put(.3,.3){$X_*$}
\put(3.3,0){$B^1$}
\put(0,3.3){$B^2$}
\put(-3.8,0){$B^3$}
\put(0,-3.8){$B^4$}
\put(-.5,-5){{ $W=X^4$}}

\put(12,0){\circle{6}}
\put(12,0){\line(0,5){3}}
\put(12,0){\line(0,-5){3}}
\put(12,0){\line(5,0){3}}
\put(11,0){\line(-5,0){2}}
\put(11,0){\line(1,2.9){1}}
\put(10,.3){$X_{*2}$}
\put(11,0){\circle*{.2}}
\put(12,0){\circle*{.2}}
\put(15,0){\circle*{.2}}
\put(9,0){\circle*{.2}}
\put(12,3){\circle*{.2}}
\put(12,-3){\circle*{.2}}
\put(12.3,.3){$X_{*1}$}
\put(15.3,0){$B^1$}
\put(12,3.3){$B^2$}
\put(8.1,0){$B^3$}
\put(12,-3.8){$B^4$}
\put(11,-5){{ $W=X^4+\lambda X^3$}}

\end{picture}

A graphical representation $\Gamma$ has the following properties \cite{brunner07a}: all the preimages of a critical value $\omega\in \mathbb{C}$ are connected on $\Gamma$;   $\Gamma \setminus\partial D$ is connected and simply connected;  $\forall i\neq j, \Gamma_i \cap \Gamma_j$ contains at most one point; and it is non-empty only if it contains an element of the fiber $f^{-1}(\infty)$;  $\Gamma_i \cap \Gamma_j\cap \Gamma_k = \emptyset$.

%%%%%%%%%%%%%%%%%%%%%%%%%%%%%
\subsection{B-type supersymmetry}

In this section we review B-type supersymmetry and boundaries which preserve it. The B-type boundary conditions on the fermionic variables at $\partial \Sigma$ is preserved by the operators
\begin{align}\label{boperators}
\xbar D_B&:= \xbar D_++\xbar D_-=-\frac{\partial}{\partial\bar \theta}+i\theta\partial_0, \\
 D_B&:=  D_++ D_-=\frac{\partial}{\partial \theta}-i\bar\theta\partial_0, \\
\xbar Q_B&:= \xbar Q_++\xbar Q_-=-\frac{\partial}{\partial\bar \theta}-i\theta\partial_0 ,\\
 Q_B&:=  Q_++ Q_-=\frac{\partial}{\partial \theta}+i\bar\theta\partial_0.
\end{align}
The general B-type variation is given by
\begin{equation}
\delta_B=\epsilon Q_B- \bar \epsilon\  \xbar Q_B.
\end{equation}
which is equivalent to the $(2,2)$ variation in Eq. (\ref{fullvar})  if in the latter we set $\epsilon_+=-\epsilon_-=:\epsilon$ and $\bar \epsilon_+ = - \bar \epsilon_-=:\bar \epsilon$. The B-type generators obey the relations
$\left\{ Q_B,\xbar Q_B \right\} =-2i(\partial_+ +\partial_-)\ , \ Q_B^2=\xbar Q_B^2=0$. 
Under B-type supersymmetry the components of the original $(2,2)$ chiral field $X$,
\begin{equation}\label{Btrans1}
X=\phi(y^\pm)+\theta^a\psi_a(y^\pm)+\theta^+\theta^- F(y^\pm),
\end{equation}
transform as
\begin{equation}\label{Btrans2}
\delta_B \phi = \epsilon \eta \ , \ \ \ \ \ \ \ \ \delta _B\bar \phi = -\bar\epsilon \bar \eta \ ,
\end{equation}
\begin{equation}\label{Btrans3}
\delta_B \eta = -2 i \bar \epsilon \partial_0 \phi \ , \ \ \ \ \ \ \ \ \delta _B\bar \eta = 2 i  \epsilon \partial_0 \bar \phi \ ,
\end{equation}
\begin{equation}
\delta_B \beta = 2i \bar \epsilon \partial_1 \phi + 2\epsilon \xbar W'(\phi) \ , \ \ \ \ \ \ \ \ \delta_B \bar \beta =- 2i  \epsilon \partial_1 \bar \phi + 2\bar \epsilon  W'(\phi),
\end{equation}
where we used the basis
\begin{equation}
\eta:=\psi_-+\psi_+ \ , \ \ \ \ \ \ \beta:=\psi_--\psi_+ .
\end{equation}
A consequence of the above B-type transformation is that the $(2,2)$ bulk chiral field $X$ rearranges into a bosonic superfield $\Phi$ and a fermionic superfield $\Theta$ under B-supersymmetry. These two fields have the $\theta$-expansions,
\begin{equation}
\Phi= \phi(y)+\theta \beta (y),
\end{equation}
\begin{equation}
\Theta= \beta(y)-2\theta F(y)+2i\bar \theta [\partial_1\phi(y)+\theta\partial_1\beta(y)],
\end{equation}
where $y=x^0-i\theta\bar \theta$ is the boundary version of the  bulk $y^\pm$ arguments.

The consideration of boundaries not only reduces by half the amount of allowed supersymmetry but it also breaks the invariance of the D-term, even after restricting to B-type variations. This is due to the appearance of boundary contributions in the integral. Recall that in this work we are interested in a LG with single chiral field $X$ and action
\begin{equation}\label{LGtotal}
S[X]=\displaystyle \int d^2 d^4 \theta \ \xbar X X + \real \displaystyle \int d^2 d^2\theta \ W(X)| _{\bar \theta ^\pm =0}.
\end{equation}
If $W=0$, then the boundary contribution to $\delta_B S$ can be compensated by the addition of the boundary term \cite{enger05}
\begin{equation}
S_{D,\partial\Sigma}=\frac{1}{2} \displaystyle \int dx^0 (\bar\beta \eta-\bar \eta \beta).
\end{equation}

Like the D-term, the F-term also contains a boundary contribution when we vary $\delta_B S_F$. Unlike the D-term, the F-term cannot be compensated by an additional boundary term. Indeed, the B-type variation of the $F$-term is given by \cite{hori00}
\begin{equation}\label{varyf}
\delta S_{ F}=2i\bar \epsilon\displaystyle\int_{\partial \Sigma}dtd\theta\  W( X)\big |_{\bar \theta=0}-2i\epsilon\displaystyle\int_{\partial \Sigma}dtd\bar \theta\ \xbar W( \xbar X)\big |_{ \theta=0}.
\end{equation}
To obtain the above result we write $S_F= S_W+ S_{\bar W}$ and note that the B-type variation of 
\begin{equation}
S_W:= \displaystyle \int _\Sigma d^2 x d^2\theta W(X)|_{\bar \theta^\pm =0},
\end{equation}
is given by
\begin{equation}\label{bvariationsteps}
\begin{split}
\delta S_{ W}&=\displaystyle\int_\Sigma d^2xd\theta^-d\theta^+\ \delta_B  W( X)\big |_{\bar\theta^\pm=0}\\
&=\displaystyle\int_\Sigma d^2xd\theta^-d \theta^+\ \left( [\epsilon( Q_++Q_-)-\bar \epsilon (\xbar Q_++\xbar Q_- )]  W( X)\right)\big |_{\bar\theta^\pm=0}\\
&=\displaystyle\int_\Sigma d^2xd\theta^-d\theta^+\  2i \epsilon\left(  (\theta^-\partial _-   +\theta^+\partial _+ )W\right)\big |_{\bar\theta^\pm=0},
\end{split}
\end{equation}
where we used  
\begin{eqnarray}
&&Q_\pm W\big |_{\bar\theta^\pm=0} =\frac{\partial}{\partial \theta^\pm} W \big |_{\bar\theta^\pm=0}\ ,\\
&&\xbar Q_\pm W\big |_{\bar\theta^\pm=0}=-2i\theta^\pm\partial_\pm W \big |_{\bar\theta^\pm=0} .
\end{eqnarray}
The last equation follows from $W(X)$ being chiral so we can write $\xbar Q_\pm W=(\xbar D_\pm -2i\theta^\pm\partial_\pm) W$. The rest of the steps follow exactly as those used to show Eq. (\ref{varytwistedf}) using the chiral field expansion of $W$ as in Eq. (\ref{twistexpansion})
\begin{equation}
 X=\phi(y)+\theta^+\psi_+( y)+\theta^-\psi_-( y)+\theta^+ \theta^-F( y).
\end{equation}
One finally obtains,
\begin{equation}
\delta_B S_{ W}=2i\epsilon\displaystyle\int_{\partial \Sigma}dtd\theta\  W( X)\big |_{\bar \theta=0}.
\end{equation}
And the antiholomorphic part $S_{\bar W}$ follows the same way by noting that 
\begin{align}
\xbar{\delta_B W}(\xbar X)=-\delta_B \xbar{W}(\xbar X).
\end{align}

To understand why no combination of bulk fields can be utilized to construct a boundary term which compensates the variation of the F-term we use the component fields. In components we see that the boundary contribution is
\begin{equation}\label{componentFvary}
\delta_B S_F= \displaystyle \int _{\partial \Sigma} dx^0 \ \left(\epsilon \bar \eta\xbar W' (\phi) +\bar \epsilon \eta W'(\phi)\right).
\end{equation}
As noted above, there is no possible boundary term whose B-type variation would can cancel $\delta_B S_F$. Such a boundary term would have to vary with term like $\epsilon \bar \eta$ but from the right-hand side of all the component variations in Eq. (\ref{Btrans1}) - Eq. (\ref{Btrans3}) one sees that such B-type variation is not possible. One approach to make the action B-type supersymmetric is consider only chiral superfields with appropriate boundary conditions such that the right-hand side of Eq. (\ref{componentFvary}) vanishes. Such approach is studied in \cite{warner95}. A more general approach that does not restrict the class of chiral superfields involves introducting a new set of boundary theory whose action variation cancels that of the bulk theory \cite{Orlov:2003yp, Lazaroiu:2003zi, kapustin02}. This second approach leads to matrix factorizations which we will explore in the next section.

%%%%%%%%%%%%%%%%%%%%%%%%%%%%%%%%%%%%%%%%%%%%%%
\section{B-type boundaries and matrix factorizations}\label{BtypeSec}

In this section we specialize to B-type supersymmetry and review the use of matrix factorizations to describe B-supersymmetric boundary conditions. 

As noted in the previous section, in order to ensure that the action for the LG model in Eq. (\ref{LGtotal}) remains invariant under B-type supersymmetry one introduces a boundary theory. To this end, one defines a boundary superfield $\Pi$ which is fermionic and not chiral. That is, 
\begin{equation}\label{bdyfield}
\xbar D \Pi=  E(X_{\partial}) \neq 0,
\end{equation}
where $X_\partial :=X|_{\partial\Sigma}$ is the boundary restriction of the bulk superfield. The $\theta$-expansion of $\Pi$ is
\begin{equation}
\Pi=\pi(y)+\theta l(y)- \bar \theta \left( E(\phi)+\theta \eta(y) E'(\phi)\right),
\end{equation}
where $E$ is the source function in Eq. (\ref{bdyfield}). The boundary action is given by a kinetic D-term and an F-term that couples the bulk fields (that is, their boundary restriction) to the boundary fields,
\begin{equation}\label{bdyaction}
S_{\partial\Sigma}=\displaystyle \int_{\partial \Sigma} dt  d^2 \theta \ \xbar \Pi \Pi +\real \left(  i\displaystyle \int_{\partial\Sigma} dt d\theta \ J \Pi\big |_{\bar \theta=0}\right),
\end{equation}
for some function $J:=J\left(X_{\partial}\right)$. A more general form for the boundary coupling of B-type topological Landau-Ginzburg models is discussed in \cite{Lazaroiu:2003zi} but we do not need it here. 

The B-supersymmetry variation of the component fields of $\Pi$ is given by \cite{enger05}
\begin{equation}
\delta_B \pi = \epsilon l -\bar \epsilon E(\phi) \ , \ \ \ \ \ \ \ \ \ \delta_B\bar \pi = \bar \epsilon \bar l - \epsilon \bar E(\bar \phi) ,
\end{equation}
\begin{equation}
\delta_B l = -2 i \bar \epsilon \partial_0 \pi +\bar \epsilon \eta E'(\phi) \ , \ \ \ \ \ \ \ \ \ \delta_B \bar l = -2 i  \epsilon \partial_0 \bar \pi -  \epsilon \bar \eta \bar E'(\bar \phi).
\end{equation}
Inserting the equation of motion $l = -i \bar J(\bar \phi)$, the action of the boundary component fields is given by
\begin{equation}
S_{\partial \Sigma}= \displaystyle \int dx^0 \left( 2i\bar \pi \partial_0 \pi - |J|^2 - |E|^2 +i \pi \eta J' + i \bar \pi \bar \eta \bar J' - \bar \pi \eta E' +\pi \bar \eta \bar E' \right).
\end{equation}
Computing the B-type variation of the above action, one gets
\begin{equation}
\delta_B S_{\partial \Sigma} = -i \displaystyle \int dx^0 \left( \epsilon \bar \eta (\bar E \bar J)' +\bar \epsilon\eta (EJ)' \right).
\end{equation}
Comparing the above integral with boundary contribution to $\delta_B S_F$ as given in Eq. (\ref{componentFvary}), we see that both terms cancel each other to give supersymmetric invariance if and only if
\begin{equation}\label{simpleFact}
W = EJ,
\end{equation}
up to an additive scalar constant. Aside from giving the necessary and sufficient condition for B-type invariance, the above result is the cornerstone upon which the theory of matrix factorizations is built.

Similar to the bulk theory where there is a BRST operator $Q$ whose cohomology catalogs the physical fields, the boundary theory also has such an operator which is labeled by $Q_\partial$. The operator $Q_\partial$ is the boundary contribution to the BRST operator for the theory defined on $\Sigma$. Using $l = -i \bar J(\bar \phi)$ the variations of the fermionic component fields are
\begin{equation}\label{BvaryPi}
\delta_B \pi = -i\epsilon\bar J(\bar\phi)  -\bar \epsilon E(\phi) \ , \ \ \ \ \ \ \ \ \ \delta_B\bar \pi = i\bar \epsilon J(\phi) - \epsilon \bar E(\bar \phi),
\end{equation}
from which we obtain the relations
\begin{equation}\label{classes}
Q\pi =E(\phi) \ , \ \ \ \ \ \ \ \ Q\bar\pi =-iJ(\phi).
\end{equation}
The BRST cohomology for the theory with a boundary can be identified from the B-type variations of the component fields and the boundary fermions as in Eq. (\ref{BvaryPi}). The equivalence classes of the boundary cohomology depend on the boundary potentials $J$ and $E$ via Eq. (\ref{classes}). Therefore, the boundary spectra of a LG on a worldsheet with boundary is determined by $(J,E)$-factorizations of the superpotential $W$ as in Eq. (\ref{simpleFact}).

The boundary contribution to the $Q_\partial$ to the BRST operator is determined from Eq. (\ref{classes}) from which it follows that,
\begin{equation}
Q_\partial =\sum_i J_i\pi_i +E_i\bar \pi_i.
\end{equation}
In the above we are allowing for a set $\left\{ \Pi_i\right\}_i$ of boundary superfields. Using the anticommutation of the fermions $\pi_i$ one obtains that 
\begin{equation}
 Q_\partial ^2 =W.
\end{equation}
Thus, just as in the bulk, the physical fields at the boundary are those which are $Q_\partial$. This observation follows from the above equation and by noting that $Q_\partial(X) = [Q_\partial, X]$ for  a boundary superfield $X$. A representation $V$ of the Clifford algebra obeyed by the fermions $\left\{\pi_i,\bar\pi_i\right\}_{i=1,\dots,r}$ is $\mathbb{Z}_2$-graded as $V=V_0\oplus V_1$. Choosing such a representation, the boundary BRST operator $Q_\partial$ obtains the form
\begin{equation}
Q_\partial= \begin{pmatrix} 0 & p_1 \\ p_0 & 0 \end{pmatrix},
\end{equation}
where the $p_i$ are $2^r \times 2^r$-matrices with polynomial entries on the chiral fields such that 
\begin{equation}
p_1p_0 = p_0p_1 = W 1_{2^r \times 2^r},
\end{equation}
which is an example of a matrix factorization. More generally,
%\begin{df}
given a polynomial $W\in A:=\mathbb{C}[X_j]$, a \emph{matrix factorization} of $W$ is a pair $(p_0, p_1)$, $p_i \in M_{k}(A)$\index{a@$M_{k}(A)$, space of $k\times k$ matrices over ring $A$}, such that $p_0 p_1 = p_1 p_0 = W 1_k$.
One denotes matrix factorizations in the following way \cite{brunner07}
\begin{equation}\label{mfactor}
P=\left(  P_0 \mathrel{\mathop{\rightleftarrows}^{\mathrm{p_0}}_{\mathrm{p_1}}} P_1\right), \ \ \  p_0 p_1 =W 1_{P_1}, \ \ p_1 p_0 =W 1_{P_0}.
\end{equation}
The \emph{rank} of matrix factorization $P$ is the rank of the maps $p_0$, $p_1$.
The simplest example of a matrix factorization is the trivial matrix factorization of the form
\begin{equation}
P=\left(  P_0 =\mathbb{C}[X_j] \mathrel{\mathop{\rightleftarrows}^{\mathrm{1}}_{\mathrm{W}}} \mathbb{C}[X_j]=P_1\right).
\end{equation}

As an example let $B$ be a boundary condition for a LG model with superpotential $W=X^d+Y^d$. Then a possible matrix factorization for such a boundary is given by the maps  
\begin{equation}
p^0_{ij}=\begin{pmatrix}  X^{d-i} & Y^j \\ -Y^{d-j} & X^i \end{pmatrix}  \ , \ \ \ \ p^1_{ij}=\begin{pmatrix}  X^{i} & -Y^j \\ Y^{d-j} & X^{d-i} \end{pmatrix}
\end{equation}
going between the rank-2 spaces $P_0\cong P_1\cong \mathbb{C}[X,Y]^2$. One checks that 
\begin{equation}
p_0p_1 = p_1p_0 = (X^d+Y^d)1_{2\times 2}.
\end{equation}

%%%%%%%%%%%%%%%%%%%%%%%%%%%%%
\section{B-type defects in LG models}

In the above section we saw that there is a correspondence between boundary data (i.e., or $Q_\partial$-cohomology classes) of a LG with boundary and matrix factorizations of the superpotential. Such a description applies also to defects between two LG models with the only difference that now the matrix factorization is of the polynomial $W=W_1(X)-W_2(Y)$, where $W_1$ and $W_2$ are the superpotentials of the LG models at either side of the defect \cite{brunner07}. That is, a defect $D$ located at $x^1=0$ on $\mathbb{R}^2$ which separates a LG with chiral superfield $X$ and superpotential $W_1(X)$ on the upper-half plane, from a LG with chiral superfield $Y$ and superpotential $W_2(Y)$ on the lower-half plane is characterized by the matrix factorization
\begin{equation}\label{}
P=\left(  P_0 \mathrel{\mathop{\rightleftarrows}^{\mathrm{p_0}}_{\mathrm{p_1}}} P_1\right),
\end{equation}
where $P_1$ and $P_0$ are $\mathbb{C}[X,Y]$-modules with
\begin{equation}
 p_0 p_1 =(W_1(X)-W_2(Y))  1_{P_1}, \ \ \ \ p_1 p_0 =(W_1(X)-W_2(Y)) 1_{P_0}.
\end{equation}

The above description of defects between LG models is consistent with the ``folding trick'' prescription of \cite{affleck}. This procedure can be extended to the LG language as follows.  Consider the defect  $D=\left\{ (x,t) | x=0\right\}$ separating two Landau-Ginzburg models  $\text{LG}_1(X_i,\xbar X_i)$ and $\text{LG}_2(Y_i,\xbar Y_i)$, where the arguments are the chiral and anti-chiral fields. The action functional is of the form
\begin{equation}
\text{LG}_1(Z_i,\xbar Z_i)=\displaystyle \int d^2x \int d^4\theta K(Z_i,\xbar Z_i) +\int d^2x \real \int d^2\theta W(Z_i),
\end{equation}
for each theory. $K$ is the K\"ahler potential and $W$ the superpotential. We do the folding by interchanging the left and right movers in the left half-plane theory. Taking the mirror along $x=0$ sends $x\rightarrow -x$, so the action for $\text{LG}_2(Y_i,\xbar Y_i)$ goes from
\begin{equation}
\text{LG}_2(Y_i,\xbar Y_i)=\displaystyle \int_{-\infty}^0dx\int_\mathbb{R} dt \int d^4\theta K_2(Y_i,\xbar Y_i) +\int_{-\infty}^0dx\int_\mathbb{R} dt \real \int d^2\theta W_2(Y_i)
\end{equation}
to
\begin{equation}
\xbar{\text{LG}}_2(Y_i,\xbar Y_i)=\displaystyle \int_0^\infty dx\int_\mathbb{R} dt \int d^4\theta K_2(Y_i,\xbar Y_i) -\int_0 ^\infty dx\int_\mathbb{R} dt \real \int d^2\theta W_2(Y_i),
\end{equation}
where we noted  $d^4\theta$ is left invariant while  $d^2\theta \rightarrow -d^2\theta$; and $K$ is real valued. Thus the folded theory is described by 
\begin{align}
\text{LG}_1(X_i,\xbar X_i)\ \otimes\  \xbar{\text{LG}}\ {}_2 (Y_i,\xbar Y_i)=&\displaystyle \int d^2x \int d^4\theta (K_1(X_i,\xbar X_i)+K_2(Y_i,\xbar Y_i))\nonumber\\
&-\int d^2x \real  \int d^2\theta (W_1(X_i)- W_2(Y_i)),
\end{align}
on the right half-plane. This description is consistent with the theory of matrix factorizations for boundaries in the Landau-Ginzburg set up. A boundary is described by a matrix factorization of the superpotential of the LG. Thus to describe the boundary of the folded theory we factorize the superpotential $W=W_1-W_2$ which is the prescribed factorization of the defect $D$ before the folding.

As an example of a matrix factorization for a defect $D$ separating two LG models with $W_1=X^d$ and $W_2=Y^d$ consider $P_0=P_1= \mathbb{C}[X,Y]$ and the maps
\begin{equation}
p_0^I=\prod_{a\in I}(X-\eta^a Y) \ \ \ , \ \ \ p_1^I=\prod_{a\in \left\{0,\dots, d-1\right\} - I}(X-\eta^a Y) \ ,
\end{equation}
where $I\subset \left\{0,\dots, d-1\right\} $, and $\eta$ primitive $d^{th}$ root of unity.

%%%%%%%%%%%%%%%%%%%%%%%%%%%%%
\subsection{Fusion of defects}

As described in the introduction, the usefulness of defects comes via the natural binary operation of fusion where defects $D_1$ and $D_2$ can be brought together to form a new defect $D_3$. This fusion of defects, denoted by $D_3=D_1*D_2$, is obtained through the tensor product of matrix factorizations \cite{brunner07}. To review this composition let us consider the upper-half plane $\Sigma$ with a defect located at $x^1=y$ which separates two LG models with superpotentials $W_1(X_i)$ and $W_2(Y_i)$. 
\setlength{\unitlength}{1.2cm}
\begin{picture}(6,6)(-3,-3)
\put(-2.5,-1){\vector(1,0){5}}
\put(2.56,-1.15){$x^0$}
\put(0,-1.0){\vector(0,1){3}}
\put(-0.35,1.72){$x^1$}
\multiput(-2.5,0.42)(0.4,0){13}
{\line(1,0){0.2}}
\put(-0.35,.27){$x^1=y$}
\put(1.5,1.1){$W_1(X_i)$}
\put(1.5,-.3){$W_2(Y_i)$}
\put(-2.5,.5){$D$}
\put(-2.2,1.5){$\Sigma$}
\put(-2.5,-.9){$B(\partial\Sigma)$}
\end{picture}

To $D$ we associate a matrix factorization $P$ and to the boundary data for the lower LG we associate a matrix factorization $Q$. The exact form of each matrix factorization depends on the on the type of defect and the boundary data, respectively. This fusion is obtained in this case by letting $y \rightarrow 0$ which produces a new boundary condition $D*B=:B'$ at $\partial \Sigma$. Since the lower LG disappears the new boundary condition is for the upper LG.

In terms of matrix factorizations, the resulting boundary condition is given by the tensor product of the matrix factorizations $P$ and $Q$. The definition of this tensor product is given below. Let $D$ be the defect above and $P$ the corresponding matrix factorization,
\begin{equation}\label{defect}
P: \ \ \  P_0 \mathrel{\mathop{\rightleftarrows}^{\mathrm{p_0}}_{\mathrm{p_1}}} P_1, \ \ \  p_0 p_1 =(W_1(X_i)-W_2(Y_i)) 1_{P_1}, \ \ p_1 p_0 =(W_1(X_i)-W_2(Y_i)) 1_{P_0}.
\end{equation}
Let  $Q$ be the corresponding matrix factorization corresponding to the boundary condition $B$,
\begin{equation}\label{boundary}
Q: \ \ \  Q_0 \mathrel{\mathop{\rightleftarrows}^{\mathrm{q_0}}_{\mathrm{q_1}}} Q_1, \ \ \  q_0 q_1 =W_2(Y_i) 1_{Q_1}, \ \ q_1 q_0 =W_2(Y_i) 1_{Q_0}
\end{equation}
Then the limit $y\rightarrow 0$ defines a new boundary condition $B'$ given by the \emph{tensor product matrix factorization} which is defined as
\begin{equation}\label{formaltensor}
Q': \ \ \  Q'_0=\left( P_0\otimes_{\mathbb{C}[Y_i]} Q_0\right)\oplus \left( P_1\otimes_{\mathbb{C}[Y_i]} Q_1\right) \mathrel{\mathop{\rightleftarrows}^{\mathrm{q'_0}}_{\mathrm{q'_1}}} \left( P_1\otimes_{\mathbb{C}[Y_i]} Q_0\right)\oplus \left( P_0\otimes_{\mathbb{C}[Y_i]} Q_1\right)=Q'_1,
\end{equation}
where,
\begin{equation}\label{mapfusion}
q'_0 = \begin{pmatrix} p_0\otimes 1_{Q_0} & 1_{P_1}\otimes q_1 \\ -1_{P_0}\otimes q_0 & p_1\otimes 1_{Q_1}  \end{pmatrix}\   , \ \ \ \ 
q'_1 = \begin{pmatrix} p_1\otimes 1_{Q_0} & -1_{P_0}\otimes q_1 \\ 1_{P_1}\otimes q_0 & p_0\otimes 1_{Q_1}  \end{pmatrix}.
\end{equation}
One can see that the matrix factorization $Q'=P\otimes Q$ resulting from the matrix factorization tensor product indeed factorizes $W_1(X_i)$. The matrix factorization $Q'$ is of infinite rank as a $\mathbb{C}[X_i]$-module. That is, the maps $q'_i$ have rank = $\infty$. Since we started with finite-rank matrix factorizations, we would like the resulting tensor product to be also of finite rank.

 If the two initial defects are of finite rank, the infinite rank of the tensor product comes from trivial matrix factorizations which can be ``peeled off'' to obtain a reduced rank matrix factorization. To obtain the reduced rank matrix factorization resulting from equation (\ref{formaltensor}) more directly, one associates to each matrix factorization $P$ a 2-periodic $\mathbb{C}[X]/W$-resolution of the space $\coker p_1$, the cokernel of the $p_1$ map. Then the problem of computing $Q'$, the matrix factorization corresponding to the tensor product of $P$ and $Q$, is translated into finding $\coker q_1'$ in its reduced form. As noted in \cite{brunner07}, at the level of $\mathbb{C}[X]/W$-modules both $\coker q_1'$ and the space
\begin{equation}\label{trick}
V=\coker(p_1\otimes1_{Q_0}, 1_{P_0}\otimes q_1),
\end{equation}
have resolutions which are identical up to the last two steps. Therefore if we can find the reduced form of $V$, we can identify the 2-periodic resolution corresponding to the matrix factorization $Q'$. It turns out that it is simpler to work out the reduced form of $V$ since its components are the known maps of the original two matrix factorizations.

It is helpful to see the technique described above using an example: Consider the matrix factorizations 
\begin{equation}\label{Xdand0}
P^{N}(X|Y)=  \left( P_1=  \mathbb{C}[X,Y] \mathrel{\mathop{\rightleftarrows}^{ X^N}_{X^{d-N}}} \mathbb{C}[X,Y]=P_0\right),
\end{equation} 
and
\begin{equation}\label{boundarylm}
Q^{L,M}(Y)=\left(  Q_1=  \mathbb{C}[Y] \mathrel{\mathop{\rightleftarrows}^{ X^M}_{0}} \mathbb{C}[Y]=Q_0\right).
\end{equation}
Using the formal expression for the tensor product given in Eq. (\ref{formaltensor}) leads to a matrix factorization $Q'$ with the factorizing maps 
\begin{equation}
q'_0 = \begin{pmatrix} X^{d-N} & Y^M \\ 0 & X^N  \end{pmatrix}\ \  , \ \ \ \
q'_1 = \begin{pmatrix} X^N & -Y^M \\0  & X^{d-N} \end{pmatrix}.
\end{equation}
We want to show that $Q'$ is equivalent to a matrix factorization of finite rank. For this we treat the spaces as $\mathbb{C}[X]$-modules by using the matrix representation where $Y^i$ corresponds to the matrix with zeros in all the entries except in for those lying in the off-diagonal $(k+i,k)$ which are set to 1. In this representation $Y^0$ is the infinite identity matrix. The map $q_1'$ is then given by the matrix

%\begin{equation}
%\sbox0{\begin{matrix}X^N& &\\ & %\ddots & \\ & & \end{matrix}}
%\sbox1{\begin{matrix}0& & &\\ \vdots %&  & & \\  -1& & &\\  & & \ddots & & %\end{matrix}}
%\sbox2{\begin{matrix}X^{d-N}& &\\ & %\ddots & \\ & & \end{matrix}}
% this line stays green
%q_1 '=\left[
%\begin{array}{c|c}
%\usebox{0}&\usebox{1}\\
%\hline
%  \vphantom{\usebox{0}}\makebox[\wd0]%{\large 0}&\usebox{2}
%\end{array}
%\right].
%\end{equation}

\begin{equation}
\sbox0{$\begin{matrix}X^N& &\\ & \ddots & \\ & & \end{matrix}$}
\sbox1{$\begin{matrix}0& & &\\ \vdots &  & & \\  -1& & &\\  & & \ddots & & \end{matrix}$}
\sbox2{$\begin{matrix}X^{d-N}& &\\ & \ddots & \\ & & \end{matrix}$}
q_1 '=\left[
\begin{array}{c|c}
\usebox{0}&\usebox{1}\\
\hline
  \vphantom{\usebox{0}}\makebox[\wd0]{\large $0$}&\usebox{2}
\end{array}
\right]
\end{equation}

Using elementary row and column operations the above matrix can be set equal to
 
%\begin{equation}\label{q1prime}
%\sbox0{\begin{matrix}X^N& & &\\& %\ddots & &\\& & 1 & \\& & &\ddots %\end{matrix}}
%\sbox1{\begin{matrix}0& & &\\ \vdots %&  & & \\  -1& & &\\  & & \ddots & & %\end{matrix}}
%\sbox2{\begin{matrix}X^d& &\\ & \ddots & \\ & & \end{matrix}}
%
%q_1 '=\left[
%\begin{array}{c|c}
%\usebox{0}& \makebox[\wd0]{\large 0}\\
%\hline
%  \vphantom{\usebox{0}}\makebox[\wd0]{\large 0}&\usebox{2}
%\end{array}
%\right],
%\end{equation}

\begin{equation}\label{q1prime}
\sbox0{$\begin{matrix}X^N& & &\\& \ddots & &\\& & 1 & \\& & &\ddots \end{matrix}$}
\sbox1{$\begin{matrix}0& & &\\ \vdots &  & & \\  -1& & &\\  & & \ddots & & \end{matrix}$}
\sbox2{$\begin{matrix}X^d& &\\ & \ddots & \\ & & \end{matrix}$}
q_1 '=\left[
\begin{array}{c|c}
\usebox{0}& \makebox[\wd0]{\large $0$}\\
\hline
  \vphantom{\usebox{0}}\makebox[\wd0]{\large $0$}&\usebox{2}
\end{array}
\right],
\end{equation}
where the first $M$ entries of the diagonal are $X^N$.

And similarly for $q'_0$,
%\begin{equation}\label{q0prime}
%\sbox0{\begin{matrix}X^{d-N}& & &\\& \ddots & &\\& & X^d & \\& & &\ddots \end{matrix}}
%\sbox1{\begin{matrix}0& & &\\ \vdots &  & & \\  1& & &\\  & & \ddots & & \end{matrix}}
%\sbox2{\begin{matrix}1& &\\ & \ddots & \\ & & \end{matrix}}
%
%q_0 '=\left[
%\begin{array}{c|c}
%\usebox{0}& \makebox[\wd0]{\large 0}\\
%\hline
%  \vphantom{\usebox{0}}\makebox[\wd0]{\large 0}&\usebox{2}
%\end{array}
%\right],
%\end{equation}
\begin{equation}\label{q0prime}
\sbox0{$\begin{matrix}X^{d-N}& & &\\& \ddots & &\\& & X^d & \\& & &\ddots \end{matrix}$}
\sbox1{$\begin{matrix}0& & &\\ \vdots &  & & \\  1& & &\\  & & \ddots & & \end{matrix}$}
\sbox2{$\begin{matrix}1& &\\ & \ddots & \\ & & \end{matrix}$}
q_0 '=\left[
\begin{array}{c|c}
\usebox{0}& \makebox[\wd0]{\large $0$}\\
\hline
  \vphantom{\usebox{0}}\makebox[\wd0]{\large $0$}&\usebox{2}
\end{array}
\right],
\end{equation}
where the first $M$ entries of the diagonal are $X^{d-N}$. From Eq. (\ref{q0prime}) and Eq. (\ref{q1prime}) we see that the matrix factorization we have obtained is equivalent to a direct sum of a rank-$M$ matrix factorization with maps
\begin{equation}
q_1'= X^N 1_{M\times M} \ , \ \ \ \ \ q_0'= X^{d-N} 1_{M\times M},
\end{equation}
plus an infinite direct sum of rank-1 trivial matrix factorizations.

On the other hand,  $P\otimes Q$ can be determined by $V = \coker m$ as in Eq. (\ref{trick}). Here $V= P_0\otimes Q_0 / \left\{p_1,q_1\right\}$. Since we want a $\mathbb{C}[X]$-module we use the $Y^M=0$ condition to treat $V$ as generated over the $M$ generators $Y^i$, $i=0,\dots, M-1$. So we have
\begin{equation}
V= {}_{\mathbb{C}[X]}\left\{Y^i \right\}_{i=0,\dots, M-1}/ \left\{ X^N=0\right\},
\end{equation}
which we recognize as $V\cong \coker q_1'$ of the matrix factorization  $Q'=Q^N_{\operatorname{rank} M}$ which denotes the rank-$M$ version of $Q^N$. So in the non-orbifolded case we obtain the simpler product $P^N*Q^M =Q^N_{\operatorname{rank} M} $.

%##### NEW SECTION   #################
%################################
\section{Describing RG flows in $\mathbb{C}/\mathbb{Z}_d$ orbifolds using defects}

In this section we describe a new way of dealing with the $\mathbb{C}/\mathbb{Z}_d$ orbifold in terms of defects. The language of matrix factorizations can be utilized to describe the RG flow between the $\mathbb{C}/\mathbb{Z}_d$ orbifolds. This can be done directly by considering the Lagrangian of the model as equivalent to that of a LG model with superpotential $W=0$. So any defects between $\mathbb{C}/\mathbb{Z}_m$ and $\mathbb{C}/\mathbb{Z}_n$ become a problem of factorizing the zero polynomial.

Since we are working with B-type supersymmetry we need to use a perturbation which preserves this type. Such a perturbation for a $\mathcal{N}=(2,2)$ theory is done using twisted chiral fields $\Psi$ in theory with the integrals
\begin{equation}
\delta S = \displaystyle \int_\Sigma d^2x d\bar x^- d\theta^+ \ \Psi \big|_{\bar \theta^+ = \theta^- =0}.
\end{equation}
But the $\mathcal{N}=(2,2)$ supersymmetry dictates that the parameters of the superpotential and twisted superpotential remain decoupled under the RG flow \cite{hori02}. This fact means that the structure of the twisted chiral sectors is independent of the specific superpotential. Especially in our case whether there is one or not. Therefore the spectrum of the twisted chiral sectors between  $\mathbb{C}/\mathbb{Z}_d$  and the $\mathbb{Z}_d$-orbifolded LG with $W=X^d$ are equivalent, and their B-type preserving perturbations can be mapped to each other.  With this observation we set out to check that the sort of defects presented in \cite{brunner07a} describing the RG flow defects coming from such perturbations over a subset of $\Sigma$, can be extended to the non-compact orbifolds and the RG flows between them.

%% %%%%%%%  SUBSECTION    ##############

\subsection{$\mathbb{C}/\mathbb{Z}_d$ as an $\text{LG}/ \mathbb{Z}_d$ with $W=0$}\label{rgflowDefects}

Superstring theory on the space $\mathbb{C}/\mathbb{Z}_d$ can be described by a chiral superfield
\begin{equation}
\Phi=\phi(y^{\pm})+\theta^\alpha\psi_\alpha(y^\pm)+ \theta^+\theta^- F(y^\pm),
\end{equation}
where $y^\pm=x^\pm-i\theta^\pm\bar\theta^\mp$.  The action takes the form
\begin{equation}\label{LGnoW}
S=\displaystyle  \int d^2 x d^4\theta \ \xbar \Phi \Phi +0,
\end{equation}
where we included the zero to emphasize that we have a LG model with superpotential $W=0$ in the D-term. In this way we can construct defects between different  $\mathbb{C}/\mathbb{Z}_d$ orbifolds and describe them in terms of matrix factorizations. Indeed, we check that when two $\mathbb{C}/\mathbb{Z}_d$  theories are related by an RG flow, we can juxtapose them with a corresponding defect which maps the boundary conditions accordingly.

Matrix factorizations of the zero polynomial work in exactly the same way as the case for any other polynomial. As an example of this we consider the fusion of a defect between two orbifolded theories; the upper one with superpotential $W_1(X)=X^d$ and the lower one with $W_2(Y)$ the zero superpotential but orbifold group $\mathbb{Z}_{d'}$. The simplest such defect is given by
\begin{equation}\label{defectNoW}
D^{m,n,N}(X|Y)=\left(  D_1=  \mathbb{C}[X,Y][m,-n] \mathrel{\mathop{\rightleftarrows}^{ X^N}_{X^{d-N}}} \mathbb{C}[X,Y][m-N,-n]=D_0\right),
\end{equation}
where $[\cdot, \cdot]$ is the $\mathbb{Z}_d \times\mathbb{Z}_{d'}$ grading. We see that $d_1 d_0 = X^d-0 = W_1(X)-W_2(Y)$. In the lower theory, the boundary conditions corresponding to rank-1 matrix factorizations are a direct sum of the irreducible matrix factorizations of the form
\begin{equation}\label{boundaryNOw}
Q^{L,M}(Y)=\left(  Q_1=  \mathbb{C}[Y][L+M] \mathrel{\mathop{\rightleftarrows}^{ Y^M}_{0}} \mathbb{C}[Y][L]=Q_0\right),
\end{equation}
where $L\in\mathbb{Z}_{d}$ labels the irreducible representations.

If the defect $D^{m,n,N}$ sits at $x^1=y$ and we take $y\rightarrow 0$ the fusion of the defect and the boundary condition is given by tensor product of both matrix factorizations. This is obtained by looking at $\coker f = D_0\otimes Q_0/ \operatorname{im} f$ where $f=(d_1\otimes1_{Q_0}, 1_{D_0} \otimes q_1)$ \cite{enger05}. We denote the $\mathbb{C}[X,Y]$-generators of $D_0$ and $Q_0$ by $e^{D_0}_{m,n}$ and $e^{Q_0}_{L}$, respectively. Then as a $\mathbb{C}[X]$-module, $\coker f$ is generated over $e^i:=Y^i e^{D_0}_{m,n}\otimes e ^{Q_0}_{L}$ modulo
\begin{equation}
X^N e^i=0\ \ \ , \ \ \ e^{i+M}= 0, \ \ \ \forall i\geq 0.
\end{equation}
The second condition means that $V$ has rank $M$. Note that $e^i$ has $\mathbb{Z}_d \times\mathbb{Z}_{d'}$-degree $[m-N,L-n+i]$, but under fusion we are left with a $\mathbb{Z}_{d}$ theory so we have to extract the $\mathbb{Z}_{d'}$-invariant subset $V^{\mathbb{Z}_{d'}}\subset V$. This means the $i$ is fixed to $i=n-L$, which means we are left with one generator with $\mathbb{Z}_{d}$-degree $m-N$ restricted to $X^N=0$.  Otherwise if $n-L \not\in [0,M-1]$ then $D^{m,n,N}*_{\text{orb}}Q^{L,M}=0$. In summary,

\begin{equation}\label{example1}
D^{m,n,N}*_{\text{orb}}Q^{L,M}= \left\{ \begin{array}{rl}
 Q^{m,N}, &\mbox{ if $n-L\leq M-1$,} \\
  0, &\mbox{ otherwise.}
       \end{array} \right.
\end{equation}

Another example of useful defects given by matrix factorizations of $W=0$ are those enforcing the action of the symmetry group. Similar to those in \cite{brunner07a} they are given by the $\mathbb{Z}_d\times\mathbb{Z}_d$- equivariant matrix factorization $T^m = (T^m_1, T^m_0; t_1, 0)$ with
\begin{equation}
T^m_1 = {}_{\mathbb{C}[X,Y]}\left\{ e^{1}_{m,k}\right\}_{(m,k) \in \mathbb{Z}_d\times\mathbb{Z}_d} \ \ , \ \ \deg e^1_{m,k} =[m+k+1,-k],
\end{equation}
\begin{equation}
T^m_0 = {}_{\mathbb{C}[X,Y]}\left\{ e^{0}_{m,k}\right\}_{(m,k) \in \mathbb{Z}_d\times\mathbb{Z}_d} \ \ , \ \ \deg e^0_{m,k} =[m+k,-k].
\end{equation}
The factorizing map is given by
\begin{equation}
t_1=\sum_{k=0}^{d-1} \left( X e_{m,k}^0\otimes {e^1}_{m,k} ^*- Y e_{m,k+1}^0\otimes {e^1}_{m,k} ^*\right),
\end{equation}
where $e^*$ is the basis dual to $e$.

One obtains the fusion rules
\begin{equation}
T^m *_{\text{orb}} T^n=T^{m+n},
\end{equation}
and
\begin{equation}
T^m *_{\text{orb}} Q^{M,N}=Q^{M+n ,N},
\end{equation}
where $D_1*_{\text{orb}} D_2$ means extracting the part of $D_1 * D_2$ which is invariant under the symmetry group of the theory between both defects $D_1$ and $D_2$. The sums are performed modulo $d$. Hence the defects $T^m$ form a representation of the symmetry group.

More importantly, we note that by also setting $p_0=0$ in the special defects introduced in \cite{brunner07a} we obtain defects which act as the interface between orbifolds sitting at opposite endpoints of the RG flow. The special defects are $\mathbb{Z}_{d'}\times\mathbb{Z}_d$ - equivariant matrix factorizations $P^{(m,\underline n)}$, with labels $m\in \mathbb{Z}_d$ and $n=(n_0,\dots, n_{{d'}-1})$ with $n_i\in\mathbb{N}_0$ such that $\sum_i n_i=d$. The $\mathbb{C}[X,Y]$-modules $P_1$ and $P_0$ and their  $\mathbb{Z}_{d'}\times\mathbb{Z}_d$-grading are given by,
\begin{equation}
P_1 = \mathbb{C}[X,Y]^{d'}
\begin{pmatrix} [1,-m] \\ [2,-m-n_1]  \\ [3,-m-n_1-n_2 ]\\ \vdots \\ [d', -m-\sum_{i=1} ^{d'-1}n_i ]\end{pmatrix} \ \ \ , \ \ \ P_0 = \mathbb{C}[X,Y]^{d'}
\begin{pmatrix} [0,-m] \\ [1,-m-n_1]  \\ [2,-m-n_1-n_2 ]\\ \vdots \\ [d'-1, -m-\sum_{i=1} ^{d'-1}n_i ]\end{pmatrix}.
\end{equation}
The factorizing maps are
\begin{equation}\label{mapwithzero}
p_1^{m,n}=Y 1_{d'}- \Xi_n(X)\ \ \ , \ \ \ p_0^{m,n} = 0 ,
\end{equation}
where $(\Xi_n(X))_{a,b}:=\delta^{(d')}_{a,b+1}X^{n_a}$.

As computed in \cite{brunner07a} the general rule for fusion of a special defect $P^{(m,n)}$ and a $\mathbb{Z}_d$-irreducible boundary condition $Q^{(M,N)}$ is
\begin{equation}\label{genfusion}
P^{(m,\underline n)}*Q^{(M,N)}=\bigoplus_{a\mathbb{Z}_{d'}:\ i(a)<\text{min}(N,n_a)} Q^{(a,k(a))},
\end{equation}
where $i(a)=\left\{n-M+\sum_{j=0}^an_j\right\}_d$.

One can check that special defects send the boundary condition $Q^{(M,1)}$ to another such boundary condition with $N=1$, $Q^{(M',1)}$.

Let $P^{(m,\underline n)}$ be a special defect and $Q^{(M,N=1)}$ an irreducible B-type boundary condition. Then their fusion is

\begin{equation}\label{defectbdy}
P^{(m,\underline n)}*Q^{(M,N=1)}=\begin{cases} 0, &  M\notin \mathcal{L}_{(m,\underline n)}\\ Q^{(a,1)}, & M=m+\sum_{i=1}^a n_i
\end{cases}
\end{equation}
where $\mathcal{L}_{(m,\underline n)}:=m+\left\{ n_0, n_0+n_1,\dots,n_0+n_1+\cdots n_{d'-1}\right\}$.

%%%%%%%%%%%%%%%%%%%%%%%%%%%
%subsection 
\subsection{Comparison with RG flow in the $\mathbb{C}/\mathbb{Z}_d$ theories}
We can compare the result for the fusion of the defects $P^{m,n}$ with boundary conditions $Q^{M,N}$ of the LG model with zero superpotential with the RG flow between the $\mathbb{C}/\mathbb{Z}_d$ orbifolds. 
 For this purpose we describe the RG flow in these models by looking at their chiral rings.

Upon bosonizing the fermionic fields of the superstring theory, one can construct the chiral operators given in \cite{harvey01}
\begin{equation}
X_j = \sigma_{j/n}\exp[i(j/n)(H-\xbar H)] \ \ , \ \ j =1,\dots, n-1,
\end{equation}
where $\sigma_{j/n}$ is the bosonic twist operator. These operators are the bosonic components of the respective chiral fields which we will also denote by $X_j$. The higher chiral fields are powers of $X:=X_1$. The chiral ring of this theory is generated by $X$ and
\begin{equation}
Y:=\frac{1}{V_2}\psi\psi= \frac{1}{V_2}\exp[i(H-\xbar H)],
\end{equation}
modulo
\begin{equation}\label{ringrel}
X^d=Y.
\end{equation}

Deformations of equation (\ref{LGnoW}) by the following F-term preserve supersymmetry since the $X_j$ fields are chiral,
\begin{equation}\label{deflangrangian}
\delta L = \sum_{j=1} ^{n-1} \lambda^j \displaystyle \int  d^2\theta \ X_j.
\end{equation}
The deformed theory has a chiral ring with the same fields as before but with relation in equation \ref{ringrel} altered to
\begin{equation}\label{deformed}
X^d+\sum_{j=1} ^{d-1}g_j(\lambda)X^j=Y,
\end{equation}
where $g_j(\lambda)$ are polynomials in the couplings \cite{harvey01}. A deformation such as in equation (\ref{deflangrangian}) induces a RG flow in the theory. By considering the case where $g_i=0$ for $i\leq d'-1$, the IR and UV limits of the ring condition above are $X^d=Y$ and $g_{d'}X^{d'}=Y$ respectively. These two are the conditions defining $\mathbb{C}/\mathbb{Z}_d$ and $\mathbb{C}/\mathbb{Z}_{d'}$, respectively.

We note that for every RG flow $\mathbb{C}/\mathbb{Z}_d \longrightarrow \mathbb{C}/\mathbb{Z}_{d'}$ there exists a matrix factorization $P^{(m,\underline n)}$ of $W=0$ representing a defect $D$ between $\mathbb{C}/\mathbb{Z}_d$ and $\mathbb{C}/\mathbb{Z}_{d'}$. Given two such bulk theories, we can juxtapose them via a defect $P^{(m,\underline n)}$ by choosing $m\in\mathbb{Z}_d$ and non-negative integers $\left\{n_0,n_1,\dots,n_{d'-1}\right\}$ subject to $n_0+\cdots n_{d'-1}=d$. The solution is a non-unique defect but that reflects the action of the overall $\mathbb{Z}_{d'}\times \mathbb{Z}_{d}$ symmetry. In the next section we will have a better description of how the boundary degrees of freedom are mapped from one theory to the other under fusion with RG flow defects.

As an example, consider the $\mathbb{Z}_5$ orbifold. In this case the chiral ring of the deformed theory is defined modulo $X^5+\sum_{j=1}^4 g_j(\lambda)X^j=Y$. If we set $g_1=g_2=0$, then the RG flow goes between $\mathbb{C}/\mathbb{Z}_5$ in the UV limit (since the theory's chiral ring has the relation $X^5=Y$) and  $\mathbb{C}/\mathbb{Z}_3$ (since in the IR limit the defining relation is $X^3=Y$). Then the defect $P^{(3,\underline n)}$ with $\underline n = (2,2,1)$ can sit at the interface  between the theories $\mathbb{C}/\mathbb{Z}_5$ and $\mathbb{C}/\mathbb{Z}_3$ such that B-type supersymmetry is preserved across the interface.

%%%%%%%%%%%%%%%%%%%%%%%%%%
%%%  SUBSECTION
\section{RG flows using mirror models}

A second strategy is to study the orbifold RG flow
in terms of the mirror of  $\mathbb{C}/\mathbb{Z}_d$ \cite{vafa01}. Using mirror symmetry we obtain the diagram below. In the following $m$ stands for mirror symmetry and $|_B$  for the B-type defects; $\text{LG}_m$ denotes the LG model with $W=X^m$;  and $\widetilde{\text{LG}}_m$ the twisted LG with $W=\widetilde X^m$. 

\begin{equation}\label{modeldiagram}
\begin{CD}
@. \text{LG}_m/\mathbb{Z}_m @ > \big |_B>> \text{LG}_n/\mathbb{Z}_n\\
@. @VV\cong V@ VV\cong V  @.\\
\mathbb{C}/\mathbb{Z}_m @ >m>> \widetilde{\text{LG}}_m @ . \ \ \ \widetilde{\text{LG}}_n @>m>> \mathbb{C}/\mathbb{Z}_n @ . \\
@. @VV m V@ VV m V  @.\\
@. \text{LG}_m @> RG >> \text{LG}_n\\
\end{CD}
\end{equation}

In the diagram above, the mirror mapping from   $\mathbb{C}/\mathbb{Z}_n$ to a twisted LG with non-vanishing potential comes from a mirror correspondence between a gauged linear sigma model (GLSM) and a more general LG theory. As detailed in \cite{vafa01, Hori:2000kt}, one considers a GLSM whose geometry is described by 
\begin{equation}
-d|X_0|^2 + \sum_{i=1}^n k_i |X_i|^2=t,
\end{equation}
where the fields $(X_0, X_i)$ come with $U(1)$ charges $(-d,k_i)$, and $t$ is the complexified Fayet-Iliopoulos (FI) parameter. Such GLSM is mirror to a LG theory with superpotential
\begin{equation}
\widetilde W=\sum_{i=1}^n Z_i^d +e^{t/d}\prod_{j=1}^n Z_j^{k_j},
\end{equation}
where the variables $Z_i$ are twisted chiral fields, and the superpotential is taken modulo $(\mathbb{Z}_d)^{n-1}$. The IR fixed point of the GLSM is obtained with the limit $t\rightarrow - \infty$. This limit breaks the $U(1)$ symmetry to $\mathbb{Z}_d$ and the geometry obtained is that of $\mathbb{C}^n/\mathbb{Z}_d$. In this note we consider the $n=1$ case, i.e. $\mathbb{C}/\mathbb{Z}_d$. On the mirror side, the $t\rightarrow - \infty$ limit gives us the LG with $\widetilde W= Z^d$. Thus we see that the RG flow between the non-compact orbifolds can be described in terms of matrix factorizations of true LG orbifolds with non-zero superpotentials.

\subsection{RG flow defects using mirror models}

The idea is that via mirror symmetry we can represent the $\mathbb{C}/\mathbb{Z}_d$ orbifold as a twisted LG model with superpotential $W=\widetilde X^d$. We denote this theory  by $\widetilde{ \text{LG}}_d$ in the above diagram. This theory is equivalent to the model $\text{LG}_d/\mathbb{Z}_d$,  the orbifold of a non-twisted LG model with superpotential $W=X^d$ by $\mathbb{Z}_d$. So we can use defects between these LG orbifolds to study the RG flow between the original $\mathbb{C}/\mathbb{Z}_d$ orbifolds.

As in the previous section we are again in the Landau-Ginzburg model so we can use the RG flows defects $P^{(m,\underline n)}$. The factorizing maps are as in equation (\ref{mapwithzero}) but with $p_0$ non-zero:

\begin{equation}
p_1^{m,n}=Y 1_{d'}- \Xi_n(X)\ \ \ , \ \ \ p_0^{m,n} = \prod_{i=1}^{d'-1}(Y 1_{d'}- \eta^i\Xi_n(X)),
\end{equation}
where $\eta$ is an elementary $d'th$ root of unity. And similarly, the irreducible matrix factorizations corresponding to these boundary conditions are of the same form as in equation (\ref{boundaryNOw}),
\begin{equation}\label{}
Q^{L,M}(Y)=\left(  Q_1=  \mathbb{C}[Y][L+M] \mathrel{\mathop{\rightleftarrows}^{ X^M}_{X^{d'-M}}} \mathbb{C}[Y][L]=Q_0\right).
\end{equation}

We review the graphical version introduced in  \cite{brunner07a} to depict the fusion of $P^{(m,\underline n)}$ with the boundary conditions $Q^{(L,M)}$. To the set $Q^{(\underline M,1)}:=\left\{Q^{(M,1)} :0\leq M\leq d-1\right\}$ the following graph is assigned: A disk divided into $d$ equal sections by segments from the origin to the boundary. One segment is decorated to start labeling the sections $S_i$ from $i=0$ to $s=d-1$. Below is such a graph for $d=4$:

\setlength{\unitlength}{5mm}
\begin{center}
\begin{picture}(10,10)(-6,-6)
\put(0,0){\circle{6}}
\put(0,0){\line(0,5){3}}
\put(0,0){\line(0,-5){3}}
\put(0,0){\vector(5,0){3}}
\put(0,0){\line(-5,0){3}}
\put(1.1,1.1){$S_0$}
\put(-1.1,1.1){$S_1$}
\put(-1.1,-1.1){$S_2$}
\put(1.1,-1.1){$S_3$}
\put(-.8,-4){{ $W=X^4 $}}
\end{picture}
\end{center}

%\setlength{\unitlength}{7mm}
%\begin{picture}(10,10)(-6,-6)
%\put(0,0){\circle{6}}
%\put(0,0){\line(1,1){2.1}}
%\put(0,0){\line(0,1){3}}
%\put(0,0){\line(-1,1){2.1}}
%\put(0,0){\line(0,-5){3}}
%\put(0,0){\line(5,0){3}}
%\put(0,0){\line(1,-1){2.1}}
%\put(0,0){\line(-5,0){3}}
%\put(.3,.3){$X_*$}
%\put(1.8,.9){$S_0$}
%\put(0,3.3){$P^2$}
%\put(-3.8,0){$P^3$}
%\put(0,-3.8){$P^4$}
%%\put(-.5,-5){{ $W=X^4$}}
%%\end{picture}

Using the graphical description described above, the special defects $P^{(m,n)}$ are represented by the operators
\begin{equation}\label{graphspecial}
\mathcal{O}^{(m,n)}:=\mathcal{T}_{-a(m,n)}\mathcal{S}_{\mathcal{L}^c(m,n)},
\end{equation}
where $\mathcal{L}_{(m,n)}$ is defined below (\ref{defectbdy}) and $a(m,n):=|\left\{0,\dots,m\right\}\cap \mathcal{L}_{(m,n)}|$. The operator $\mathcal{S}_{\left\{s_1,\dots, s_k\right\}}$  deletes the sectors $S_{s_j}$ by merging the segments which bound them. The operator $\mathcal{T}_k$ acts as the $\mathbb{Z}_d$-symmetry by shifting $M\rightarrow M+k$ in $Q^{(M,1)}$. So just like $P^{(m,\underline n)}$, the operator $\mathcal{O}^{(m,n)}$ annihilates the sectors associated to boundary conditions whose label $M$ does not belong in $\mathcal{L}_{(m,n)}$. Then it relabels the remaining sectors by setting the $S_{m}$ to $S_0$.

The above pictorial representation generalizes to boundary conditions $Q^{(M,N)}$ with $N>1$ as well. In this case, $Q^{(M,N)}$ corresponds to the union $S_M\cup S_{M+1}\cup\dots\cup S_{M+N-1}$. We want to show that the operators in the definition (\ref{graphspecial}) still represent the action of special defects on the boundary conditions in this $N>1$ case.

Represent $Q^{(M,N)}$ by $S^{(M,N)}:=S_{M}\cup\dots\cup S_{M+N-1}$ and assume that $\mathcal{S}_{\mathcal{L}^c_{(m,n)}}$ shrinks $S^{(M,N)}$ to nothing. Then $\left\{M,M+1,\dots,M+N-1\right\}\subset \mathcal{L}^c_{(m,n)}$. Thus, $M+k\neq m +\sum_{i=1}^a n_i \ \forall a \in \mathbb{Z}_{d'},\ N-1\geq k\geq0$. This means, $k\neq m-M +\sum_{i=1}^a n_i=i(a), \ N-1\geq k\geq0$. Therefore, $i(a)>N-1$ which means $i(a)\geq N$. By equation (\ref{genfusion}), one has $P^{(m,\underline n)}*Q^{(M,N)}=0$. Here $P^{(m,n)}$ is the defect with the set $(m,n)$  a solution to $\mathcal{L}^c_{(m,n)}=\left\{ M,\dots, M+N-1\right\}$; and $Q^{(M,N)}$ such that $M= \min\left\{ M,\dots, M+N-1\right\}$, and $N =|\left\{ M,\dots, M+N-1\right\}|$.

Now if $\mathcal{S}_{\mathcal{L}^c_{(m,n)}}$ does not delete the full union  $S^{(M,N)}$, then
\begin{equation}\label{diskintersection}
\begin{split}
\left\{ M,\dots, M+N-1\right\}\cap \mathcal{L}^c_{(m,n)}&=\left\{ 0,\dots, N-1\right\}\cap m -M+\left\{n_0,n_0+n_1,\dots, n_0+n_1+\cdots +n_{d'-1}\right\}\\
&=\left\{ 0,\dots, N-1\right\}\cap \mathcal{J}\\
&=\left\{i_1,\dots,i_l\right\}\neq \emptyset,
\end{split}
\end{equation}
where $\mathcal{J}:=\left\{i(a)\ |\ a\in \mathbb{Z}_{d'}\right\}$. Hence, there exists $a\in \mathbb{Z}_{d'}$ such that $i(a)< N$ and by equation (\ref{genfusion}) the corresponding fusion $P^{(m,n)}*Q^{(M,N)}$ is not zero. As previously discussed this fusion is then $P^{(m,n)}*Q^{(M,N)}=Q^{(a_1,k(a_1)}$ where $a_1$ minimizes $i(a)$ and
\begin{equation}
k(a)=\min\left\{j>0 \ | \ \sum_{l=1}^j n_{a+l} \leq N\right\}.
\end{equation}
Since we have restricted to the case $n_i\geq 1 \ \forall i$, $k(a_1)=l$ is the number of sections of $S^{(M,N)}$ not annihilated by $S_{\left\{\cdots\right\}}$. Thus, $P^{(m,\underline n)}*Q^{(M,N)}=Q^{(a_1,l)}$. {One notes that $a_1$ is the number of $Q^{(M',1)}$ with $M'\in\left\{m,\dots, M\right\}$ not annihilated by $P$.} Hence, the operators $\mathcal{O}$ represent the $P$  action on all B-type boundary conditions \cite{brunner07a}.

%%%%%%%%%%%%%%%%%%%%%%%%%%%%
\subsection{Comparison with RG flow}\label{comparison}

The RG flows between the $\mathbb{C}/\mathbb{Z}_d$ orbifolds can be studied in terms of the mirror picture as well. As we previously mentioned, mirror symmetry relates these orbifolds and the twisted Landau-Ginzburg model with twisted superpotential $\widetilde W = \widetilde X^d$. These twisted model can be related via mirror symmetry to a Landau-Ginzburg model with superpotential $ W =  X^d$. Therefore we can frame the RG flow of interest $\mathbb{C}/\mathbb{Z}_d\longrightarrow \mathbb{C}/\mathbb{Z}_{d'}$ as the RG flow $\text{LG}_d \longrightarrow \text{LG}_{d'}$ in the presence of A-supersymmetry.

The RG flows in the Landau-Ginzburg models are encoded in the behavior of the deformed superpotential $W_\lambda$ of the respective model. That is, we consider perturbations 
\begin{equation}
W_{\lambda_0}=X^d+\lambda_0 X^{d'}\ \ , \ \ d'<d,
\end{equation}
of $W=X^d$. The RG flow affects the superpotential by scaling it
\begin{equation}\label{rgscale}
W_{\lambda_0}\rightarrow \Lambda^{-1} W_{\lambda_0}.
\end{equation}
Upon a field redefinition, $X\rightarrow \Lambda X$, we obtain
\begin{equation}\label{deformedW}
\Lambda^{-1}W_{\lambda_0}=X^d+\lambda_0\Lambda^{\frac{d'-d}{d}}X^{d'}=:W_\lambda(X),
\end{equation}
where $\lambda(\Lambda):=\lambda_0\Lambda^{\frac{d'-d}{d}}$ is the running parameter:
\begin{equation}
\lim_{\Lambda \rightarrow \infty} \lambda = 0\ \ (\textit{UV}) \ , \ \ \lim_{\Lambda \rightarrow 0} \lambda = \infty \ \ (\textit{IR}).
\end{equation}
So at either end of the flow we end up with a homogeneous potential. We assume that the imaginary parts of the critical values of $W_\lambda$ stay different $\forall \ \lambda$.

Since we are interested in Landau-Ginzburg models on the half-plane with a non-zero boundary, we refer to the language of A-branes discussed in section \ref{abraneslg}. The RG flow has a description in terms of the A-branes and the respective deformations \cite{hori00, brunner07} under non-zero $\lambda$ in equation (\ref{deformedW}). Each A-brane formed by segments from $X_{*i}$ to the boundary points $B_a$ and $B_b$ is denoted by $\overline{B_a X_{*i} B_b}$. As the deformed superpotential flows into the IR, the critical points $X_{*i}$, $i>0$, flow to infinity, while the critical point $X_{*1}=0$ associated with the homogeneous superpotential remains. The A-branes associated with the points $X_{*i}$ then decouple from the theory since the respective Lagrangian submanifolds $\gamma_{X_{*i}}$ disappear.  Therefore the IR A-branes are labeled by the equivalent classes $([B_i],[B_j])$ of the relationship $B_k\sim B_l$ when connected on $\Gamma\setminus\Gamma_1$. A generic A-brane in the UV might be composed of segments which are part of $\Gamma_1$ and $\Gamma_i$ in the deformed potential $(\lambda\neq0)$. In this case the A-brane decays into the sum of an A-brane which decouples in the IR and an A-brane which flows to an IR A-brane.

To illustrate, let us consider the example we discussed in section \ref{abraneslg} with $W=X^4$ and the deformation $W_\lambda=X^4+\lambda X^3$. $W=X^4$ corresponds to the $\mathbb{C}/\mathbb{Z}_4$ orbifold. The deformed $W_\lambda$ has critical points $X_{*1}=0$ of order $n=2$, and $X_{*2}=-3\lambda$ of order $n=1$. We see that we flow to the IR $X_{*2}\rightarrow \partial D$ so the A-brane $\overline{B_3 X_{*2}B_2}$ decouples. So the endpoint of the flow is the $\mathbb{C}/\mathbb{Z}_3$ orbifold. As an example of the decay of the UV A-branes when $\lambda\neq 0$, consider $\overline{B_3 X_{*}B_1}$. As we turn on $\lambda$ this A-brane decays to $\overline{B_3 X_{*2}B_2} +\overline{B_2 X_{*1}B_1}$.

One can map the A-brane diagrams to the disk diagrams representing the B-type boundary conditions \cite{brunner07}; and hence there is a correspondence between the flow of the A-brane deformations and the action of the special defects on the disk diagrams of B-type boundary conditions. As noted above, in the IR only those preimages of $\infty$ which are not connected on $\Gamma \setminus\Gamma_1$ survive. These are precisely the points in the set
\begin{equation}\label{pointset}
\mathcal{L}=\left\{a\in\mathbb{Z}_d | B_a \nsim  B_{a+1}\right\}.
\end{equation}
In terms of the graphical disk operations for the B-type defects, this is equivalent to starting with disk partitioned into $S_i$ sectors representing the $Q^{M,N}$ B-type boundary conditions; and acting on this disk with the $\mathcal{S}_{\mathcal{L}^c}$ operator with $\mathcal{L}$ as in Eq. (\ref{pointset}).

%%%%%%%%%%%%%%%%%%%%%%%%%%%%%%%%%%%%%%%%%%%%%%%%%%%
%
%  New template code for TAMU Theses and Dissertations starting Fall 2016.
%
%  Author: Sean Zachary Roberson
%	 Version 3.17.01
%  Last updated 1/10/2017
%
%%%%%%%%%%%%%%%%%%%%%%%%%%%%%%%%%%%%%%%%%%%%%%%%%%%
%%%%%%%%%%%%%%%%%%%%%%%%%%%%%%%%%%%%%%%%%%%%%%%%%%%%%%%%%%%%%%%%%%%%%%
%%                           SECTION III
%%%%%%%%%%%%%%%%%%%%%%%%%%%%%%%%%%%%%%%%%%%%%%%%%%%%%%%%%%%%%%%%%%%%%

\chapter{\uppercase{Boundary theory of $S^1/\mathbb{Z}_2 \times S^1/\mathbb{Z}_2$} }

In this chapter we explore possible D-brane constructions for the $c=2$ free bosonic theory with target space $S^1/\mathbb{Z}_2 \times S^1/\mathbb{Z}_2$. The $S^1/\mathbb{Z}_2$ orbifold constitutes another archetypal model in string theory rich enough to test ideas and find applications. Indeed, almost half of all $c=1$, 2D CFTs can be realized as instances of this model \cite{Ginsparg:1988ui}. To obtain the orbifold construction we start with a free bosonic field $X$ on a circle
\begin{align}
     X = X + 2\pi R,
\end{align}
and perform the identification by $\mathbb{Z}_2$ action
\begin{align}
     \rho: X(z,\bar z) \rightarrow{} -X(z,\bar z).
\end{align}
The D-branes found here are described as  elements of the boundary CFT. We start with a review of the boundary CFT formalism in the case of the $S^1/\mathbb{Z}_2$ orbifold as developed in \cite{affleck}. Then we move on to find the allowed boundary states in $S^1/\mathbb{Z}_2 \times S^1/\mathbb{Z}_2$. In chapter 4, these D-branes are mapped to defects between the single orbifold theories at different radii.

\section{Review of the boundary states in the circle orbifold}\label{review}

In this section we review the boundary conformal field theory (BCFT) for the bosonic free theory following \cite{affleck}. The action is
\begin{equation}\label{freeTheory}
    S=\frac{1}{2\pi}\displaystyle\int dzd\bar z \ \partial X\bar\partial X,
\end{equation}
with $X: \mathbb{C}\rightarrow {S}^1_R/\mathbb{Z}_2$ where $R$ is radius of the unorbifolded circle. 

There are two types of solutions to above variation problem: One is the field $X$ which satisfies $X(e^{i2\pi}z,e^{-i2\pi}\bar z) = X(z,\bar z)$ and thus it is called ``untwisted''. The other solution is ``twisted'' with respect to the $\mathbb{Z}_2$ action, $Y(e^{i2\pi}z,e^{-i2\pi}\bar z) = -Y(z,\bar z)$. The field $X$ has the Fourier expansion \cite{affleck}
\begin{equation}\label{xExpansion}
X(\tau,\sigma)=\hat x_0 +\frac{\widehat N}{2 R}\tau + \widehat M R\sigma+\sum_{n=1}^\infty\frac{i}{2\sqrt{n}} (a_n e^{-in(\tau+\sigma)}+\tilde a _n e^{-in(\tau-\sigma)}- h.c.),
\end{equation}
where $\hat x_0$ is the zero-mode operator; $a_n$ and $\tilde a_n$ are lowering operators and $a_n^\dag = a_{-n}$, $\tilde a_n ^\dag = \tilde a_{-n}$ are the corresponding raising operators; and $\widehat N$ and $\widehat M$ are the momentum and winding operators. These operators satisfy
 \begin{equation}\label{aalgebra2}
 [a_n,a_m]  = \delta_{m+n}=  [\tilde a_n, \tilde a_m]  = \delta_{m+n} \ , \ \ \ \left[\hat x_0, \widehat N \right] = iR\ , \ \ \ \left[\widehat{\tilde x}_0, \widehat M\right] = -\frac{i}{R}.
\end{equation}
In the above, $\widehat{\tilde x}_0$ is the variable conjugate to the winding number operator.

The Hamiltonian in the untwisted sector is given by $H_c=L_0+\widetilde L_0$ and it has the mode expansion
\begin{equation}\label{hamiltonian}
H_c =\frac{\widehat N^2}{4R^2}+\widehat M^2R^2+\sum_{n>0}n(a_n^\dag a_n +\tilde a_n^\dag \tilde a_n)-\frac{1}{12}.
\end{equation}
In the twisted sector the boson $Y$ has mode expansion
\begin{equation}\label{yExpansion}
Y(\tau,\sigma)=\hat y_0+\sum_{n>0}\frac{i}{2\sqrt{(n-\frac{1}{2})}}\left( b_ne^{-i(n-\frac{1}{2})(\tau+\sigma)}+\tilde b_ne^{-i(n-\frac{1}{2})(\tau-\sigma)}-  h.c. \right),
\end{equation}
where the $b_n$ modes satisfy the same canonical commutations as the $a_n$ modes, and $y_0\in\left\{0,\pi R\right\}$. The last condition means that the twisted field is restricted to the endpoints of the orbifold, i.e., the fixed points of the $\mathbb{Z}_2$ action. The respective Hamiltonian $H_t$ is given by
\begin{equation}\label{ht2}
H_t = \sum_n\left( \left( n-\frac{1}{2}\right) b_n^\dag b_n + \left( n-\frac{1}{2}\right)  \tilde b_n^\dag\tilde  b_n\right)+\frac{1}{24}.
\end{equation}
Given a 2D CFT on a subspace $\Sigma \subset \mathbb{C}$ with non-trivial boundary $\partial \Sigma\neq \displaystyle \emptyset$, the following condition must hold along the boundary 
\begin{equation}\label{TbarT}
T=\xbar T,
\end{equation}
which is a requirement for the conformal Ward identity to hold in the presence of boundaries \cite{cardy}. The Hilbert space of a BCFT contains elements which are consistent with Eq. (\ref{TbarT}), that is there is a boundary CFT whose elements $|v\rangle\rangle$ satisfy
\begin{equation}\label{VirasorDefining}
    (L_n - \widetilde L_{-n})|v\rangle\rangle=0.
\end{equation}
The operator in the above equation follows by taking the Fourier expansion of both sides of $T=\xbar T$. For the free theory of Eq. (\ref{freeTheory}), boundary states solving Eq. (\ref{VirasorDefining}) can be obtained as solutions to the systems of equations
\begin{equation}\label{Ndefining}
(a_n +\tilde a_{-n})|k,w\rangle\rangle_N =0 ,
\end{equation}
\begin{equation}\label{Ddefining}
    (a_n -\tilde a_{-n})|k,w\rangle\rangle_D=0,
\end{equation}
where $n\in \mathbb{Z}$. The $(k,w)$ labels of the boundary states are winding mode and momentum eigenvalues furnishing the elements of direct sum of the bulk $u(1)^2$ representations
\begin{equation}
\mathcal{H}(R)=\bigoplus_{k,w\in\mathbb{Z}} \mathcal{H}_{q_{m,w}(R)} \otimes \widetilde{ \mathcal{H}}_{\tilde q_{m,w}(R)},
\end{equation}
where the charges $(q,\tilde q)$ are the eigenvalues of $(a_0,\tilde a_0)$. That is,
\begin{equation}\label{alpha0}
a_0 |k, w\rangle\rangle = \left(\frac{k}{R} -\frac{Rw}{2}\right) |k, w\rangle\rangle,
\end{equation}
\begin{equation}\label{baralpha0}
\tilde a_0 |k, w\rangle = \left(\frac{k}{R} + \frac{Rw}{2}\right) |k, w\rangle\rangle.
\end{equation}
The solutions to the system in Eq. (\ref{Ndefining}) correspond to Neumann boundary states while those for Eq. (\ref{Ddefining}) correspond to Dirichlet ones. Each state encodes the corresponding type of boundary conditions. 

The Cardy-consistent boundary states for the orbifolded theory which are invariant under the action of $\mathbb{Z}_2$ are built as symmetric combinations of the boundary states of the circle theory. We refer to these invariant states as ``untwisted''. The Neumann untwisted state is give by
\begin{equation}\label{orbifoldn}
|N_O(\tilde x_0)\rangle\rangle = \frac{1}{\sqrt{2}}\left(|N(\tilde x_0)\rangle\rangle+|N(-\tilde x_0)\rangle\rangle \right),
\end{equation}
where
\begin{equation}\label{generaln}
|N(\tilde x_0) \rangle\rangle:=\sqrt{R}\sum_{w\in\mathbb{Z}}e^{iMR\tilde x_0}\exp\left(-\sum_{n=1} ^\infty \frac{1}{n} a_{-n}\tilde a_{-n} \right) |0,w\rangle,
\end{equation}
is the Neumann boundary state for the $S_R^1$ theory. The invariant Dirichlet counterpart is 
\begin{equation}\label{orbifoldd}
|D_O(x_0)\rangle\rangle = \frac{1}{\sqrt{2}}\left(|D(x_0)\rangle\rangle+|D(-x_0)\rangle\rangle \right),
\end{equation}
with the circle Dirichlet expression being
\begin{equation}\label{generald}
|D(x_0)\rangle\rangle:=\frac{1}{\sqrt{2R}}\sum_{k\in\mathbb{Z}} e^{ikx_0/R}\exp\left(\sum_{n=1} ^\infty \frac{1}{n} a_{-n}\tilde a_{-n} \right) |k,0\rangle.
\end{equation}
The choice of coefficients $e^{ikx_0/R}/\sqrt{2R}$
in the Dirichlet solution follows from the requirement that the state is Cardy-consistent with itself and the Neumann state. That is, the amplitudes among these states transform to partition functions under a modular $S$-transformation. This requirement fixes the given coefficients for the Neumann state as well. The two vectors $|D(x_0)\rangle\rangle$ and $|N(\tilde x_0)\rangle\rangle$ encode the Dirichlet and Neumann boundary conditions of the free compact field $X$ \cite{bachas07}. This characteristic can be seen via the following two relationships.
\begin{equation}
X(0,\sigma)|D(x_0)\rangle\rangle=x_0|D(x_0)\rangle\rangle ,
\end{equation}
\begin{equation}
\partial_\tau X(0,\sigma)|N(\tilde x_0)\rangle\rangle=0.
\end{equation}

In the twisted sector, there are two systems of equations similar to those in Eq. (\ref{Ndefining}) and Eq. (\ref{Ddefining}) but with the $b_n$ oscillator modes:
\begin{equation}\label{NdefiningT}
(b_n +\tilde b_{-n})|v\rangle\rangle_N =0 , 
\end{equation}
\begin{equation}\label{DdefiningT}
    (b_n -\tilde b_{-n})|v\rangle\rangle_D=0.
\end{equation}
The Dirichlet solution is given by
\begin{equation}\label{twistedD}
|D_{O}(y_0),T\rangle\rangle =e^{\sum_{n>0} b^\dag _n \tilde b_n ^\dag}|y_0,T\rangle,
\end{equation}
with the discreet parameter $y_0\in\left\{0,\pi R\right\}$ taking value at the fixed points of the $\mathbb{Z}_2$ action. This state satisfies the twisted Dirichlet condition 
\begin{equation}\label{dirichletequation}
Y(0,\sigma)|D_O(y_0),T\rangle\rangle=y_0|D_O(y_0),T\rangle\rangle.
\end{equation}
The solution to the Neumann-type system of Eq. (\ref{NdefiningT}) is
\begin{equation}
|N_O(\tilde y_0),T\rangle\rangle=e^{-\sum_{n>0} b^\dag _n \tilde b_n ^\dag}\frac{1}{\sqrt{2}} (|0,T\rangle+e^{i 2R\tilde y_0}|\pi R,T\rangle),
\end{equation}
where $\tilde y_0\in \left\{0, \frac{\pi}{2R}\right\}$ is the variable T-dual to $y_0$. 

The states $|N_O(\tilde y_0),T\rangle\rangle$ and $|D_O(y_0),T\rangle\rangle$ are not Cardy consistent. Instead, in the twisted sector boundary states come as elements $\mathcal{H}_{\text{circle}}\oplus \mathcal{H}_{\text{twisted}}$, where $\mathcal{H}_{\text{circle}}$ is the boundary Hilbert space of the circle theory; and $ \mathcal{H}_{\text{twisted}}$ is the Hilbert space of states which satisfy the systems of equations of Eq. (\ref{DdefiningT}) and Eq. (\ref{NdefiningT}). The consistent boundary states were first written by \cite{affleck}:
\begin{equation}\label{Dgenerator}
|D_O^\pm(y_0)\rangle\rangle= \frac{1}{\sqrt{2}}( |D(y_0)\rangle\rangle e_c \pm 2^{1/4} |D_O(y_0),T\rangle\rangle e_t),
\end{equation}
\begin{equation}\label{Ngenerator}
|N_O^\pm(\tilde y_0)\rangle\rangle= \frac{1}{\sqrt{2}}( |N (\tilde y_0)\rangle\rangle e_c \pm 2^{1/4} |N_O(\tilde y_0),T\rangle\rangle e_t),
\end{equation}
where $e_c$ and $e_t$ are the left and right generators of the direct sum $\mathcal{H}_{\text{circle}}\oplus \mathcal{H}_{\text{twisted}}$. That is, we write a generic element $a\in \mathcal{H}_{\text{circle}}\oplus \mathcal{H}_{\text{twisted}}$ as $a =a^c e_c + a^t e_t$. We will omit the generators except in places where they help to clarify the computations. With this notation, the boundary states in Eq. (\ref{Dgenerator}) and Eq. (\ref{Ngenerator}) satisfy the following equations
\begin{equation}
((a_n-\tilde a_n)e_c\otimes e_c^*+ (b_n-\tilde b_n)e_t\otimes e_t^*)|D_O^\pm(y_0)\rangle\rangle=0,
\end{equation}
\begin{equation}
((a_n+\tilde a_n)e_c\otimes e_c^*+ (b_n+\tilde b_n)e_t\otimes e_t^*)|N_O^\pm(\tilde y_0)\rangle\rangle=0.
\end{equation}

%ssssssssssssssssssssssssssssssssssssssssssssssssssssssssssssssssss
\section{D-branes for $(S^1/\mathbb{Z}_2)^2$}

In this section we proceed to find possible D-branes for the free bosonic theory with target space $(S^1_R/\mathbb{Z}_2)^2$. We parametrize the target space by two bosons $(Z^1,Z^2)\in S^1_{R_1}/\mathbb{Z}_2\times S^1_{R_1}/\mathbb{Z}_2$ where each $Z^i$ stands for the untwisted field $X^i$ or twisted $Y^i$. To obtain more general D-branes we allow for a target-space rotation by angle $\phi$ and we denote the rotated target space by $({}^R Z^1,{}^RZ^2)$. This target-space transformation leaves the conformal requirement $T= \xbar T$ for points at the boundary invariant in the case of the free boson.

We proceed below by following the same procedure to find D-branes but in the product theory.  By solving equations defining boundary conformal states we find families of D-branes describing possible boundary conditions for open strings. First, we present the untwisted sector composed of the D-branes which remain after projecting out those in the $\mathbb{S}_{R_1}\times \mathbb{S}_{R_2}$ theory which are not $\mathbb{Z}_2 \times \mathbb{Z}_2$-invariant. Then we present the twisted D-branes which contain those D-branes which arise as a tensor product of twisted boundary states. We are mainly interested on finding solutions for the rotated D-branes, i.e., D-branes for the target space $({}^R Z^1,{}^RZ^2)$.

%sssssssssssssssssssssssssssssssssssssssssssssssssssssssssssssssssssssssssssssssssss
\subsection{Untwisted boundary states}\label{UntwistedBraneSection}

The untwisted boundary states for the general rotated D-branes are obtained as solutions to the equations 
\begin{align}
    ({}^R\underline a_n\pm  {}^R\tilde{ \underline a}_{-n}) |v\rangle\rangle&=0, \label{rotated1}\\
    ({}^R\underline a_n\pm \Omega {}^R\tilde{ \underline a}_{-n}) |v\rangle\rangle&=0,\label{rotated2}
\end{align}
where $\Omega =\operatorname{diag}(1,-1)$, and ${}^R\underline a_n:= R(\phi) \underline a_n$ with $\underline a_n:=(a^1_n,a^2_n)^t$. The angle $\phi$ is given by
\begin{equation}
\phi=\tan^{-1}\left( \frac{k_2R_2}{k_1R_1}\right),
\end{equation}
where $k_1$, $k_2$ are coprime integers.

As in the $S^1/\mathbb{Z}_2$ case in \cite{affleck}, we directly construct untwisted D-branes in the product quotient theory by symmetrizing the boundary states of the $\mathbb{S}^1_{R_1} \times \mathbb{S}^1_{R_2}$- theory. The boundary theory of  $\mathbb{S}^1_{R_1} \times \mathbb{S}^1_{R_2}$ is given in \cite{bachas07} and we use their notation here. The most general untwisted D-branes fall into two large classes. One of them is a D1-brane wrapping $k_1$-times one direction of the orbifold, and $k_2$-times the other direction and satisfies Eq. (\ref{rotated2}). Up to a normalization factor $C$, the boundary state for such a D-brane is
\begin{align}\label{d1orbifold}
|D1_O(\alpha,\beta)\rangle\rangle_\phi = C & \left(|D1(\alpha,\beta)\rangle\rangle_\phi + |D1(-\alpha,-\beta)\rangle\rangle_\phi \right. \nonumber \\
&\left.+|D1(\alpha,-\beta)\rangle\rangle_\phi +|D1(-\alpha,\beta)\rangle\rangle_\phi \right),
\end{align}
where $|D1(\alpha,\beta)\rangle\rangle_\phi$ was found in \cite{bachas07} and it is given by
\begin{equation}\label{d1bachas}
|D1,(\alpha,\beta)\rangle\rangle_\phi :=\prod_{n>0}e^{-\Omega_\phi^{ij}a^{i\dag}_n\tilde a_n^{j\dag}}\left(g^{+} \sum_{M,N}e^{iN\alpha-iM\beta}|k_2N,k_1M\rangle\rangle_1\otimes|-k_1N,k_2M\rangle\rangle _2\right),
\end{equation}
where 
\begin{equation}\label{omegaPhi}
\Omega_\phi=R^t(\phi)\Omega R(\phi)=\left( \begin{array}{cc}
\cos(2\phi) & \sin(2\phi)\\
\sin(2\phi)&-\cos(2\phi)  \end{array}\right), \ \ \ \ g^{+}=\sqrt{\frac{k_1k_2}{\sin2\phi}}.
\end{equation}
To fix the overall constant we note that at $\phi = n\pi$ the above state reduces to
\begin{equation}\label{d101}
|D1(\alpha,\beta)\rangle\rangle_{n\pi} =|N(\alpha)^{k_1}\rangle\rangle\otimes|D(\beta)^{k_1}\rangle\rangle ,
\end{equation}
where the $k_1$ superscript means $k_1$ copies of the state. At this angle, the symmetrized boundary state in Eq. (\ref{d1orbifold}) becomes
\begin{equation}\label{rightd1}
|D1_O(\alpha,\beta)\rangle\rangle_{n\pi}= 2C |N_O(\alpha) ^{k_1}\rangle\rangle \otimes  |D_O(\beta)^{k_1} \rangle\rangle ,
\end{equation}
which is the orbifolded version of the previous state if $C=1/2$. Simplifying the sum in Eq. (\ref{d1orbifold}) we can write the $ |D1_O(\alpha,\beta)\rangle\rangle_\phi$ state as
\begin{align}\label{d1orbifoldcosine}
 |D1_O(\alpha,\beta)\rangle\rangle_\phi=2g^{+} \prod_{n>0}e^{(-\Omega_\phi^{ij} a^i_n\tilde a_n^j)^\dag} \sum_{M,N}(&\cos(N\alpha)\cos(M\beta)  \nonumber\\
 &|k_2N,k_1M\rangle_1\otimes|-k_1N,k_2M \rangle _2).
\end{align}
The second general type of untwisted D-branes is a bound system of $k_1$ D2-branes and $k_2$ D0-branes. Such state is the $\mathbb{Z}_2 \times\mathbb{Z}_2$-symmetric solution to Eq. (\ref{rotated2}) and it can also be obtained by T-dualizing the right-movers in $|D1_O(\alpha,\beta)\rangle\rangle_\phi$. This D2/D0 state is given by
\begin{align}\label{d2d0orbifoldcosine}
|D2/D0_O(\alpha,\beta)\rangle\rangle_\theta:= 2 g^{(-)}\prod_{n>0}e^{-\widetilde \Omega_\theta ^{ij} a^{i\dag}_n \tilde a_n^{j\dag}}\sum_{M,N}&(\cos(N\alpha)\cos(M\beta)\nonumber\\
& |k_1M,k_2N\rangle_1\otimes|-k_1N,k_2M\rangle _2),
\end{align}
with
\begin{equation}
\widetilde \Omega_\theta: =\Omega_\phi\left( \begin{array}{cc}
-1 & 0\\
0& 1  \end{array}\right)\Big |_{\phi = \theta}\ , \ \ \ \ g^{(-)}=\sqrt{\frac{k_1k_2}{\sin2\theta}},
\end{equation}
where 
\begin{equation}\label{thetatheta}
\theta=\tan^{-1}\left(\frac{2k_2R_1R_2}{k_1}\right).
\end{equation}
is the T-dualized rotation angle.

At values which are multiples of $\pi/2$ we obtain the orbifolded version of the D2 and D0-branes of \cite{bachas07},
\begin{align}
    |D2/D0_O(\alpha,\beta)\rangle\rangle_{n\pi}&=|D_O(\alpha)^{k_1}\rangle\rangle_1\otimes |D_O(\beta)^{k_1}\rangle\rangle_2=:|D0_O^{k_1}\rangle\rangle,\\
    |D2/D0_O(\alpha,\beta)\rangle\rangle_{(2n+1)\pi/2}& =|N_O(\alpha)^{k_2}\rangle\rangle_1\otimes |N_O(\beta)^{k_2}\rangle\rangle_2=:|D2_O^{k_2}\rangle\rangle.
\end{align}

%sssssssssssssssssssssssssssssssssssssssssssssssssssssssssssssssssssssssssssssssss
\subsection{Twisted D-branes}

Aside from the $\mathbb{Z}_2\times \mathbb{Z}_2$-symmetric boundary states obtained as projections from the $T^2$ boundary theory, the orbifold carries twisted D-branes. Without considering the general rotated case at first, we write down boundary states which are tensor products of the single theory boundary theory. We will use these states as a guide to find the general types in the next section. The non-rotated elements are
\begin{equation}\label{diagonaldd}
|DD^{\pm\pm}(y_0^1,y_0^2)\rangle\rangle:=|D_O^\pm(y_0^1)\rangle\rangle_1\otimes |D_O^\pm(y_0^2)\rangle\rangle_2,
\end{equation}
\begin{equation}\label{diagonalnn}
|NN^{\pm\pm}(\tilde y_0^1,\tilde y_0^2)\rangle\rangle:=|N_O^\pm(\tilde y_0^1)\rangle\rangle_1\otimes |N_O^\pm(\tilde y_0^2)\rangle\rangle_2,
\end{equation}
\begin{equation}\label{diagonaldn}
|DN^{\pm\pm}(y_0^1,\tilde y_0^2)\rangle\rangle:=|D_O^\pm(y_0^1)\rangle\rangle_1\otimes |N_O^\pm(\tilde y_0^2)\rangle\rangle_2,
\end{equation}
\begin{equation}\label{diagonalnd}
|ND^{\pm\pm}(\tilde y_0^1, y_0^2)\rangle\rangle:=|N_O^\pm(\tilde y_0^1)\rangle\rangle_1\otimes |D_O^\pm( y_0^2)\rangle\rangle_2.
\end{equation}
The first two states above can are solutions to the equations
\begin{equation}\label{twistedConditionDiag}
((\underline a_n\pm\tilde{ \underline a}_{-n}) + (\underline b_n\pm\tilde{ \underline b}_{-n})) |V\rangle\rangle=0,
\end{equation}
while the latter two can be obtained as solutions to 
\begin{equation}\label{TconditionNoDiag}
((\underline a_n\pm \Omega \ \tilde{ \underline a}_{-n}) + (\underline b_n \pm\Omega \ \tilde{ b}_{-n})) |V\rangle\rangle=0,
\end{equation}
where $\Omega =\operatorname{diag}(1,-1)$, $\underline a_n:= (a^1_n,a^2_n)^t$,  $\underline b_n:= (b^1_n,b^2_n)^t$. In the above equations we are using the implicit identity operators acting on the left or the right elements of the tensor product as needed. We are also implicitly using the generators $e_c^i$ and $e_t^i$ over which the modules $\mathcal{H}_{\text{circle}}$ and $\mathcal{H}_{\text{twisted}}$ are built, both for the left and right elements of the tensor products.

In order to find the boundary states for the twisted D-branes in the rotated case, it helps to write out the above solutions in a compact manner. Just like $\underline a_n:=(a^1_n,a^2_n)^t$ and $\underline b_n:=(b^1_n,b^2_n)^t$ it helps to define two new sets of 2-vectors of oscillators given by
\begin{equation}\label{oscillatorvectors}
 \ \ \underline c_n:=(a^1_n,b^2_n)^t \ \ , \ \ \underline d_n:=(b^1_n,a^2_n)^t,
\end{equation}
and similarly for the antiholomorphic oscillators. We will label the set of all such pairs of oscillators by $\underline s_n$. Let us start with the DN and ND tensors; inserting the expressions for single boundary states give in Eq. (\ref{Dgenerator}) and Eq. (\ref{Ngenerator}) we obtain
\begin{equation}\label{dnNoRotated}
|DN^{\pm\pm}(y_0,\tilde y_0)\rangle\rangle = B^{(+)}[V_{DN}^{\pm\pm}(y_0,\tilde y_0)],
\end{equation}
\begin{equation}\label{ndNoRotated}
|ND^{\pm\pm}(y_0,\tilde y_0)\rangle\rangle = B^{(-)}[V_{ND}^{\pm\pm}(y_0,\tilde y_0)],
\end{equation}
where the operators are
\begin{equation}
B^{(\pm)}:=\prod_{n>0}\left( e^{\pm \underline a_n^{\dag t}\Omega \tilde{ \underline a}_n^\dag} +  e^{\pm \underline c_n^{\dag t}\Omega \tilde{ \underline c}_n^\dag} +  e^{\pm \underline d_n^{\dag t}\Omega \tilde{ \underline d}_n ^\dag} +  e^{\pm \underline b_n^{\dag t}\Omega \tilde{ \underline b}_n^\dag}\right),
\end{equation}
and the lattice sums are 
\begin{equation}
\begin{split}
V_{DN}^{\pm\pm}=& \left(2^{-1/2}\frac{1}{\sqrt{2R_1}}\sum_M e^{iMy_0/R_1}|M,0\rangle_1 \pm 2^{-1/4} |y_0,T\rangle_1\right)\\
&\otimes \left( 2^{-1/2} \sqrt{R_2}\sum_N e^{iNR_2\tilde y_0}|0,N\rangle_2 \pm 2^{-1/4}|\tilde y_0,T\rangle_2 \right),
\end{split}
\end{equation}
\begin{equation}
\begin{split}
V_{ND}^{\pm\pm}=&\left( 2^{-1/2} \sqrt{R_1}\sum_N e^{iNR_1\tilde y_0}|0,N\rangle_1 \pm 2^{-1/4}|\tilde y_0,T\rangle_1 \right) \\
&\otimes \left(2^{-1/2}\frac{1}{\sqrt{2R_2}}\sum_M e^{iMy_0/R_2}|M,0\rangle_2 \pm 2^{-1/4} |y_0,T\rangle_2\right).
\end{split}
\end{equation}

Aside from the fully twisted solutions in Eq. (\ref{diagonaldd}) - Eq. (\ref{diagonalnd}), in the twisted sector we can also find boundary states which are untwisted in one direction and twisted along the other direction of the D-brane. We call such boundary states ``partially twisted''. The NN or DD combinations are
\begin{equation}\label{partialT1}
|A1^{\pm}(y_0,x_0)\rangle\rangle := |D^{\pm}(y_0)\rangle\rangle_1 \otimes |D_O(x_0)\rangle\rangle_2,
\end{equation}
\begin{equation}
|A4^{\pm}(\tilde y_0,\tilde x_0)\rangle\rangle := |N^{\pm}(\tilde y_0)\rangle\rangle_1 \otimes |N_O(\tilde x_0)\rangle\rangle_2,
\end{equation}
\begin{equation}
|B1^{\pm}(x_0,y_0)\rangle\rangle :=  |D_O(x_0)\rangle\rangle_1 \otimes |D^{\pm}(y_0)\rangle\rangle_2,
\end{equation}
\begin{equation}\label{partialT4}
|B4^{\pm}(\tilde x_0, \tilde y_0)\rangle\rangle :=  |N_O(\tilde x_0)\rangle\rangle_1 \otimes |N^{\pm}(\tilde  y _0)\rangle\rangle_2,
\end{equation}
which solutions to the systems of equations
\begin{equation}\label{PartialNNDD}
((\underline a_n\pm\tilde{ \underline a}_{-n}) +P^i (\underline b_n\pm \tilde{ \underline b}_{-n})) |V\rangle\rangle=0,
\end{equation}
where the $P^i$ are projectors defined on the twisted sector by $P^i \underline b_n = b_n^i$. 
There are also the DN and ND tensor products,
\begin{equation}
|A2^{\pm}(y_0,x_0)\rangle\rangle := |D^{\pm}(y_0)\rangle\rangle_1 \otimes |N_O(\tilde x_0)\rangle\rangle_2,
\end{equation}
\begin{equation}
|A3^{\pm}(\tilde y_0,x_0)\rangle\rangle := |N^{\pm}(\tilde y_0)\rangle\rangle_1 \otimes |D_O(x_0)\rangle\rangle_2,
\end{equation}
\begin{equation}
|B2^{\pm}(x_0,\tilde y_0)\rangle\rangle :=  |D_O(x_0)\rangle\rangle_1 \otimes |N^{\pm}(\tilde y_0)\rangle\rangle_2 ,
\end{equation}
\begin{equation}
|B3^{\pm}(\tilde x_0,y_0)\rangle\rangle :=  |N_O(\tilde x_0)\rangle\rangle_1 \otimes |D^{\pm}(y_0)\rangle\rangle_2 ,
\end{equation}
which solve the defining equation
\begin{equation}\label{}
((\underline a_n\pm\Omega\  \tilde{ \underline a}_{-n}) +P^i (\underline b_n\pm \Omega \ \tilde{ \underline b}_{-n})) |V\rangle\rangle=0.
\end{equation}

As in the case for the fully twisted boundary states, it is helpful to write out some of the above states in a way that gives us insight when solving the general rotated case. Inserting the expressions for the single states from the $S^1/\mathbb{Z}_2$ boundary theory we obtain,
\begin{align}
    |A2^{\pm}(y_0,\tilde x_0)\rangle\rangle &= \prod_{n>0}\left( e^{\underline a_n^{\dag t}\Omega \tilde{ \underline a}_n^\dag } +  e^{\underline d_n^{\dag t}\Omega \tilde{ \underline d}_n ^\dag}   \right) V_{A2}^{\pm}(y_0,\tilde x_0),\label{A2nonRotated}\\
    |A3^{\pm}(\tilde y_0,x_0)\rangle\rangle &=  \prod_{n>0}\left( e^{-\underline a_n^{\dag t}\Omega \tilde{ \underline a}_n ^\dag} +  e^{-\underline d_n^{\dag t}\Omega \tilde{ \underline d}_n ^\dag} \right) V_{A3}^{\pm}(\tilde y_0, x_0),\\
    |B3^{\pm}(\tilde x_0,y_0)\rangle\rangle &=  \prod_{n>0}\left( e^{-\underline a_n^{\dag t}\Omega \tilde{ \underline a}_n ^\dag} +  e^{-\underline c_n^{\dag t}\Omega \tilde{ \underline c}_n ^\dag}   \right) V_{B3}^{\pm}(\tilde x_0,y_0),\\
    |B4^{\pm}(\tilde x_0,\tilde y_0)\rangle\rangle &=  \prod_{n>0}\left( e^{-\underline a_n^{\dag t} \tilde{ \underline a}_n ^\dag} +  e^{-\underline c_n^{\dag t} \tilde{ \underline c}_n ^\dag}   \right) V_{B4}^{\pm}(\tilde x_0,\tilde y_0),
\end{align}
where $ V_{A2}^{\pm}(y_0,\tilde x_0)$, $ V_{A3}^{\pm}(\tilde y_0, x_0)$,  $V_{B3}^{\pm}(\tilde x_0,y_0)$, and $V_{B4}^{\pm}(\tilde x_0,\tilde y_0)$  contain the vacuum expressions coming from $|D^{\pm}(y_0)\rangle\rangle_i$, $|N^{\pm}(\tilde y_0)\rangle\rangle_i$, $|D_O(x_0)\rangle\rangle_i$, and $|N_O(\tilde x_0)\rangle\rangle_i$.

%ssssssssssssssssssssssssssssssssssssssssssssssssssssssssssssssss
\subsection{Rotated twisted D-branes}

In this section we look for solutions to the boundary defining equations for the rotated target space. We restrict to the twisted sector because the untwisted rotated D-branes are already found in Subsection \ref{UntwistedBraneSection} as projections from their counterparts in the $T^2$ theory developed in \cite{bachas07}. For the rotated case we have the oscillators 
\begin{equation}
{}^R \underline a_n:= R(\phi)\underline a_n\ , \ \ \ {}^R \underline b_n:= R(\phi)\underline b_n,
\end{equation}
where  $R(\phi)$ is the same rotation matrix which acts on target-space coordinates. The twisted states $|DD^{\pm\pm}(y_0^1,y_0^2)\rangle\rangle$ and $|NN^{\pm\pm}(\tilde y_0^1,\tilde y_0^2)\rangle\rangle$ are $R(\phi)$-invariant while the DN and ND are not. In the rotated frame, the latter two are the solutions to the defining equations
\begin{equation}\label{RtwistedDef}
(({}^R\underline a_n\pm \Omega {}^R\tilde{ \underline a}_{-n}) + ({}^R\underline b_n\pm \Omega {}^R\tilde{ \underline b}_{-n})) |V\rangle\rangle_\phi=0,
\end{equation}
Noting that the rotation $R$ preserves the oscillator algebras and the Virasoro algebra of the modes of the energy-momentum tensor we 
observe that the solutions to Eq. (\ref{RtwistedDef}) have the form as the solutions to non-rotated defining equations of Eq. (\ref{TconditionNoDiag}). That is, the rotated versions for $\phi \neq n\pi$ of $|DN^{\pm\pm}(y_0^1,\tilde y_0^2)\rangle\rangle$ and $|ND^{\pm\pm}(y_0^1,\tilde y_0^2)\rangle\rangle$ are given by
\begin{equation}\label{twistedDNlong}
|DN^{\pm\pm}(y_0^1,\tilde y_0^2)\rangle\rangle_\phi = B_\phi^{(+)}[\widetilde  V_{DN}^{\pm\pm}(y_0^1,\tilde y_0^2)],
\end{equation}
\begin{equation}\label{twistedNDlong}
|ND^{\pm\pm}(y_0^1,\tilde y_0^2)\rangle\rangle_\phi = B_\phi^{(-)}[\widetilde  V_{ND}^{\pm\pm}(\tilde y_0^1, y_0^2)],
\end{equation}
where
\begin{equation}\label{boundaryOps}
B^{(\pm)}_\phi:=\prod_{n>0}\left( e^{\pm \underline a_n^{\dag t}\Omega_\phi \tilde{ \underline a}_n^\dag} +  e^{\pm \underline c_n^{\dag t}\Omega_\phi \tilde{ \underline c}_n^\dag} +  e^{\pm \underline d_n^{\dag t}\Omega_\phi \tilde{ \underline d}_n ^\dag} +  e^{\pm \underline b_n^{\dag t}\Omega_\phi \tilde{ \underline b}_n^\dag}\right),
\end{equation}
and $\widetilde  V_{DN}^{\pm\pm}(y_0^1,\tilde y_0^2)$ and $\widetilde  V_{ND}^{\pm\pm}(\tilde y_0^1, y_0^2)$ are the vacuum expressions determined to be
\begin{equation}\label{VacTwistDN}
\widetilde  V_{DN}^{\pm\pm}(y_0^1,\tilde y_0^2) =\left(2^{-1/2}|0\rangle_1 \pm 2^{-1/4}|y_0^1,T\rangle_1\right)\otimes \left( 2^{-1/2} |0\rangle_2 \pm 2^{-1/4}|\tilde y_0^2,T\rangle_2 \right),
\end{equation}
\begin{equation}
\widetilde  V_{ND}^{\pm\pm}(\tilde y_0^1, y_0^2)=\left(2^{-1/2}|0\rangle_1 \pm 2^{-1/4}|\tilde y_0^1,T\rangle_1\right)\otimes \left( 2^{-1/2} |0\rangle_2 \pm 2^{-1/4}| y_0^2,T\rangle_2 \right).
\end{equation}
In order to obtain the two vacuum expressions above we use the $n=0$ case in Eq. (\ref{RtwistedDef}). Here we focus on the DN case, but similar steps lead to the ND expression as well. Since $a_0$ commutes with the higher modes we have,
\begin{equation}\label{}
(\underline a_0+ \Omega_\phi \  \tilde{ \underline a}_{0})\widetilde  V_{DN}^{\pm\pm}(y_0^1,\tilde y_0^2)=0.
\end{equation}
We note that the DN vacuum expression a priori would be a (possibly infinite) linear combination of states $|\underline k; \underline w;\pm\pm\rangle\rangle$ which have the schematic shape (up to coefficients)
\begin{equation}
\begin{split}
|\underline k; \underline w;\pm\pm\rangle\rangle \sim &|k_1,w_1\rangle_1 |k_1,w_1\rangle_2 e_1^c\otimes e^2_c \pm |k_1,w_1\rangle |\tilde y_0,T\rangle_2 e_1^c\otimes e^2_t\\
& \pm |\tilde y_0,T\rangle_1|k_2,w_2\rangle_2 e_1^t\otimes e^2_c + | y_0,T\rangle_1 |\tilde y_0,T\rangle_2 e_1^t\otimes e^2_t.
\end{split}
\end{equation}
Then the $\underline k$ and $\underline w$ labels are fixed by
\begin{equation}\label{}
(\underline a_0+ \Omega_\phi \  \tilde{ \underline a}_{0})|\underline k; \underline w;\pm\pm\rangle =0.
\end{equation}
Writing out the above equation we obtain
\begin{equation}
\begin{split}
0=&\left(\left(\begin{array}{c}
\frac{k_1}{2R_1}+w_1R_1 \\ \frac{k_2}{2R_2}+w_2R_2 
\end{array}\right) + \Omega_\phi \left( \begin{array}{c}
\frac{k_1}{2R_1}-w_1R_1 \\ \frac{k_2}{2R_2}-w_2R_2 
\end{array}      \right)\right)|\underline k; \underline w\rangle e_1^c\otimes e^2_c \\
&\pm \left(\begin{array}{c}
(\frac{k_1}{2R_1}+w_1R_1) +\Omega_\phi^{11}( \frac{k_1}{2R_1}-w_1R_1 )\\ \Omega_\phi^{21}(\frac{k_1}{2R_1}-w_1R_1 )\end{array}\right)  |k_1,w_1\rangle_1\otimes |\tilde y_0,T\rangle_2  e_1^c\otimes e^2_t\\
&\pm \left(\begin{array}{c}
\Omega_\phi^{12}(\frac{k_2}{2R_2}-w_2R_2) \\ (\frac{k_2}{2R_2}+w_2R_2 )+\Omega_\phi^{22}(\frac{k_2}{2R_2}-w_2R_2 )\end{array}\right)  | y_0,T\rangle_1\otimes|k_2,w_2\rangle_2  e^1_t\otimes e^2_c.
\end{split}
\end{equation}
The unique solution to the above system of equations is
\begin{equation}
|\underline k; \underline w;\pm\pm\rangle=|\underline 0; \underline 0;\pm\pm\rangle,
\end{equation}
which gives us the vacuum part of our solutions.

Observe that there is a discontinuity in these boundary states as $\phi \rightarrow n\pi$. For $\phi=n\pi$, we have $R(\phi)=(-1)^n 1_{2\times 2}$ so the defining equations in Eq. (\ref{RtwistedDef}) are left invariant. Therefore the solutions in this case are the non-rotated D-branes, yet these solution are part of the $\phi$-families given by $|DN^{\pm\pm}(y_0^1,\tilde y_0^2)\rangle\rangle_\phi$ and $|ND^{\pm\pm}(y_0^1,\tilde y_0^2)\rangle\rangle_\phi$. The non-rotated twisted D-branes are singletons in the space of twisted solutions.

The specific set of coefficients taken in Eq. (\ref{VacTwistDN}) are taken to match the overall coefficients from the non-rotated twisted solutions. This ansatz makes the boundary state Cardy consistent with itself. To see this we need to compute the amplitude
\begin{equation}\label{ampScheme}
A_{DN,\phi}:= {}_\phi\langle \langle DN^{\pm\pm}(y_0^1,\tilde y_0^2)| e^{-\pi T H} |DN^{\pm\pm}(y_0^1,\tilde y_0^2)\rangle\rangle_\phi,
\end{equation}
and perform a modular $S$-transformation. The $H$ in the above expression stands for the total Hamiltonian of the bulk theory which is a sum over the twisted and untwisted sector Hamiltonians, as well as tensor products between the left and right Hamiltonians. This amplitude is the sum of four components,
\begin{equation}
A_{DN,\phi}= \frac{1}{4} A_{cc}+ \frac{1}{2} A_{tt}+ \frac{1}{2^{3/2}} A_{ct}+ \frac{1}{2^{3/2}} A_{tc}.
\end{equation}
The untwisted term is given by 
\begin{equation}\label{untwistedAmp2}
\begin{split}
A_{cc}&=\prod_{n>0} \langle 0| e^{ \underline a_n^{ t}\Omega \tilde{ \underline a}_n}| e^{-2\pi T H} |e^{ \underline a_n^{\dag t}\Omega \tilde{ \underline a}_n^\dag} |0\rangle\\
&=\prod_{n>0} \langle 0| e^{ a_n^1  \tilde a^1_n}e^{- a_n^2  \tilde a^2_n} | e^{-\pi T  H_{c,1}} e^{-\pi T  H_{c,2}}| e^{ a_n^{\dag 1}  \tilde a^{\dag 1}_n}e^{- a_n^{2\dag}  \tilde a^{\dag 2}_n}|0\rangle\\
&=\prod_{n>0} \langle 0| e^{ a_n^1  \tilde a^1_n} e^{-\pi T  H_{c,1}} e^{ a_n^{\dag 1}  \tilde a^{\dag 1}_n} |0\rangle  \langle 0|  e^{- a_n^2  \tilde a^2_n} e^{-\pi T  H_{c,2}}   e^{- a_n^{2\dag}  \tilde a^{\dag 2}_n}|0\rangle
\end{split}
\end{equation}
where we used $\langle v_1\otimes w_1 | v_2\otimes w_2 \rangle := \langle v_1|w_1\rangle \langle v_2 |w_2\rangle$. $H_{c,i}$ are the untwisted Hamiltonian expressions from Eq. (\ref{hamiltonian}). Formally, this amplitude is computed with the rotated oscillators ${}^R\underline a_n$ and the respectively rotated Hamiltonian ${}^R H$ but again used the fact that the oscillator algebra is left invariant by the the $R(\phi)$ action. So the untwisted contribution is,
\begin{equation}\label{AccFinal}
A_{cc}=\frac{1}{\eta(q)^2} \ , \ \ q := e^{-2\pi T}.
\end{equation}
The twisted term is
\begin{equation}\label{twistedAmp2}
\begin{split}
A_{tt} =\prod_{n>0} \langle 0| e^{  b_n^1  \tilde{ b}_n^1} e^{-\pi T  H_{t,1}} e^{  b_n^{\dag 1}  \tilde{ b}_n^{\dag1} }|0\rangle \langle 0| e^{ - b_n^2  \tilde{ b}_n^2} e^{-\pi T H_{t,2}} |e^{- b_n^{\dag 2}  \tilde{b}_n^{2 \dag}} |0\rangle,\\
\end{split}
\end{equation}
where $H_{t,i}$ is the Hamiltonian in the twisted sectors, given in Eq. (\ref{ht2}). This term provides the following contribution,
\begin{equation}\label{twistedAmp3}
\begin{split}
A_{tt} = \left(q^{\frac{1}{48}} \prod_{n>0} \frac{1}{1-q^{{(n-\frac{1}{2})}}}\right)^2.
\end{split}
\end{equation}
The cross terms $A_{ct}$ and $A_{tc}$ are computed are the partially twisted amplitudes
\begin{equation}
\begin{split}
A_{ct}:=&\prod_{n>0} \langle 0| e^{ \underline c_n^{ t}\Omega_\phi \tilde{ \underline c}_n}e^{-2\pi T H} e^{ \underline c_n^{\dag t}\Omega_\phi \tilde{ \underline c}_n^\dag}| 0\rangle \ , \ \ A_{tc}:=\prod_{n>0} \langle 0| e^{ \underline d_n^{ t}\Omega_\phi \tilde{ \underline d}_n} e^{-2\pi T H} e^{ \underline d_n^{\dag t}\Omega_\phi \tilde{ \underline d}_n^\dag} |0\rangle.
\end{split}
\end{equation}
These two terms have the same contribution of
\begin{equation}\label{Act}
A_{ct}=A_{tc} =\left(q^{\frac{1}{48}} \prod_{n>0} \frac{1}{1-q^{n-\frac{1}{2}}}\right) \frac{1}{\eta(q)} = A_{tt}^{1/2} A_{cc}^{1/2}.
\end{equation}
Thus combining all the terms we have that the amplitude $A_{DN,\phi}$ of Eq. (\ref{ampScheme}) is given by the total expression
\begin{equation}\label{treeLevel}
A_{DN,\phi}(q) = \frac{1}{4}\frac{1}{\eta(q)^2} + \frac{1}{2} \left(q^{\frac{1}{48}} \prod_{n>0} \frac{1}{1-q^{{(n-\frac{1}{2})}}}\right)^2 + \frac{1}{2^{1/2}} \left(q^{\frac{1}{48}} \prod_{n>0} \frac{1}{1-q^{n-\frac{1}{2}}}\right) \frac{1}{\eta(q)},
\end{equation}
with $q := e^{-2\pi T}$. 

Now, using a modular $S$-transformation we take the tree-level channel amplitude $A_{DN,\phi}$ to a loop-channel partition function. Under the transformation $T\rightarrow 1/T=:t$ the first summand of Eq. (\ref{treeLevel}) transforms as
\begin{equation}\label{sqrtAccTilde}
A_{cc}(\tilde q)=  \frac{1}{t \ \eta\left(\tilde q \right)^2} \ , \ \ \ \tilde q := e^{-2\pi t},
\end{equation}
where we are abusing notation by writing $A_{cc}(\tilde q)$. To transform the twisted term $A_{tt}$ we use the identity found in \cite{affleck}
\begin{equation}\label{IdentityAffleck}
 q ^{1/48}\prod_{n>0} \frac{1}{1-  q^{n-1/2}} = \frac{\theta_2 ( q^{1/2}) }{2\eta( q)},
\end{equation}
which we use to rewrite Eq. (\ref{twistedAmp3}) as
\begin{equation}
A_{tt}^{1/2} = \frac{\theta_2 ( q^{1/2}) }{2\eta( q)} .
\end{equation}
By using the S-transformation of the theta-function $\theta_2(e^{2\pi i \tau})$
\begin{equation}
\theta_4 \left(-\frac{1}{\tau}\right) = \sqrt{-i \tau} \theta_2 (\tau),
\end{equation}
we obtain the transformed twisted term
\begin{equation}
A_{tt}(\tilde q)  =\frac{1}{2}\left( \frac{\theta_4(\tilde q^2)}{ \eta(\tilde q)}\right)^2.
\end{equation}
Inserting the $\tilde q$ expressions into the full amplitude
\begin{equation}\label{}
A_{DN,\phi}(\tilde q) = \frac{1}{4}\frac{1}{ t\ \eta\left(\tilde q\right)^2} + \frac{1}{4} \left( \frac{\theta_4(\tilde q^2)}{\ \eta(\tilde q)}\right)^2 + \frac{1}{2} \frac{\theta_4(\tilde q^2)}{\sqrt{ t} \ \eta\left(\tilde q\right)^2},
\end{equation}
which shows self-consistency. That is, the modular $S$-transformation maps the $DN^{\pm\pm}-DN^{\pm\pm}$ amplitude to a partition function with a unique vacuum. Given that the $DN^{\pm\pm}$ state would single out the vacuum term in any other boundary state, a similar computation shows that these rotated twisted states are not consistent with the untwisted states we already built. Therefore, twisted boundary states are allowed as long as they are not rotated.

In the next chapter, we use the spectrum we have found for D-branes in the product orbifold to obtain possible defects between the $S^1/\mathbb{Z}_2$ theories. The structure of the D-branes listed here is reminiscent of those in the torus theory except that we have catalogued the twisted degrees of freedom as well. The set of D-branes found here are given as Cardy-consistent elements of the boundary CFT. It is important to note that the boundary states presented here are more than just means to obtaining defects. There has been a prolific amount of work on D-branes on $\mathbb{C}^2/\mathbb{Z}_n$ orbifolds \cite{Douglas:1996sw} but much less number of examples in theories with curved or compact orbifold backgrounds \cite{Brunner:1999} such as the solutions we have presented in this work.

%ssssssssssssssssssssssssssssssssssssssssssssssssssssssssssssssssss
\chapter{\uppercase{Defects between $S^1/\mathbb{Z}_2$ theories}}

In this chapter we answer the question of how to glue together two $S^1_R/\mathbb{Z}_2$ bosonic theories by cataloguing the possible defects between such models. These defects are obtained from the D-branes in the product theory $({S}^1_{R}/\mathbb{Z}_2)^2$ via the unfolding mapping. Unfolding as developed in \cite{bachas07} is the inverse procedure of the ``folding trick'' used in 2D field theories. 
Given two CFTs separated by a 1-dimensional interface, the total theory $\text{CFT}_1\oplus \text{CFT}_2$ is equivalent to the theory $\text{CFT}_1\otimes \xbar{\text{CFT}}_2$ defined on the worldsheet folded such that the interface is mapped to a boundary for the tensor theory. $\xbar{\text{CFT}}_2$ refers to ${\text{CFT}}_2$ but with the left- and right-movers interchanged. Such a correspondence is referred to as the folding trick. The inverse of the previous procedure involves starting with a larger theory on a worldsheet with boundary. Under the assumption that the theory has two non-interacting sectors in the interior of the worldsheet, the boundary can be mapped to a defect separating the two sectors which are now the full theories to either side of the interface. That is, unfolding is the linear map
\begin{equation}
\mathcal{H}^{(1\otimes 2)}_{\partial \Sigma} \xrightarrow{\text{unfolding}} \Hom(\mathcal{H}^{(2)},\mathcal{H}^{(1)}) \cong \mathcal{H}^{(2)*}\otimes\mathcal{H}^{(1)},
\end{equation}
 from the boundary states in the product theory to  the space of linear maps between the Hilbert spaces on each side of the defect.

To unfold we define mirror fields by taking $\tau\rightarrow-\tau$ in in the expressions for $X$ and $Y$ in Eq. (\ref{xExpansion}) and Eq. (\ref{yExpansion}). The results are also solutions to the variation problem but sending
\begin{equation}\label{unfoldoscillators}
(\widehat N ,a_n,\widetilde a_n)\rightarrow (-\widehat N,-\widetilde a_n^\dag,-a_n^\dag),
\end{equation}
in the mode expansions. Using the above mapping we take the $(S^1/\mathbb{Z}_2)^2$ boundary states
\begin{equation}
|B\rangle\rangle =\sum B_{\lambda_1,\widetilde \lambda_1,\lambda_2,\widetilde \lambda_2}|\lambda_1,\widetilde \lambda_1\rangle|\lambda_2,\widetilde \lambda_2\rangle ,
\end{equation}
to oriented defects given by
\begin{equation}
I^{1\leftarrow 2}= \sum B_{\lambda_1,\widetilde \lambda_1,\lambda_2,\widetilde \lambda_2} |\lambda_1,\widetilde \lambda_1\rangle \langle\widetilde \lambda_2,\lambda_2|.
\end{equation}

%sssssssssssssssssssssssssssssssssssssssssssssssssssssssssssssssss
\section{Untwisted defects}

There are two main classes of untwisted defects coming from the rotated classes of boundaries states $|D1_O(\alpha,\beta)\rangle\rangle_\phi$ and $|D2/D0_O(\alpha,\beta)\rangle\rangle_\theta$ in the untwisted sector. Unfolding the former, given in Eq. (\ref{d1orbifoldcosine}), we obtain the conformal defect
\begin{equation}\label{evendefect}
E^{(R_1\leftarrow R_2)}_{(k_1,k_2)}(\alpha,\beta):= \mathcal{E}_{(k_1,k_2)}(\alpha,\beta) \prod_{n>0}\exp ( -\Omega_\phi ^{11}a^{\dag 1}_n\tilde a_n^{\dag 1} + \Omega_\phi^{12}a^{\dag 1}_n a_n^{2} -\Omega_\phi ^ {22}a^{ 2}_n\tilde a_n^{ 2} + \Omega_\phi ^{21}\tilde a^{ 2}_n\tilde a_n^{\dag 1}),
\end{equation}
where
\begin{equation}\label{vacuumLatSum}
\mathcal{E}_{(k_1,k_2)}(\alpha,\beta)= 2 g^{+}\sum_{M,N}\cos(N\alpha)\cos(M\beta)|k_2N,k_1M\rangle_1\langle k_1 N, k_2 M|_2.
\end{equation}
The D2/D0-brane give in Eq. (\ref{d2d0orbifoldcosine}) unfolds to
\begin{equation}\label{oddDefect}
O^{(R_1\leftarrow R_2)}_{(k_1,k_2)}(\alpha,\beta):=\mathcal{O}_{(k_1,k_2)}(\alpha,\beta) \prod_{n>0}\exp ( - \widetilde \Omega_\theta^{11}a^{\dag 1}_n\tilde a_n^{\dag 1}+ \widetilde \Omega_\theta^{12}a^{\dag 1}_n a_n^{2} -  \widetilde \Omega_\theta^{22}a^{ 2}_n\tilde a_n^{ 2} + \widetilde \Omega_\theta^{21}\tilde a^{ 2}_n\tilde a_n^{\dag 1}),
\end{equation}
where
\begin{equation}
\mathcal{O}_{(k_1,k_2)}(\alpha,\beta)= 2 g^{(-)}\sum_{M,N}\cos(N\alpha)\cos(M\beta)
 |k_1M,k_2N\rangle_1\langle k_1N, k_2M| _2.
\end{equation}
The two defects above are the $\mathbb{Z}_2\times\mathbb{Z}_2$-symmetrized version of their counterparts in the circle theory of \cite{bachas07}.

At different values of the parameter $\phi$ we obtain defects which fall somewhere in the spectrum between being totally reflective or totally transmissive. A defect is called totally reflective when no information can flow across the interface which means the theories flanking the defect decouple entirely. A defect is totally transmissive when it commutes with the field insertion of the free fields.

The totally reflective defects in the untwisted sector appear when  $\phi$ or $\theta$ are multiples of $\pi/2$. At these values we obtain the following four varieties of totally reflectivity,
\begin{equation}
E^{(R_1\leftarrow R_2)}_{(0,k_2)}(\alpha,\beta)= |D_O(\alpha)^{k_2}\rangle\rangle \langle\langle N_O(\beta)^{k_2}|,
\end{equation}
\begin{equation}
E^{(R_1\leftarrow R_2)}_{(k_1,0)}(\alpha,\beta)=|N_O(\alpha)^{k_2}\rangle\rangle \langle\langle D_O(\beta)^{k_2}|,
\end{equation}
\begin{equation}\label{}
O^{(R_1\leftarrow R_2)}_{(0,k_2)}(\alpha,\beta)= |N_O(\alpha)^{k_2}\rangle\rangle \langle\langle N_O(\beta)^{k_2}|,
\end{equation}
\begin{equation}
O^{(R_1\leftarrow R_2)}_{(k_1,0)}(\alpha,\beta)=|D_O(\alpha)^{k_2}\rangle\rangle \langle\langle D_O(\beta)^{k_2}|.
\end{equation}

To obtain defects which are totally transmissive set the rotation angles to odd multiples of $\pi/4$. At these values, the gluing matrices are
\begin{equation}\label{}
\Omega_{(2m+1)\pi/4}=\left( \begin{array}{cc}
0 & (-)^m\\
(-)^m &0  \end{array}\right) \ \ \ \ , \ \ \ \ \widetilde \Omega_{(2m+1)\pi/4}=\left( \begin{array}{cc}
0 & (-)^m\\
-(-)^m&0  \end{array}\right).
\end{equation}
Using $\Omega_{(2m+1)\pi/4}$ and $\widetilde \Omega_{(2m+1)\pi/4}$ as above we can obtain the perfectly transmissive untwisted defects. From the defect class in Eq. (\ref{evendefect}) we get,
\begin{equation}\label{tranUntwisted}
E^{(R_1\leftarrow R_2)}_{(p_1,p_2)}(\alpha,\beta)= \mathcal{E}_{(p_1,p_2)}(\alpha,\beta) \prod_{n>0}e^{(-)^m( a^{\dag 1}_n a_n^{2}  + \tilde a^{ 2}_n\tilde a_n^{\dag 1})} ,
\end{equation}
where
\begin{equation}
\mathcal{E}_{(p_1,p_2)}(\alpha,\beta)= 2 g^{+}\sum_{M,N}\cos(N\alpha)\cos(M\beta)|p_2N,p_1M\rangle_1\langle p_1 N, p_2 M|_2,
\end{equation}
and the integers $(p_1,p_2)$ satisfy
\begin{equation}\label{evenP}
(-1)^m\frac{p_1}{p_2}=\frac{R_2}{R_1}.
\end{equation}
From defect class in Eq. (\ref{oddDefect}) we obtain,
\begin{equation}\label{OddTransmissive}
O^{(R_1\leftarrow R_2)}_{(q_1,q_2)}(\alpha,\beta):=\mathcal{O}_{(q_1,q_2)}(\alpha,\beta) \prod_{n>0}e^{(-)^m( a^{\dag 1}_n a_n^{2}-\tilde a^{ 2}_n\tilde a_n^{\dag 1})},
\end{equation}
where
\begin{equation}
\mathcal{O}_{(q_1,q_2)}(\alpha,\beta)= 2 g^{(-)}\sum_{M,N}\cos(N\alpha)\cos(M\beta)
 |q_1M,q_2N\rangle_1\langle q_1 N, q_2 M| _2.
\end{equation}

At the level of the oscillator modes, a defect is totally transmissive if it commutes with oscillators up to a phase factor. One can check that for the defect $E^{(R_1\leftarrow R_2)}$  in Eq. (\ref{tranUntwisted}) the following holds,
\begin{equation}\label{trans1}
E\ a_m^{2\dag} = (-)^l a_m^{1\dag}E \ , \ \ \ \ E\ a_m^{2} =(-)^{l}a_m^1 E,
\end{equation}
\begin{equation}\label{trans3}
E\  \widetilde a_m^{2\dag} =(\pm) (-)^{l}\widetilde a_m^{1\dag} E \ , \ \ \ E\ \widetilde a_m^{2} =(\pm) (-)^{l}\widetilde a_m^{1\dag} E.
\end{equation}
\begin{equation}\label{zeroCommute}
E\ \widehat N^2 = \frac{R_2}{R_1}\widehat N^1 E \ , \ \ \ \ E \widehat M^2 = \frac{R_1}{R_2}\widehat M^1 E,
\end{equation}
which shows that indeed the defect is totally transmissive. Furthermore it follows that the defect also satisfies $L_n^1 E = E\ L_n^2$ and $\bar L_n^1 E= E \ \bar L_n^2$ which means that it is topological. Similar relationships hold for the defect $O^{(R_1\leftarrow R_2)}$ in Eq. (\ref{oddDefect}).

%ssssssssssssssssssssssssssssssssssssssssssssssssssssssssssssssss
\section{Fully twisted defects}

Now we apply the unfolding map to the rotated D-branes in the twisted sector thus obtaining defects which are twisted. We first focus on those defects coming from the fully twisted boundary states $|DN^{\pm\pm}(y_0^1,\tilde y_0^2)\rangle\rangle_\phi$ and $|ND^{\pm\pm}(\tilde y_0^1, y_0^2)\rangle\rangle_\phi$. In the next subsection we list those defects arising from the unfolding of the partially twisted D-branes.

The defects corresponding to the isolated D-branes $|DN^{\pm\pm}(y_0^1,\tilde y_0^2)\rangle\rangle$ and $|ND^{\pm\pm}(\tilde y_0^1, y_0^2)\rangle\rangle$ (in equations (\ref{dnNoRotated}) and (\ref{ndNoRotated})) are, respectively,
\begin{equation}\label{LDN}
L_{DN_{\pm\pm},0}^{(R_1\leftarrow R_2)}(y_0^1,\tilde y_0^2):=\mathcal{V}^{\pm\pm}_{DN,0}(y_0^1,\tilde y_0^2)\prod_{n>0}\sum_{s\in\left\{a,b,c,d\right\}}  e^{ s^{\dag 1}_n\tilde s_n^{\dag 1} - s^{ 2}_n\tilde s_n^{ 2} },
\end{equation}
\begin{equation}
L_{ND_{\pm\pm},0}^{(R_1\leftarrow R_2)}(y_0^1,\tilde y_0^2) := \mathcal{V}^{\pm\pm}_{ND,0}(\tilde y_0^1, y_0^2)\prod_{n>0}\sum_{s\in\left\{a,b,c,d\right\}}  e^{ - s^{\dag 1}_n\tilde s_n^{\dag 1}  + s^{ 2}_n\tilde s_n^{ 2} }.
\end{equation}
where,
\begin{equation}
\begin{split}
\mathcal{V}^{\pm\pm}_{DN,0}(y_0^1,\tilde y_0^2) =& \frac{1}{2^{3/2}}\sqrt{\frac{R_2}{R_1}}\sum_{M,N} e^{iMy^1_0/R_1 +iNR_2 \tilde y^2_0}|M,0\rangle_1 \langle 0,-N|_2 \pm\pm   \frac{1}{2^{1/2}} |y^1_0,T\rangle_1 \langle \tilde y^2_0,T |_2\\
&\pm \frac{1}{2^{5/4}}\frac{1}{\sqrt{R_1}}\sum_M e^{iMy^1_0/R_1}|M,0\rangle_1 \langle \tilde y^2_0,T |_2\\
&\pm \frac{1}{2^{3/4}} \sqrt{R_2}\sum_N e^{iNR_2\tilde y^2_0} |y^1_0,T\rangle_1 \langle 0,-N |_2,
\end{split}
\end{equation}

\begin{equation}
\begin{split}
\mathcal{V}^{\pm\pm}_{ND,0}(\tilde y_0^1, y_0^2) =&  \frac{1}{2^{3/2}}\sqrt{\frac{R_1}{R_2}}\sum_{M,N} e^{iNR_1 \tilde y^1_0+ iMy^2_0/R_2}|0,N\rangle_1 \langle M,0|_2 \pm\pm \frac{1}{2^{1/2}} |\tilde y^1_0,T\rangle_1 \langle  y^2_0,T |_2 \\
& \pm \frac{1}{2^{3/4}}\sqrt{R_1}\sum_N e^{iN\tilde y^1_0 R_1}|0,N\rangle_1 \langle  y^2_0,T |_2 \\
&\pm  \frac{1}{2^{5/4}}\frac{1}{\sqrt{R_2}}\sum_M e^{iMy^2_0/R_2}| \tilde y^1_0,T\rangle_1 \langle  M,0 |_2 .
\end{split}
\end{equation}
We note that differently from the untwisted sector in the product theory, there are no twisted boundary states which represent a bound system of D2- and D0-branes. So the $R(\phi)$-invariant DD and NN boundary states in Eq.(\ref{diagonaldd}) and (\ref{diagonalnn}) have to unfolded separately, giving the two additional defects
\begin{equation}\label{phiInvariant1}
L_{DD_{\pm\pm}}^{(R_1\leftarrow R_2)}(y_0^1, y_0^2):= \mathcal{V}^{\pm\pm}_{DD}(y_0^1, y_0^2)\prod_{n>0}\sum_{s\in\left\{a,b,c,d\right\}}  e^{s_n^{\dag 1}\tilde s_n^{\dag 1}+\tilde s_n^{ 2} s_n^{ 2}},
\end{equation}
\begin{equation}\label{phiInvariant2}
L_{NN_{\pm\pm}}^{(R_1\leftarrow R_2)}(\tilde y_0^1, \tilde y_0^2):= \mathcal{V}^{\pm\pm}_{NN}(\tilde y_0^1, \tilde y_0^2)\prod_{n>0}\sum_{s\in\left\{a,b,c,d\right\}}  e^{-s_n^{\dag 1}\tilde s_n^{\dag 1}-\tilde s_n^{ 2} s_n^{ 2}},
\end{equation}
where
\begin{equation}
\begin{split}
\mathcal{V}^{\pm\pm}_{DD}(y_0^1, y_0^2):= & \frac{1}{4\sqrt{R_1R_2}}\sum_{M_1,M_2} e^{iM_1y_0^1/R_1+iM_2y_0^2/R_2 }|M_1,0\rangle_1 \langle M_2,0| \pm \pm\frac{1}{ \sqrt{2}} |y_0^1 ,T\rangle_1 \langle y_0^2, T|_2 \\
&\pm  \frac{1}{2^{5/4}\sqrt{R_1}}\sum_{M_1} e^{iM_1y_0^1/R_1}|M_1,0\rangle_1\langle y_0^2, T|_2\\
&\pm \frac{1}{2^{5/4}\sqrt{R_2}}\sum_{M_2} e^{iM_2y_0^2/R_2} |y_0^1, T\rangle_1 \langle M_2,0|_2
\end{split}
\end{equation}
\begin{equation}
\begin{split}
 \mathcal{V}^{\pm\pm}_{NN}(\tilde y_0^1, \tilde y_0^2):= & \frac{\sqrt{R_1 R_2}}{2}\sum_{N_1, N_2} e^{iN_1\tilde y_0^1R_1+iM_2\tilde y_0^2 R_2 }|0, N_1\rangle_1 \langle 0, - N_2 |_2 \pm \pm \frac{1}{\sqrt{2}} |\tilde y_0^1 ,T\rangle_1  \langle \tilde y_0^2 ,T|_2\\
&\pm \frac{\sqrt{R_1}}{2^{3/4}}\sum_{N_1} e^{iN_1\tilde y_0^1 R_1}|0,N_1,\rangle_1  \langle \tilde y_0^2 ,T| _2\\
&\pm \frac{\sqrt{R_2}}{2^{3/4}}\sum_{N_2} e^{iN_2\tilde y_0^2 R_2}|\tilde y_0^1 ,T\rangle_1 \otimes \langle 0, -N_2|_2 .
\end{split}
\end{equation}

The fully twisted sector contains elements which are completely reflective. The defects $L_{DD_{\pm\pm}}^{(R_1\leftarrow R_2)}$ and $L_{NN_{\pm\pm}}^{(R_1\leftarrow R_2)}$ given by equations (\ref{phiInvariant1}) and (\ref{phiInvariant2}) are fully reflective, and so are  $L_{DN_{\pm\pm},0}^{(R_1\leftarrow R_2)}$ and $L_{ND_{\pm\pm},0}^{(R_1\leftarrow R_2)}$.

%sssssssssssssssssssssssssssssssssssssssssssssssssssssssssssssss
\section{Partially twisted defects}
We call ``partially twisted defects'' those defects coming via unfolding the partially twisted boundary states $|Ai^{\pm}(y_0,x_0)\rangle\rangle$ and $|Bi^{\pm}(x_0,y_0)\rangle\rangle$.
The defects corresponding to those D-branes which are $R(\phi)$-invariant are given below,
\begin{equation}\label{pt1}
L_{A1_{\pm}}^{(R_1\leftarrow R_2)}(y_0, x_0):= \mathcal{V}^{\pm}_{A1}(y_0, x_0)\prod_{n>0}\sum_{s\in\left\{a,d\right\}}  e^{s_n^{\dag 1}\tilde s_n^{\dag 1}+\tilde s_n^{ 2} s_n^{ 2}},
\end{equation}
\begin{equation}
L_{A4_{\pm}}^{(R_1\leftarrow R_2)}(\tilde y_0,\tilde  x_0):= \mathcal{V}^{\pm}_{A4}(\tilde y_0, \tilde x_0)\prod_{n>0}\sum_{s\in\left\{a,d\right\}}  e^{-s_n^{\dag 1}\tilde s_n^{\dag 1}-\tilde s_n^{ 2} s_n^{ 2}},
\end{equation}
\begin{equation}
L_{B1_{\pm}}^{(R_1\leftarrow R_2)}( x_0,  y_0):= \mathcal{V}^{\pm}_{B1} (x_0,  y_0)\prod_{n>0}\sum_{s\in\left\{a,c\right\}}  e^{s_n^{\dag 1}\tilde s_n^{\dag 1}+\tilde s_n^{ 2} s_n^{ 2}},
\end{equation}
\begin{equation}\label{pt4}
L_{B4_{\pm}}^{(R_1\leftarrow R_2)}(\tilde x_0, \tilde y_0):= \mathcal{V}^{\pm}_{B1} (\tilde x_0,  \tilde y_0)\prod_{n>0}\sum_{s\in\left\{a,c\right\}}  e^{-s_n^{\dag 1}\tilde s_n^{\dag 1}-\tilde s_n^{ 2} s_n^{ 2}},
\end{equation}
where
\begin{equation}
\begin{split}
\mathcal{V}^{\pm}_{A1}(y_0, x_0) = &\frac{1}{2\sqrt{R_1}}\sum_{M_1, M_2} e^{iM_1 y_0/R_1}\cos\left(\frac{M_2x_0}{R_2}\right) |M_1 ,0\rangle_1 \langle M_2,0|_2 \\
&\pm \frac{1}{ 2^{1/4} \sqrt{R_2}}\sum_{M_2} \cos\left(\frac{M_2x_0}{R_2}\right)|y_0,T\rangle_1 | \langle M_2, 0|_2,
\end{split}
\end{equation}
\begin{equation}
\begin{split}
\mathcal{V}^{\pm}_{A4}(\tilde y_0, \tilde x_0) =&\sqrt{R_1 R_2}\sum_{N_1, N_2} e^{iN_1R_1\tilde y_0}\cos\left(N_2 R_2\tilde x_0\right) |0,N_1\rangle_1\langle 0 , -N_2|_2\\
& \pm  2^{1/4}\sqrt{R_2}\sum_{N_2} \cos\left(N_2 R_2\tilde x_0\right) |\tilde y_0,T\rangle_1  \langle 0,-N_2|_2,
\end{split}
\end{equation}

\begin{equation}
\begin{split}
\mathcal{V}^{\pm}_{B1}(x_0, y_0)= & \frac{1}{2 \sqrt{R_1 R_2 }}\sum_{M_1, M_2} e^{iM_2y_0/R_2} \cos\left(\frac{M_1 x_0}{R_1}\right)| M_1,0\rangle_1 \langle M_2,0 |_2 \\
&\pm \frac{1}{2^{1/4}\sqrt{R_1}}\sum_{M_1} \cos\left(\frac{M_1 x_0}{R_1}\right)| M_1,0\rangle_1\langle y_0, T |_2,
\end{split}
\end{equation}

\begin{equation}
\begin{split}
\mathcal{V}^{\pm}_{B4}(\tilde x_0, \tilde y_0) =  &\sqrt{ R_1 R_2}\sum_{N_1, N_2} e^{iN_2 R_2\tilde y_0}\cos\left(N_1 R_1\tilde x_0\right)|0,N_1  \rangle_1  \langle 0, -N_2|_2\\
&\pm 2^{1/4} \sqrt{ R_1}\sum_{N_1} \cos\left(N_1 R_1\tilde x_0\right)|0,N_1 \rangle_1 \langle \tilde y_0, T|_2 .
\end{split}
\end{equation}
%%%%%%%%%%%%%%%%%%%%%%%%%%%%%%%%%%%%%%%%%
\section{Fusion algebra}

Whether topological or not, defects can be added. In the context of the unfolding map, this statement follows because boundary states can be added. Central to defects, they can also be fused together to obtain a new defect. This operation is well defined without any need for regularization when working with topological defects. For two generic defects $D$ and $D'$, their fusion may be defined as
\begin{align}\label{interfaceFuse}
    D*D' = \lim_{\epsilon \rightarrow{0}} e^{2\pi d/\epsilon} D \ e^{-\epsilon H} D',
\end{align}
where $H$ is the Hamiltonian of the bulk theory wedged between the two defects, and $e^{d/\epsilon}$ is a self-energy counter-term with a free parameter $d$ which must be fixed \cite{bachas07}.

In this section we restrict to those defects which are topological. Recalling that topological defects commute with elements of the Virasoro algebra, we have $\left[ D, H\right] = 0$ since $H = L_0 + \xbar L_0$. Therefore we pay no penalty by moving $e^{-\epsilon H}$ outside the product in the right-hand side of Eq. (\ref{interfaceFuse}) and taking the limit $\epsilon \rightarrow{0}$. Since no regularization is required in this case, we will set $d=0$ in the products we compute below. That is, the fusion products for the topological cases here are given by the composition of the operators representing such defects.

%%%%%%%%%%%%%%%%%%%%%%%%%
\section{Fusion between totally transmissive defects with $R_2=R_1=R$}

We first consider the totally transmissive defects which are untwisted. At generic radii $R$, the class of defects ${O}^{(R\leftarrow R)}_{(q_1,q_2)}$ do not contribute any defects since the condition
\begin{equation}\label{blah}
(-1)^m\frac{q_1}{q_2}=2R_1 R_2.
\end{equation}
coming from Eq. (\ref{thetatheta}) is not satisfied unless we are at the self-dual radius, $R_1 = R_2 = R_*$. There are two defects coming from ${E}^{(R\leftarrow R)}_{(p_1,p_2)}$ since there are only two solutions to Eq. (\ref{evenP}), which is satisfied by the integers $(p_1,p_2)$ such that $p_1 = (-)^m p_2=p$. But like in the $S^1$-defects, only those defects with $p=1$ are invertible. To check this, we only have to work with the vacuum sums since the higher terms annihilate the vacua $|N,M\rangle$:
\begin{equation}
\begin{split}
E^{(R\leftarrow R)}_{(p,p)}(\alpha,\beta)|N',M'\rangle&=  2 g^{+}\sum_{M,N}\cos(N\alpha)\cos(M\beta)|pN,pM\rangle \langle p N, p M| N',M'\rangle\\
&=  2 g^{+}\sum_{M,N}\cos(N\alpha)\cos(M\beta)|pN,pM\rangle \delta_{pN,N'} \delta_{pM, M'},
\end{split}
\end{equation}
but this expression will be zero for any $N'\neq 0 \mod p$, or $M'\neq 0 \mod p$. Hence the only $E^{(R\leftarrow R)}_{(p,p)}$ which is not many-to-one are those with $p=1$. Same argument holds for $E^{(R\leftarrow R)}_{(p,-p)}$.

Recalling that the normalization expression $g^{+}$ for untwisted defect is
\begin{equation}
g^{+}_{(p_1,p_2)}=\sqrt{\frac{p_1p_2}{\sin2\phi}},
\end{equation}
we see that at $\phi=(2m+1)\pi/4$, we have 
\begin{equation}
g^{+}_{(1,1)}=g^{+}_{(1,-1)}=1.
\end{equation}

At a generic radius $R$ only two of our defects are topological. These are the defects $E^{(R\leftarrow R)}_{(1,1)}$ and $E^{(R\leftarrow R)}_{(1, -1)}$, where the general expression is given in Eq. (\ref{tranUntwisted}). Their fusion obeys
\begin{equation}\label{fusionEven++}
E_{(1,1)}(\alpha,\beta)E_{(1,1)}(\alpha ',\beta ') = E_{(1,-1)}(\alpha,\beta)E_{(1,-1)}(\alpha ',\beta '),
\end{equation}

\begin{equation}\label{fusionEven+-}
E_{(1,1)}(\alpha,\beta)E_{(1,-1)}(\alpha ',\beta ') = E_{(1,-1)}(\alpha,\beta)E_{(1,1)}(\alpha ',\beta ') ,
\end{equation}
as we show below.
\begin{equation}
\begin{split}
&E_{(1,-1)}(\alpha,\beta)E_{(1,-1)}(\alpha ',\beta ')\\
&=\left(\mathcal{E}_{(1,-1)}(\alpha,\beta)\prod_{n>0} e ^{-a_n^\dag a_n - \tilde a _n^\dag \tilde a_n}\right)\left(\mathcal{E}_{(1,-1)}(\alpha',\beta') \prod_{n>0}e ^{-a_n^\dag a_n - \tilde a _n^\dag \tilde a_n}\right)\\
&=\prod_{n>0}\sum_{s_n,t_n} \sum_{i_n,j_n} \frac{(-)^{s_n}}{s_n!}\frac{(-)^{t_n}}{t_n!} \frac{(-)^{i_n}}{i_n!}\frac{(-)^{j_n}}{j_n!}\nonumber\\
& \ \ \ \ \ (a_n^\dag)^{s_n} (\tilde a_n^\dag)^{t_n }\mathcal{E}_{(1,-1)}(\alpha,\beta) (a_n)^{s_n} (\tilde a_n)^{t_n } (a_n^\dag)^{i_n} (\tilde a_n^\dag)^{j_n }\mathcal{E}_{(1,-1)}(\alpha',\beta') (a_n)^{i_n} (\tilde a_n)^{j_n } \\
&=\prod_{n>0}\sum_{s_n,t_n}  \frac{1}{s_n!}\frac{1}{t_n!}  (a_n^\dag)^{s_n} (\tilde a_n^\dag)^{t_n }\mathcal{E}_{(1,-1)}(\alpha,\beta)\mathcal{E}_{(1,-1)}(\alpha',\beta') (a_n)^{s_n} (\tilde a_n)^{t_n }\\
&=\left( \mathcal{E}_{(1,-1)}(\alpha,\beta)\mathcal{E}_{(1,-1)}(\alpha',\beta') \right)\prod_{n>0} e ^{a_n^\dag a_n + \tilde a _n^\dag \tilde a_n},
\end{split}
\end{equation}
in the above we used 
\begin{equation}
\langle 0|a_p ^{m_p} a_{-p}^{m_p'}|0\rangle = m_p! \delta_{m_p,m_p'},
\end{equation}
 which holds for the canonical algebra. Now we observe that,
\begin{equation}
\begin{split}
&\mathcal{E}_{(1,-1)}(\alpha,\beta)\mathcal{E}_{(1,-1)}(\alpha',\beta') \\
&=4 \sum_{M,N} \sum_{M',N'} \cos(N \alpha)\cos (M\beta) \cos(N '\alpha') \cos (M' \beta ')\\
& \ \ \ \ \ \ \ \ \ \ \ |-N,M\rangle\langle N,-M| -N',M'\rangle\langle N',-M'|\\
 &=4 \sum_{M,N} \cos(N \alpha)\cos (M\beta) \cos(N \alpha') \cos (M\beta ') |-N,M\rangle\langle -N,M|\\
 &=4 \sum_{M,N} \cos(N \alpha)\cos (M\beta) \cos(N \alpha') \cos (M\beta ') |N,M\rangle\langle N,M|.
\end{split}
\end{equation}
Using the trigonometric identity
\begin{align}
    2\cos x \cos y = \cos(x+y) + \cos(x-y),
\end{align}
we rewrite the vacuum expression as
\begin{align}
   & \mathcal{E}_{(1,-1)}(\alpha,\beta)\mathcal{E}_{(1,-1)}(\alpha',\beta') \nonumber\\
&= \sum_{M,N} \left( \cos[N(\alpha- \alpha')] \cos[M(\beta- \beta')] + \cos[N(\alpha- \alpha')] \cos[M(\beta + \beta')]\right. \nonumber\nonumber \\
    &\left. + \cos[N(\alpha + \alpha')] \cos[M(\beta - \beta')] + \cos[N(\alpha + \alpha')] \cos[M(\beta + \beta')]\right)|N,M\rangle\langle N,M|\nonumber\\
    & = \frac{1}{2}\mathcal{E}_{(1,1)}(\alpha-\alpha',\beta-\beta') + \frac{1}{2}\mathcal{E}_{(1,1)}(\alpha-\alpha',\beta+ \beta')\nonumber \\
    & + \frac{1}{2}\mathcal{E}_{(1,1)}(\alpha + \alpha',\beta-\beta') + \frac{1}{2}\mathcal{E}_{(1,1)}(\alpha + \alpha',\beta + \beta').
\end{align}
So we obtain,
\begin{align}
E_{(1,-1)}(\alpha,\beta)E_{(1,-1)}(\alpha ',\beta ') &= \frac{1}{2}{E}_{(1,1)}(\alpha-\alpha',\beta-\beta') + \frac{1}{2}{E}_{(1,1)}(\alpha-\alpha',\beta+ \beta') \nonumber\\
    & + \frac{1}{2}{E}_{(1,1)}(\alpha + \alpha',\beta-\beta') + \frac{1}{2}{E}_{(1,1)}(\alpha + \alpha',\beta + \beta').
\end{align}
To compute $ E_{(1,1)}(\alpha,\beta)E_{(1,1)}(\alpha ',\beta ')$,
\begin{align}\label{EEplusPlus}
    E_{(1,1)}(\alpha,\beta)E_{(1,1)}(\alpha ',\beta ') & =\left(\mathcal{E}_{(1,1)}(\alpha,\beta)\prod_{n>0} e ^{a_n^\dag a_n + \tilde a _n^\dag \tilde a_n}\right)\left(\mathcal{E}_{(1,1)}(\alpha',\beta') \prod_{n>0}e ^{a_n^\dag a_n  \tilde a _n^\dag \tilde a_n}\right)\nonumber\\
    &=\left(\mathcal{E}_{(1,1)}(\alpha,\beta) \mathcal{E}_{(1,1)}(\alpha',\beta') \right)\prod_{n>0} e ^{a_n^\dag a_n + \tilde a _n^\dag \tilde a_n}.
\end{align}
The vacuum part of the product is obtained via
\begin{align}
  &  \mathcal{E}_{(1,1)}(\alpha,\beta)\mathcal{E}_{(1,1)}(\alpha',\beta')\nonumber\\
 &= 4 \sum_{M,N} \sum_{M',N'} \cos(N \alpha)\cos (M\beta) \cos(N '\alpha') \cos (M' \beta ') |N,M\rangle\langle N,M| N',M'\rangle\langle N',M'|\nonumber\\
 &=4 \sum_{M,N} \cos(N \alpha)\cos (M\beta) \cos(N \alpha') \cos (M\beta ') |N,M\rangle\langle N,M|\nonumber\\
    & = \frac{1}{2}\mathcal{E}_{(1,1)}(\alpha-\alpha',\beta-\beta') + \frac{1}{2}\mathcal{E}_{(1,1)}(\alpha-\alpha',\beta+ \beta')\nonumber \\
    & + \frac{1}{2}\mathcal{E}_{(1,1)}(\alpha + \alpha',\beta-\beta') + \frac{1}{2}\mathcal{E}_{(1,1)}(\alpha + \alpha',\beta + \beta').
\end{align}
Inserting the above into Eq. (\ref{EEplusPlus}) we get
\begin{align}
        E_{(1,1)}(\alpha,\beta)E_{(1,1)}(\alpha ',\beta ') &= \frac{1}{2}{E}_{(1,1)}(\alpha-\alpha',\beta-\beta') + \frac{1}{2}{E}_{(1,1)}(\alpha-\alpha',\beta+ \beta')\nonumber \\
    & + \frac{1}{2}{E}_{(1,1)}(\alpha + \alpha',\beta-\beta') + \frac{1}{2}{E}_{(1,1)}(\alpha + \alpha',\beta + \beta').
\end{align}
The other two combinations are,
\begin{align}
E_{(1,1)}(\alpha,\beta)E_{(1,-1)}(\alpha ',\beta ')& =\frac{1}{2}{E}_{(1,-1)}(\alpha-\alpha',\beta-\beta') + \frac{1}{2}{E}_{(1,-1)}(\alpha-\alpha',\beta+ \beta') \nonumber\\
    & + \frac{1}{2}{E}_{(1,-1)}(\alpha + \alpha',\beta-\beta') + \frac{1}{2}{E}_{(1,-1)}(\alpha + \alpha',\beta + \beta'),
\end{align}
and
\begin{align}
  {E}_{(1,-1)}(\alpha,\beta){E}_{(1,1)}(\alpha',\beta') &=   \frac{1}{2}{E}_{(1,-1)}(\alpha-\alpha',\beta-\beta') + \frac{1}{2}{E}_{(1,-1)}(\alpha-\alpha',\beta+ \beta') \nonumber\\
    & + \frac{1}{2}{E}_{(1,-1)}(\alpha + \alpha',\beta-\beta') + \frac{1}{2}{E}_{(1,-1)}(\alpha + \alpha',\beta + \beta').
\end{align}

We see that the topological defects at a generic radius $R$ form an algebra. The fusion products built out of $E_{(1,1)}$ and $E_{(1,-1)}$ in Eq. (\ref{fusionEven++}) and Eq. (\ref{fusionEven+-}) are expected if we note that their counterpart in the circle theory form the symmetry group $U(1)^2 \rtimes\mathbb{Z}_2$. Since we mod out the action of the $\mathbb{Z}_2$ subgroup, the twisting disappears giving $E_{(1,1)}E_{(1,1)}=E_{(1,-1)}E_{(1,-1)}$ and $E_{(1,-1)}E_{(1,1)}=E_{(1,1)}E_{(1,-1)}$. 

%%%%%%%%%%%%%%%%%%%%%%%%%%%%%%%%%%%%%%%%%%%%%%%%%%%%%%%%%%%%%%%%%%%%%%%%%%%%%5
\section{At the self-dual radius $R=R_*$}

At the self-dual radius we obtain two new defects which satisfy the condition in Eq. (\ref{blah}) and are invertible:  $O_{(1,1)}^{(R_*\leftarrow R_*)}$ and $O_{(1,-1)}^{(R_*\leftarrow R_*)}$ for $m$ even or odd, respectively. The enlarged set of topological defects forms an enhanced algebra. This algebra is related to the $(U(1) \ltimes \mathbb{Z}_2)^2$ symmetries of the circle self-dual symmetries, again by breaking the $\mathbb{Z}_2$-twisting.

To compute $O_{(1,1)}(\alpha,\beta) O_{(1,1)}(\alpha',\beta')$ we use Eq. (\ref{OddTransmissive}) with $m=0$,

\begin{equation}\label{EEplusplus}
\begin{split}
&O_{(1,1)}(\alpha,\beta) O_{(1,1)}(\alpha',\beta')=\left(\mathcal{O}_{(1,1)}(\alpha,\beta) \prod_{n>0}e^{ a^{\dag }_n a_n^{}-\tilde a^{ }_n\tilde a_n^{\dag }}\right)\left(\mathcal{O}_{(1,1)}(\alpha,\beta) \prod_{n>0}e^{ a^{\dag }_n a_n^{}-\tilde a^{ }_n\tilde a_n^{\dag }}\right)\\
&=\prod_{n>0}\sum_{s_n,t_n} \sum_{i_n,j_n} \frac{1}{s_n!}\frac{(-)^{t_n}}{t_n!} \frac{1}{i_n!}\frac{(-)^{j_n}}{j_n!}\\
& (a_n^\dag)^{s_n} (\tilde a_n^\dag)^{t_n }\mathcal{O}_{(1,1)}(\alpha,\beta) (a_n)^{s_n} (\tilde a_n)^{t_n } (a_n^\dag)^{i_n} (\tilde a_n^\dag)^{j_n }\mathcal{O}_{(1,1)}(\alpha',\beta') (a_n)^{i_n} (\tilde a_n)^{j_n } \\
&=\prod_{n>0}\sum_{s_n,t_n}  \frac{1}{s_n!}\frac{1}{t_n!}  (a_n^\dag)^{s_n} (\tilde a_n^\dag)^{t_n }\mathcal{O}_{(1,1)}(\alpha,\beta)\mathcal{O}_{(1,1)}(\alpha',\beta') (a_n)^{s_n} (\tilde a_n)^{t_n }\\
&=\left( \mathcal{O}_{(1,1)}(\alpha,\beta) \mathcal{O}_{(1,1)}(\alpha',\beta') \right)\prod_{n>0} e ^{a_n^\dag a_n + \tilde a _n^\dag \tilde a_n}.
\end{split}
\end{equation}
The vacuum part is given by,
\begin{equation}
\begin{split}
&\mathcal{O}_{(1,1)}(\alpha,\beta)\mathcal{O}_{(1,1)}(\alpha',\beta') \\
&= 4 \sum_{M,N} \sum_{M',N'} \cos(N \alpha)\cos (M\beta) \cos(N '\alpha') \cos (M' \beta ') |M,N\rangle\langle N,M| M',N'\rangle\langle N',M'|\\
 &=4\sum_{M,N} \cos(N \alpha)\cos (M\beta) \cos(M\alpha') \cos (N\beta ') |M,N\rangle\langle M,N|\\
 & =4\sum_{M,N} \cos(M \alpha)\cos (N\beta) \cos(N\alpha') \cos (M\beta ') |N,M\rangle\langle N,M|\\
 & =4\sum_{M,N} \cos (N\beta) \cos(N\alpha')  \cos(M \alpha) \cos (M\beta ') |N,M\rangle\langle N,M|\\
  & = \frac{1}{2}\mathcal{E}_{(1,1)}(\beta-\alpha',\alpha-\beta') + \frac{1}{2}\mathcal{E}_{(1,1)}(\beta-\alpha',\alpha + \beta') \\
    & + \frac{1}{2}\mathcal{E}_{(1,1)}(\beta + \alpha',\alpha-\beta') + \frac{1}{2}\mathcal{E}_{(1,1)}(\beta + \alpha',\alpha + \beta').
\end{split}
\end{equation}
Inserting the above into the Eq. (\ref{EEplusplus}) we obtain
\begin{align}
     O_{(1,1)}(\alpha,\beta) O_{(1,1)}(\alpha',\beta')& = \frac{1}{2} {E}_{(1,1)}(\beta-\alpha',\alpha-\beta') + \frac{1}{2} {E}_{(1,1)}(\beta-\alpha',\alpha + \beta') \nonumber\\
    & + \frac{1}{2} {E}_{(1,1)}(\beta + \alpha',\alpha-\beta') + \frac{1}{2} {E}_{(1,1)}(\beta + \alpha',\alpha + \beta').
\end{align}

 To compute $O_{(1,-1)}(\alpha,\beta) O_{(1,-1)}(\alpha',\beta')$ we use Eq. (\ref{OddTransmissive}) with $m=1$,
\begin{equation}\label{OO}
\begin{split}
&O_{(1,-1)}(\alpha,\beta) O_{(1,-1)}(\alpha',\beta')\\
&=\left(\mathcal{O}_{(1,-1)}(\alpha,\beta) \prod_{n>0}e^{- a^{\dag }_n a_n^{}+ \tilde a^{ }_n\tilde a_n^{\dag }}\right)\left(\mathcal{O}_{(1,-1)}(\alpha,\beta) \prod_{n>0}e^{ - a^{\dag }_n a_n^{}+ \tilde a^{ }_n\tilde a_n^{\dag }}\right)\\
&=\left( \mathcal{O}_{(1,-1)}(\alpha,\beta)  \mathcal{O}_{(1,-1)}(\alpha',\beta') \right)\prod_{n>0} e ^{a_n^\dag a_n + \tilde a _n^\dag \tilde a_n}.
\end{split}
\end{equation}
The vacuum part is given by,
\begin{equation}
\begin{split}
&\mathcal{O}_{(1,-1)}(\alpha,\beta)\mathcal{O}_{(1,-1)}(\alpha',\beta') \\
&= 4 \sum_{M,N} \sum_{M',N'} \cos(N \alpha)\cos (M\beta) \cos(N '\alpha') \cos (M' \beta ') |M,-N\rangle\langle N,-M| M',-N'\rangle\langle N',-M'|\\
 &=4\sum_{M,N} \cos(N \alpha)\cos (M\beta) \cos(M\alpha') \cos (N\beta ') |M,-N\rangle\langle M,-N|\\
&=4\sum_{M,N} \cos(N \alpha)\cos (M\beta) \cos(M\alpha') \cos (N\beta ') |M,N\rangle\langle M,N|\\
&= \sum_{M,N} \left( \cos[N(\alpha - \beta')] \cos[M(\beta- \alpha')] + \cos[N(\alpha - \beta')] \cos[M(\beta + \alpha')]\right.  \\
    &\left. + \cos[N(\alpha + \beta')] \cos[M(\beta - \alpha')] + \cos[N(\alpha + \beta')] \cos[M(\beta + \alpha')]\right)|N,M\rangle\langle N,M|\nonumber\\
    & = \frac{1}{2}\mathcal{E}_{(1,1)}(\alpha-\beta',\beta-\alpha') + \frac{1}{2}\mathcal{E}_{(1,1)}(\alpha-\beta',\beta+ \alpha') \\
    & + \frac{1}{2}\mathcal{E}_{(1,1)}(\alpha + \beta',\beta-\alpha') + \frac{1}{2}\mathcal{E}_{(1,1)}(\alpha + \beta',\beta + \alpha').
\end{split}
\end{equation}
Inserting the above into Eq. (\ref{OO}) we get 
\begin{align}
     O_{(1,-1)}(\alpha,\beta) O_{(1,-1)}(\alpha',\beta') & = \frac{1}{2}{E}_{(1,1)}(\alpha-\beta',\beta-\alpha') + \frac{1}{2}E_{(1,1)}(\alpha-\beta',\beta+ \alpha')\nonumber \\
    & + \frac{1}{2}{E}_{(1,1)}(\alpha + \beta',\beta-\alpha') + \frac{1}{2}{E}_{(1,1)}(\alpha + \beta',\beta + \alpha').
\end{align}

Using a similar procedure we obtain the other two odd-odd fusion products at the self-dual radius.
\begin{align}
     O_{(1,1)}(\alpha,\beta) O_{(1,-1)}(\alpha',\beta') & =  \frac{1}{2}{E}_{(1,-1)}(\beta -\alpha',\alpha-\beta') + \frac{1}{2}{E}_{(1,-1)}(\beta -\alpha',\alpha + \beta')\nonumber \\
    & + \frac{1}{2}{E}_{(1,-1)}(\beta + \alpha',\alpha -\beta') + \frac{1}{2}{E}_{(1,-1)}(\beta + \alpha',\alpha + \beta').
\end{align}
\begin{align}
     O_{(1,-1)}(\alpha,\beta) O_{(1,1)}(\alpha',\beta') & = \frac{1}{2}{E}_{(1,-1)}(\beta -\alpha',\alpha-\beta') + \frac{1}{2}{E}_{(1,-1)}(\beta -\alpha',\alpha + \beta')\nonumber \\
    & + \frac{1}{2}{E}_{(1,-1)}(\beta + \alpha',\alpha -\beta') + \frac{1}{2}{E}_{(1,-1)}(\beta + \alpha',\alpha + \beta').
\end{align}

We also have fusion products between the odd and even defects as given below.
\begin{align}
     E_{(1,1)}(\alpha,\beta)O_{(1,1)}(\alpha',\beta ') &=  \frac{1}{2}{O}_{(1,1)}(\beta+\alpha',\alpha+ \beta') + \frac{1}{2}{O}_{(1,1)}(\beta - \alpha',\alpha+ \beta')\nonumber \\
    & + \frac{1}{2}{O}_{(1,1)}(\beta + \alpha',\alpha-\beta') + \frac{1}{2}{O}_{(1,1)}(\beta - \alpha',\alpha - \beta').
\end{align}
\begin{align}
     {O}_{(1,1)}(\alpha,\beta){E}_{(1,1)}(\alpha',\beta') & =  \frac{1}{2}{O}_{(1,1)}(\alpha+\alpha',\beta + \beta') + \frac{1}{2}{O}_{(1,1)}(\alpha + \alpha',\beta - \beta')\nonumber \\
    & + \frac{1}{2}{O}_{(1,1)}(\alpha - \alpha',\beta + \beta') + \frac{1}{2}{O}_{(1,1)}(\alpha - \alpha',\beta - \beta').
\end{align}
\begin{align}
     E_{(1,1)}(\alpha,\beta)O_{(1,-1)}(\alpha',\beta ') &=  \frac{1}{2}{O}_{(1,-1)}(\beta+\alpha',\alpha+ \beta') + \frac{1}{2}{O}_{(1,-1)}(\beta - \alpha',\alpha+ \beta')\nonumber \\
    & + \frac{1}{2}{O}_{(1,-1)}(\beta + \alpha',\alpha-\beta') + \frac{1}{2}{O}_{(1,-1)}(\beta - \alpha',\alpha - \beta').
\end{align}
\begin{align}
     {O}_{(1,-1)}(\alpha,\beta){E}_{(1,1)}(\alpha',\beta')&=  \frac{1}{2}{O}_{(1,1)}(\alpha+\alpha',\beta + \beta') + \frac{1}{2}{O}_{(1,1)}(\alpha + \alpha',\beta - \beta')\nonumber \\
    & + \frac{1}{2}{O}_{(1,1)}(\alpha - \alpha',\beta + \beta') + \frac{1}{2}{O}_{(1,1)}(\alpha - \alpha',\beta - \beta').
\end{align}
\begin{align}
     E_{(1,-1)}(\alpha,\beta)O_{(1,1)}(\alpha',\beta ') &=  \frac{1}{2}{O}_{(1,-1)}(\beta+\alpha',\alpha+ \beta') + \frac{1}{2}{O}_{(1,-1)}(\beta - \alpha',\alpha+ \beta')\nonumber \\
    & + \frac{1}{2}{O}_{(1,-1)}(\beta + \alpha',\alpha-\beta') + \frac{1}{2}{O}_{(1,-1)}(\beta - \alpha',\alpha - \beta').
\end{align}
\begin{align}
     {O}_{(1,1)}(\alpha,\beta){E}_{(1,-1)}(\alpha',\beta')& = \frac{1}{2}{O}_{(1,-1)}(\alpha+\alpha',\beta+ \beta') + \frac{1}{2}{O}_{(1,-1)}(\alpha + \alpha',\beta - \beta')\nonumber \\
    & + \frac{1}{2}{O}_{(1,-1)}(\alpha - \alpha',\beta+ \beta') + \frac{1}{2}{O}_{(1,-1)}(\alpha - \alpha',\beta - \beta').
\end{align}
\begin{align}
     E_{(1,-1)}(\alpha,\beta)O_{(1,-1)}(\alpha',\beta ') &=  \frac{1}{2}{O}_{(1,1)}(\beta+\alpha',\alpha+ \beta') + \frac{1}{2}{O}_{(1,1)}(\beta - \alpha',\alpha+ \beta')\nonumber \\
    & + \frac{1}{2}{O}_{(1,1)}(\beta + \alpha',\alpha-\beta') + \frac{1}{2}{O}_{(1,1)}(\beta - \alpha',\alpha - \beta').
\end{align}
\begin{align}
     {O}_{(1,-1)}(\alpha,\beta){E}_{(1,-1)}(\alpha',\beta')
     & = \frac{1}{2}{O}_{(1,1)}(\alpha+\alpha',\beta+ \beta') + \frac{1}{2}{O}_{(1,1)}(\alpha + \alpha',\beta - \beta') \nonumber\\
    & + \frac{1}{2}{O}_{(1,1)}(\alpha - \alpha',\beta+\beta') + \frac{1}{2}{O}_{(1,1)}(\alpha - \alpha',\beta - \beta').
\end{align}

%%%%%%%%%%%%%%%%%%%%%%%%%%%%%%%%%%%%%%%%%%%%%%%%%%%
%
%  New template code for TAMU Theses and Dissertations starting Fall 2016.
%
%  Author: Sean Zachary Roberson
%	 Version 3.17.01
%  Last updated 1/10/2017
%
%%%%%%%%%%%%%%%%%%%%%%%%%%%%%%%%%%%%%%%%%%%%%%%%%%%
%%%%%%%%%%%%%%%%%%%%%%%%%%%%%%%%%%%%%%%%%%%%%%%%%%%%%%%%%%%%%%%%%%%%%%
%%                           SECTION IV
%%%%%%%%%%%%%%%%%%%%%%%%%%%%%%%%%%%%%%%%%%%%%%%%%%%%%%%%%%%%%%%%%%%%%

\chapter{SUMMARY AND CONCLUSIONS \label{cha:Summary}}

The non-supersymmetric space $\mathbb{C}/\mathbb{Z}_d$ is the exemplary model for the study of tachyon condensation in string theory. By studying topological defects between these non-compact orbifolds we have found defects which encode the bulk RG flow that drives the process of tachyon condensation. Besides the bulk RG flow, the algebraic language of the defects found here provides a simplified way to tackle the boundary RG flow. The latter is a topic not yet well understood in string theory in general and our work gives more evidence to the adequacy of defects in this area. The discussion of describing the boundary RG flow in terms of defects was first exploited in \cite{brunner07a} for the superconformal models $\mathcal{M}_{d-2}/\mathbb{Z}_d$ but their treatment needed spacetime supersymmetry to avoid dealing with regularization. Our work shows that one can do away with such an assumption without having to invoke a  regularization scheme when working with the non-compact orbifolds.

In \cite{hori00a}, Hori and Vafa argued  that the twisted chiral
sector of LG orbifolds is independent of the concrete superpotential
term. Starting from there, one would first expect that in the cases of
$\mathbb{C}/\mathbb{Z}_d$  and a LG with superpotential $W=X^d$, the spectrum of the twisted sectors (i.e., the $(a,c)$-rings) of both theories agree. Hence, one can map the perturbations of the two models to each
other. The work presented in this dissertation shows that the previous conclusion can be taken further, namely that the flows between the flat models follow a similar pattern to the ones with superpotential. 

In the bosonic, compact orbifolds at $c=1$ we have provided a fuller picture of possible defects for these theories which now include the twisted part of the spectrum. By mapping out what are the possible defects in this important class of 2D theories other questions can be tackled such as the RG flow in these models. Our work shows that the topological defects form a closed algebra. By computing this fusion algebra, we have explicitly shown the way in which the symmetry breaks from the quantum groups upon orbifolding.

In this dissertation we have explored the spectrum and applications of conformal and topological defects in   $\mathbb{C}/\mathbb{Z}_d$  and  $S^1_R/\mathbb{Z}_2$ orbifold theories. We have constructed topological defects which implement the action of the RG flow between $\mathbb{C}/\mathbb{Z}_n$  theories.  The language we have employed to describe the RG flow defects is the natural description for such objects in the framework of Landau-Ginzburg models and their orbifolds. As we reviewed in section \ref{BtypeSec}, this description involves factorizing the superpotentials of the given theories over different polynomial rings. 

The language of matrix factorizations for boundaries and defects has been shown to carry over to the case of a zero superpotential. The matrix factorizations we used in this case were obtained by setting $p_0=0$ in those given in \cite{brunner07a}. This is a very natural choice since it relates matrix factorizations in the $\mathbb{C}/\mathbb{Z}_d$ models to another method of characterizing D-branes.  Indeed, a common description of D-branes in geometric spaces (when there is no superpotential) is via chain complexes of vector bundles, with a differential $d$ built from the BRST operator $Q$ \cite{aspinwall}. On the other hand, out of the matrix factorizations associated with the D-branes in the Landau-Ginzburg models one obtains 2-periodic twisted complexes by taking the differentials to be the factorizing maps $p_1$ and $p_0$. Therefore with $p_0=0$,  $W\rightarrow 0$ produces an ordinary complex which coincides with above description for the D-branes.  It would be interesting to make this connection precise in a more general context\footnote{We thank Ilka Brunner for emphasizing this connection to us.}.

We have put forth two different ways of checking that the defects we posit here indeed enforce the RG flow between the non-compact orbifolds. One method uses the chiral rings of the theories at hand, and their deformations. The other method is a geometrical description of A-branes which are the equivalent representation of B-type boundary conditions in the mirror theory. Both methods keep track of the RG flow and show that the endpoints are $\mathbb{C}/\mathbb{Z}_n$ orbifolds. The defects $P^{(m,\underline n)}$ of subsection \ref{rgflowDefects} are shown to be appropriate interfaces between any two such orbifolds. 

By studying the fusion rules we showed that we can use these defects to tackle the question of the boundary RG flow when the theory has a nontrivial worldsheet boundary. In this note we provided evidence that the defects $P^{(m,\underline n)}$ successfully map the boundary conditions associated with the IR theory $\mathbb{C}/\mathbb{Z}_n$, to those of the UV theory $\mathbb{C}/\mathbb{Z}_n'$, $n'<n$. We established such correspondence by working with the mirror theory of the non-compact orbifolds. In this picture, we can compare the action of the RG flow defects on the B-type D-branes with the action of the RG flow on the dual A-type D-branes, i.e., A-branes. In comparing with the work of \cite{brunner07a}, we have shown that the RG flows between the $\mathbb{C}/\mathbb{Z}_d$ models follow a similar pattern to that of the LG orbifolds with a superpotential turned on.

Although we checked that RG flow defects properly describe the bulk-induced boundary RG flow by going to the mirror description in subsection \ref{comparison}, a similar comparison can be done between the result of the fusion rules and the flow of the deformed relation of the chiral ring given in equation (\ref{deformed}). This can be done by considering the quotient relation of the chiral ring in equation (\ref{deformed}) as a branched covering of the complex plane. Such a description would provide an equivalent geometrical formalism to that of the deformed A-branes, so that an analysis could be done along the lines of the one done in Subsection \ref{comparison} for the A-branes.

A different approach to building conformal defects in these non-compact orbifolds is via the unfolding procedure employed here for the compact orbifold case. In this method one constructs the boundary states corresponding to D-branes in the target space  $\mathbb{C}/\mathbb{Z}_n \times \mathbb{C}/\mathbb{Z}_{n'}$. These states can be mapped to defects between the theories $\mathbb{C}/\mathbb{Z}_{n}$ and $\mathbb{C}/\mathbb{Z}_{n'}$ via the unfolding procedure. An interesting question would be to find an equivalent description of the RG flow defects presented here in terms of a representation as Hilbert space operators. 

For the compact orbifold $S^1/\mathbb{Z}_2$ we have identified the spectrum of possible conformal defects which glue together these theories. The first step was to obtain solutions for the boundary states of the product theory $(S^1/\mathbb{Z}_2)^2$, and secondly to unfold these elements to the defects. Although not exhaustive, this work and that of \cite{bachas07, fuchs07} taken together map out the possible defects between the $c=1$, 2D CFTs. This classification makes this family of field theories the one whose defects are best classified. 

In our construction of D-branes for the product theory $(S^1/\mathbb{Z}_2)^2$ we encountered that the boundary states come in three varieties: untwisted, partially twisted, and twisted depending on whether we used twisted fields in none, one, or both of the directions of the orbifold. These same varieties translate to the defects upon unfolding. The untwisted defects presented here are the $\mathbb{Z}_2 \times \mathbb{Z}_2$-symmetrized versions of defects for the circle theory in \cite{bachas07}. The analysis of the untwisted D-branes and defects guides our analysis for the fully twisted ones.

The obvious next step to explore the full algebra between the defects and all possible boundary conditions cataloged in this note. Given the prolific amount of these objects upon orbifolding the circle, we have left this step out of our project. Furthermore, defects have been shown to feel a Casimir force between themselves, and between a defect and a boundary \cite{Brunner:2010xm, bachas02}. It would be interesting to study this attribute in the presence of the twisted and partially twisted defects presented here. Lastly, now that defects have been written for both branches of the $c=1$, 2D CFTs, and that we have an understanding of twisted degrees of freedom on defects it would be very interesting to explore defects gluing $S^1$ and $S^1/\mathbb{Z}_2$ theories.
%The next line is the format for inserting new sections.
%Replace the name "newsection"  with the name of your
%new section file.
%\include{data/newsection}

%fix spacing in bibliography, if any...
%%%%%%%%%%%%%%%%%%%%%%%%%%%%%%%%%%%%%%%%%%%%%%%%%%%%%%%%%%%%%
\let\oldbibitem\bibitem
\renewcommand{\bibitem}{\setlength{\itemsep}{0pt}\oldbibitem}
%%%%%%%%%%%%%%%%%%%%%%%%%%%%%%%%%%%%%%%%%%%%%%%%%%%%%%%%%%%%%%%	
%The bibliography style declared is the IEEE format. If
%you require a different style, see the document
%bibstyles.pdf included in this package. This file,
%hosted by the University of Vienna, shows several
%bibliography styles and examples of in-text citation
%and a references page.

%%%%%%%%%%%%ADDED
%\bibliographystyle{ieeetr}

\phantomsection
\addcontentsline{toc}{chapter}{REFERENCES}

\renewcommand{\bibname}{{\normalsize\rm REFERENCES}}

%This file is a .bib database that contains the sources.
%This removes the dependency on the previous file
%bibliography.tex.

\bibliographystyle{utphys}
\bibliography{BibliographyDefectsLandauGinzburg}

\providecommand{\href}[2]{#2}\begingroup\raggedright\begin{thebibliography}{10}

\bibitem{belavin}
A.~Belavin and A.~Polyakov, ``{Infinite conformal symmetry in two-dimensional
  quantum field theory},''
{\em Nucl. Phys.} {\bfseries B241} (1984) .
%%CITATION = COND-MAT/9612187;%%.

\bibitem{cardy}
J.~Cardy, ``{Conformal symmetry and critical surface behavior},''
{\em Nucl. Phys.} {\bfseries B240} (1984) .
%%CITATION = COND-MAT/9612187;%%.

\bibitem{Recknagel:1998ih}
A.~Recknagel and V.~Schomerus, ``{Boundary deformation theory and moduli spaces
  of D-branes},'' \href{http://dx.doi.org/10.1016/S0550-3213(99)00060-7}{{\em
  Nucl. Phys.} {\bfseries B545} (1999) 233--282},
\href{http://arxiv.org/abs/hep-th/9811237}{{\ttfamily arXiv:hep-th/9811237
  [hep-th]}}.
%%CITATION = HEP-TH/9811237;%%.

\bibitem{Brunner:1999}
I.~Brunner, R.~Entin, and C.~Romelsberger, ``{D-branes on T**4 / Z(2) and T
  duality},'' \href{http://dx.doi.org/10.1088/1126-6708/1999/06/016}{{\em JHEP}
  {\bfseries 06} (1999) 016},
\href{http://arxiv.org/abs/hep-th/9905078}{{\ttfamily arXiv:hep-th/9905078
  [hep-th]}}.
%%CITATION = HEP-TH/9905078;%%.

\bibitem{Gaberdiel:2001zq}
M.~R. Gaberdiel and A.~Recknagel, ``{Conformal boundary states for free bosons
  and fermions},'' \href{http://dx.doi.org/10.1088/1126-6708/2001/11/016}{{\em
  JHEP} {\bfseries 11} (2001) 016},
\href{http://arxiv.org/abs/hep-th/0108238}{{\ttfamily arXiv:hep-th/0108238
  [hep-th]}}.
%%CITATION = HEP-TH/0108238;%%.

\bibitem{Gaberdiel:2008rk}
M.~R. Gaberdiel and O.~Schlotterer, ``{Bulk induced boundary perturbations for
  N=1 superconformal field theories},''
  \href{http://dx.doi.org/10.1088/1751-8113/42/11/115209}{{\em J. Phys.}
  {\bfseries A42} (2009) 115209},
\href{http://arxiv.org/abs/0810.4719}{{\ttfamily arXiv:0810.4719 [hep-th]}}.
%%CITATION = ARXIV:0810.4719;%%.

\bibitem{affleck}
M.~Oshikawa and I.~Affleck, ``{Boundary conformal field theory approach to the
  critical two-dimensional Ising model with a defect line},''
  \href{http://dx.doi.org/10.1016/S0550-3213(97)00219-8}{{\em Nucl. Phys.}
  {\bfseries B495} (1997) 533--582},
\href{http://arxiv.org/abs/cond-mat/9612187}{{\ttfamily arXiv:cond-mat/9612187
  [cond-mat]}}.
%%CITATION = COND-MAT/9612187;%%.

\bibitem{bachas02}
C.~Bachas, J.~de~Boer, R.~Dijkgraaf, and H.~Ooguri, ``{Permeable conformal
  walls and holography},''
  \href{http://dx.doi.org/10.1088/1126-6708/2002/06/027}{{\em JHEP} {\bfseries
  06} (2002) 027},
\href{http://arxiv.org/abs/hep-th/0111210}{{\ttfamily arXiv:hep-th/0111210
  [hep-th]}}.
%%CITATION = HEP-TH/0111210;%%.

\bibitem{bachas07}
C.~Bachas and I.~Brunner, ``{Fusion of conformal interfaces},''
  \href{http://dx.doi.org/10.1088/1126-6708/2008/02/085}{{\em JHEP} {\bfseries
  02} (2008) 085},
\href{http://arxiv.org/abs/0712.0076}{{\ttfamily arXiv:0712.0076 [hep-th]}}.
%%CITATION = ARXIV:0712.0076;%%.

\bibitem{fuchs07}
J.~Fuchs, M.~R. Gaberdiel, I.~Runkel, and C.~Schweigert, ``{Topological defects
  for the free boson CFT},''
  \href{http://dx.doi.org/10.1088/1751-8113/40/37/016}{{\em J. Phys.}
  {\bfseries A40} (2007) 11403},
\href{http://arxiv.org/abs/0705.3129}{{\ttfamily arXiv:0705.3129 [hep-th]}}.
%%CITATION = ARXIV:0705.3129;%%.

\bibitem{gaiotto12}
D.~Gaiotto, ``{Domain walls for two-dimensional renormalization group flows},''
  \href{http://dx.doi.org/10.1007/JHEP12(2012)103}{{\em JHEP} {\bfseries 12}
  (2012) 103},
\href{http://arxiv.org/abs/1201.0767}{{\ttfamily arXiv:1201.0767 [hep-th]}}.
%%CITATION = ARXIV:1201.0767;%%.

\bibitem{Quella:2002CT}
T.~Quella and V.~Schomerus, ``{Symmetry breaking boundary states and defect
  lines},'' \href{http://dx.doi.org/10.1088/1126-6708/2002/06/028}{{\em JHEP}
  {\bfseries 06} (2002) 028},
\href{http://arxiv.org/abs/hep-th/0203161}{{\ttfamily arXiv:hep-th/0203161
  [hep-th]}}.
%%CITATION = HEP-TH/0203161;%%.

\bibitem{brunner03}
I.~Brunner, M.~Herbst, W.~Lerche, and B.~Scheuner, ``{Landau-Ginzburg
  realization of open string TFT},''
  \href{http://dx.doi.org/10.1088/1126-6708/2006/11/043}{{\em JHEP} {\bfseries
  11} (2006) 043},
\href{http://arxiv.org/abs/hep-th/0305133}{{\ttfamily arXiv:hep-th/0305133
  [hep-th]}}.
%%CITATION = HEP-TH/0305133;%%.

\bibitem{brunner07}
I.~Brunner and D.~Roggenkamp, ``{B-type defects in Landau-Ginzburg models},''
  \href{http://dx.doi.org/10.1088/1126-6708/2007/08/093}{{\em JHEP} {\bfseries
  0708} (2007) 093},
\href{http://arxiv.org/abs/0707.0922}{{\ttfamily arXiv:0707.0922 [hep-th]}}.
%%CITATION = ARXIV:0707.0922;%%.

\bibitem{Konechny:2015qla}
A.~Konechny, ``{Fusion of conformal interfaces and bulk induced boundary RG
  flows},'' \href{http://dx.doi.org/10.1007/JHEP12(2015)114}{{\em JHEP}
  {\bfseries 12} (2015) 114},
\href{http://arxiv.org/abs/1509.07787}{{\ttfamily arXiv:1509.07787 [hep-th]}}.
%%CITATION = ARXIV:1509.07787;%%.

\bibitem{Graham:2003nc}
K.~Graham and G.~M.~T. Watts, ``{Defect lines and boundary flows},''
  \href{http://dx.doi.org/10.1088/1126-6708/2004/04/019}{{\em JHEP} {\bfseries
  04} (2004) 019},
\href{http://arxiv.org/abs/hep-th/0306167}{{\ttfamily arXiv:hep-th/0306167
  [hep-th]}}.
%%CITATION = HEP-TH/0306167;%%.

\bibitem{Fuchs:2015ska}
J.~Fuchs and C.~Schweigert, ``{Surface defects and symmetries},''
\href{http://dx.doi.org/10.1088/1742-6596/597/1/012002}{{\em J. Phys. Conf.
  Ser.} {\bfseries 597} no.~1, (2015) 012002}.
%%CITATION = 00462,597,012002;%%.

\bibitem{frohlich09}
J.~Frohlich, J.~Fuchs, I.~Runkel, and C.~Schweigert, ``{Defect lines,
  dualities, and generalised orbifolds},'' in {\em {Proceedings, 16th
  International Congress on Mathematical Physics (ICMP09): Prague, Czech
  Republic, August 3-8, 2009}}.
\newblock 2009.
\newblock
\href{http://arxiv.org/abs/0909.5013}{{\ttfamily arXiv:0909.5013 [math-ph]}}.
\newblock
%%CITATION = ARXIV:0909.5013;%%.

\bibitem{dorey00}
P.~Dorey, M.~Pillin, R.~Tateo, and G.~M.~T. Watts, ``{One point functions in
  perturbed boundary conformal field theories},''
  \href{http://dx.doi.org/10.1016/S0550-3213(00)00622-2}{{\em Nucl. Phys.}
  {\bfseries B594} (2001) 625--659},
\href{http://arxiv.org/abs/hep-th/0007077}{{\ttfamily arXiv:hep-th/0007077
  [hep-th]}}.
%%CITATION = HEP-TH/0007077;%%.

\bibitem{keller07}
C.~A. Keller, ``{Brane backreactions and the Fischler-Susskind mechanism in
  conformal field theory},''
  \href{http://dx.doi.org/10.1088/1126-6708/2007/12/046}{{\em JHEP} {\bfseries
  12} (2007) 046},
\href{http://arxiv.org/abs/0709.1076}{{\ttfamily arXiv:0709.1076 [hep-th]}}.
%%CITATION = ARXIV:0709.1076;%%.

\bibitem{hori04}
K.~Hori, ``{Boundary RG flows of N=2 minimal models},'' in {\em {Calabi-Yau
  varieties and mirror symmetry. Proceedings, Workshop, Mirror Symmetry 5,
  Banff, Canada, December 6-11, 2003}}, pp.~381--405.
\newblock 2004.
\newblock
\href{http://arxiv.org/abs/hep-th/0401139}{{\ttfamily arXiv:hep-th/0401139
  [hep-th]}}.
\newblock
%%CITATION = HEP-TH/0401139;%%.

\bibitem{fredenhagen06}
S.~Fredenhagen, M.~R. Gaberdiel, and C.~A. Keller, ``{Bulk induced boundary
  perturbations},'' \href{http://dx.doi.org/10.1088/1751-8113/40/1/F03}{{\em J.
  Phys.} {\bfseries A40} (2007) F17},
\href{http://arxiv.org/abs/hep-th/0609034}{{\ttfamily arXiv:hep-th/0609034
  [hep-th]}}.
%%CITATION = HEP-TH/0609034;%%.

\bibitem{brunner07a}
I.~Brunner and D.~Roggenkamp, ``{Defects and bulk perturbations of boundary
  Landau-Ginzburg orbifolds},''
  \href{http://dx.doi.org/10.1088/1126-6708/2008/04/001}{{\em JHEP} {\bfseries
  04} (2008) 001},
\href{http://arxiv.org/abs/0712.0188}{{\ttfamily arXiv:0712.0188 [hep-th]}}.
%%CITATION = ARXIV:0712.0188;%%.

\bibitem{enger05}
H.~Enger, A.~Recknagel, and D.~Roggenkamp, ``{Permutation branes and linear
  matrix factorisations},''
  \href{http://dx.doi.org/10.1088/1126-6708/2006/01/087}{{\em JHEP} {\bfseries
  0601} (2006) 087},
\href{http://arxiv.org/abs/hep-th/0508053}{{\ttfamily arXiv:hep-th/0508053
  [hep-th]}}.
%%CITATION = HEP-TH/0508053;%%.

\bibitem{hori00}
K.~Hori, ``{Linear models of supersymmetric D-branes},'' in {\em {Symplectic
  geometry and mirror symmetry. Proceedings, 4th KIAS Annual International
  Conference, Seoul, South Korea, August 14-18, 2000}}, pp.~111--186.
\newblock 2000.
\newblock
\href{http://arxiv.org/abs/hep-th/0012179}{{\ttfamily arXiv:hep-th/0012179
  [hep-th]}}.
\newblock
%%CITATION = HEP-TH/0012179;%%.

\bibitem{vafa01}
C.~Vafa, ``{Mirror symmetry and closed string tachyon condensation},''
\href{http://arxiv.org/abs/hep-th/0111051}{{\ttfamily arXiv:hep-th/0111051
  [hep-th]}}.
%%CITATION = HEP-TH/0111051;%%.

\bibitem{hori00a}
K.~Hori and C.~Vafa, ``{Mirror symmetry},''
\href{http://arxiv.org/abs/hep-th/0002222}{{\ttfamily arXiv:hep-th/0002222
  [hep-th]}}.
%%CITATION = HEP-TH/0002222;%%.

\bibitem{harvey01}
J.~A. Harvey, D.~Kutasov, E.~J. Martinec, and G.~W. Moore, ``{Localized
  tachyons and RG flows},''
\href{http://arxiv.org/abs/hep-th/0111154}{{\ttfamily arXiv:hep-th/0111154
  [hep-th]}}.
%%CITATION = HEP-TH/0111154;%%.

\bibitem{adams01}
A.~Adams, J.~Polchinski, and E.~Silverstein, ``{Don't panic! Closed string
  tachyons in ALE space-times},''
  \href{http://dx.doi.org/10.1088/1126-6708/2001/10/029}{{\em JHEP} {\bfseries
  10} (2001) 029},
\href{http://arxiv.org/abs/hep-th/0108075}{{\ttfamily arXiv:hep-th/0108075
  [hep-th]}}.
%%CITATION = HEP-TH/0108075;%%.

\bibitem{Ginsparg:1988ui}
P.~H. Ginsparg, ``{Applied conformal field theory},'' in {\em {Les Houches
  Summer School in Theoretical Physics: Fields, Strings, Critical Phenomena Les
  Houches, France, June 28-August 5, 1988}}, pp.~1--168.
\newblock 1988.
\newblock
\href{http://arxiv.org/abs/hep-th/9108028}{{\ttfamily arXiv:hep-th/9108028
  [hep-th]}}.
\newblock
%%CITATION = HEP-TH/9108028;%%.

\bibitem{hori02}
K.~Hori, ``{Trieste lectures on mirror symmetry},'' in {\em {Superstrings and
  related matters. Proceedings, Spring School, Trieste, Italy, March 18-26,
  2002}}, pp.~109--202.
\newblock
2002.
\newblock
%%CITATION = INSPIRE-608888;%%.

\bibitem{Hori0005}
K.~Hori, A.~Iqbal, and C.~Vafa, ``{D-branes and mirror symmetry},''
\href{http://arxiv.org/abs/hep-th/0005247}{{\ttfamily arXiv:hep-th/0005247
  [hep-th]}}.
%%CITATION = HEP-TH/0005247;%%.

\bibitem{warner95}
N.~P. Warner, ``{Supersymmetry in boundary integrable models},''
  \href{http://dx.doi.org/10.1016/0550-3213(95)00402-E}{{\em Nucl. Phys.}
  {\bfseries B450} (1995) 663--694},
\href{http://arxiv.org/abs/hep-th/9506064}{{\ttfamily arXiv:hep-th/9506064
  [hep-th]}}.
%%CITATION = HEP-TH/9506064;%%.

\bibitem{Orlov:2003yp}
D.~Orlov, ``{Triangulated categories of singularities and D-branes in
  Landau-Ginzburg models},''
\href{http://arxiv.org/abs/math/0302304}{{\ttfamily arXiv:math/0302304
  [math-ag]}}.
%%CITATION = MATH/0302304;%%.

\bibitem{Lazaroiu:2003zi}
C.~I. Lazaroiu, ``{On the boundary coupling of topological Landau-Ginzburg
  models},'' \href{http://dx.doi.org/10.1088/1126-6708/2005/05/037}{{\em JHEP}
  {\bfseries 05} (2005) 037},
\href{http://arxiv.org/abs/hep-th/0312286}{{\ttfamily arXiv:hep-th/0312286
  [hep-th]}}.
%%CITATION = HEP-TH/0312286;%%.

\bibitem{kapustin02}
{Kapustin, Anton and Li, Yi}, ``{D branes in Landau-Ginzburg models and
  algebraic geometry},''
  \href{http://dx.doi.org/10.1088/1126-6708/2003/12/005}{{\em JHEP} {\bfseries
  12} (2003) 005},
\href{http://arxiv.org/abs/hep-th/0210296}{{\ttfamily arXiv:hep-th/0210296
  [hep-th]}}.
%%CITATION = HEP-TH/0210296;%%.

\bibitem{Hori:2000kt}
K.~Hori and C.~Vafa, ``{Mirror symmetry},''
\href{http://arxiv.org/abs/hep-th/0002222}{{\ttfamily arXiv:hep-th/0002222
  [hep-th]}}.
%%CITATION = HEP-TH/0002222;%%.

\bibitem{Douglas:1996sw}
M.~R. Douglas and G.~W. Moore, ``{D-branes, quivers, and ALE instantons},''
\href{http://arxiv.org/abs/hep-th/9603167}{{\ttfamily arXiv:hep-th/9603167
  [hep-th]}}.
%%CITATION = HEP-TH/9603167;%%.

\bibitem{aspinwall}
P.~S. Aspinwall, ``{D-branes on Calabi-Yau manifolds},'' in {\em {Progress in
  string theory. Proceedings, Summer School, TASI 2003, Boulder, USA, June
  2-27, 2003}}, pp.~1--152.
\newblock 2004.
\newblock
\href{http://arxiv.org/abs/hep-th/0403166}{{\ttfamily arXiv:hep-th/0403166
  [hep-th]}}.
\newblock
%%CITATION = HEP-TH/0403166;%%.

\bibitem{Brunner:2010xm}
I.~Brunner and D.~Roggenkamp, ``{Attractor flows from defect lines},''
  \href{http://dx.doi.org/10.1088/1751-8113/44/7/075402}{{\em J. Phys.}
  {\bfseries A44} (2011) 075402},
\href{http://arxiv.org/abs/1002.2614}{{\ttfamily arXiv:1002.2614 [hep-th]}}.
%%CITATION = ARXIV:1002.2614;%%.

\end{thebibliography}\endgroup
%%%%%%%%%%%%%%%%%%%%%%%%%%%%%%%%%

%This next line includes appendices. The file
%appendix.tex contains commands pointing to
%the appendix files; be sure to change these
%pointers if you end up changing the filenames.
%Leave this commented if you will not need
%appendix material.
%\include{data/appendices}

\end{document}